\documentclass[11pt]{article}
\usepackage{booktabs}
\usepackage{slashed}
\usepackage{jheppub}
\usepackage{orcidlink}
\usepackage{graphicx}
\usepackage{amsfonts}
\usepackage{amssymb}
\usepackage{xcolor}
\usepackage{amsmath}
\usepackage{multirow}
\usepackage{appendix}
\usepackage{cancel}
\usepackage{float}

\preprint{CERN-TH-2026-046}

\title{Constraining the real scalar singlet extension of the SM}
\author[a]{Moritz Bosse~\orcidlink{0009-0001-6339-0478},}
\author[b,a]{Gudrun Hiller~\orcidlink{0000-0002-4419-3612},}
\author[c]{Daniel Litim~\orcidlink{0000-0001-9963-5345},}
\author[b,d,e]{Fabio Maltoni~\orcidlink{0000-0003-4890-0676},}
\author[f,g,h]{Michael J. Ramsey-Musolf~\orcidlink{0000-0001-8110-2479},}
\author[d]{Simone Tentori~\orcidlink{0000-0002-2398-9056},}
\author[f,g]{Guotao Xia~\orcidlink{0009-0006-1847-4935}}

\affiliation[a]{TU Dortmund University, Department of Physics, Otto-Hahn-Str.4, D-44221 Dortmund, Germany}

\affiliation[b]{Theoretical Physics Department, CERN, 1211 Geneva 23, Switzerland}

\affiliation[c]{{Department of Physics and Astronomy, University of  Sussex, Brighton, BN1 9QH, U.K.}}

\affiliation[d]{Centre for Cosmology, Particle Physics and Phenomenology (CP3),
Universit\'{e} Catholique de Louvain, B-1348 Louvain-la-Neuve, Belgium}

\affiliation[e]{Dipartimento di Fisica e Astronomia, Universit\`{a} di Bologna and INFN, Sezione di Bologna, Via Irnerio 46, 40126 Bologna, Italy}

\affiliation[f]{Tsung-Dao Lee Institute \& School of Physics and Astronomy, Shanghai Jiao Tong University, Shanghai 200240, China}
\affiliation[g]{Shanghai Key Laboratory for Particle Physics and Cosmology, Key Laboratory for Particle Astrophysics and Cosmology (MOE), Shanghai Jiao Tong University, Shanghai 200240, China}
\affiliation[h]{Kellogg Radiation Laboratory, California Institute of Technology, Pasadena, CA 91125, USA}

\emailAdd{moritz.bosse@tu-dortmund.de}
\emailAdd{gudrun.hiller@cern.ch}
\emailAdd{d.litim@sussex.ac.uk}
\emailAdd{fabio.maltoni@uclouvain.be}
\emailAdd{mjrm@sjtu.edu.cn}
\emailAdd{simone.tentori@uclouvain.be}
\emailAdd{xiagt-summer@sjtu.edu.cn}

\newcommand{\abs}[1]{\left\lvert #1 \right\rvert}

\newcommand{\mone}{m_{h_1}}
\newcommand{\mtwo}{m_{h_2}}
\newcommand{\omuH}{\overline{\mu_H}}
\newcommand{\omuS}{\overline{\mu_S}}
\newcommand{\lH}{\lambda_{H}}
\newcommand{\olH}{\overline{\lambda_{H}}}

\newcommand{\lS}{b_4}
\newcommand{\olS}{\overline{b_4}}

\newcommand{\vS}{v_{S}}
\newcommand{\lM}{a_2}
\newcommand{\olM}{\overline{a_2}}
\newcommand{\mthree}{b_{3}}
\newcommand{\omthree}{\overline{b_{3}}}
\newcommand{\mfour}{\frac{a_1}{2}}
\newcommand{\omfour}{\frac{\overline{a_1}}{2}}

\abstract{The real scalar singlet extension of the standard model provides a minimal framework in which the Higgs sector can realise a strong first-order electroweak phase transition and improve the stability of the electroweak vacuum. We combine the electroweak phase transition and high-scale vacuum stability with current and projected collider probes, including precision Higgs measurements, the Higgs trilinear coupling, EWPO  and resonant searches for a heavy singlet-like scalar in $ZZ$ and di-Higgs final states. Focusing on a singlet heavier than the standard model Higgs, we find that there is parameter space compatible with a strong first-order electroweak phase transition for singlet-like scalar masses up to nearly 1 TeV. Deviations in the Higgs self-coupling can be  larger than those in the Higgs--$Z$ coupling, making Higgs-potential measurements a key probe. We find that the HL-LHC will test a large fraction of the parameter space, while the FCC will provide ultimate  discovery and model-discrimination capabilities. }

\begin{document}
\maketitle

 
\section{Introduction}

Among  the simplest and most motivated extensions of the Standard Model (SM) is the addition of a real scalar singlet field,   coupled to the Higgs sector via the so-called Higgs portal.  Despite its minimality, this model, henceforth denoted by xSM, exhibits a rich phenomenology, both from a collider and a cosmological perspective. It allows a strong first-order electroweak phase transition (SFOEWPT),  thereby offering a potential mechanism for successful EW baryogenesis and generation of primordial gravitational waves. In the presence of a $Z_2$ symmetry,  
the singlet scalar can also contribute to the dark matter relic density.
The cosmological requirements on the scalar singlet mass and Higgs portal couplings render the xSM eminently testable, with a plethora of associated collider signatures~\cite{Profumo:2007wc,Barger:2007im}. Moreover, the model can stabilise the SM vacuum against high-scale instabilities, addressing the near-criticality of the EW vacuum~\cite{Gonderinger:2009jp}. 

In the SM, the Higgs boson mass and self-coupling parameters result in a  crossover EW symmetry-breaking transition. However, a strong first-order EW phase transition is a key ingredient for successful EW baryogenesis, potentially providing a dynamical origin of the cosmic matter-antimatter asymmetry. Furthermore, the SM predicts a metastable vacuum, raising questions about the ultimate stability of the EW ground state, particularly when extrapolated to very high energies~\cite{Buttazzo:2013uya}.
Introducing a scalar singlet modifies both the  transition dynamics and the vacuum structure of the theory. The singlet coupling to the Higgs sector can catalyse the occurrence of an EW phase transition and make it strongly first order, and contribute to stabilising the vacuum up to the Planck scale.

For these reasons, the xSM has been extensively investigated from both phenomenological and cosmological perspectives. Ref.~\cite{Ramsey-Musolf:2019lsf} provides a broad overview of the model and a compendium of references. Early studies addressed its LHC signatures and the connection between collider probes and the electroweak phase transition at future facilities~\cite{OConnell:2006rsp,Barger:2007im,Profumo:2014opa,Curtin:2014jma,Kotwal:2016tex,Hashino:2016xoj}. More recent analyses have revisited the collider and gravitational-wave phenomenology of the xSM with updated theoretical tools and experimental projections~\cite{Hashino:2016xoj,Zhang:2023jvh,Ramsey-Musolf:2024ykk,Braathen:2025svl,Aboudonia:2024frg}. Finally, non-perturbative lattice studies of the electroweak phase transition and its phenomenological implications have been carried out in Refs.~\cite{Gould:2019qek,Niemi:2024axp}.

In this work, we systematically revisit the constraints on this model, focusing on the most recent Higgs couplings determinations~\cite{CMS:2018uag,ATLAS:2022vkf} and prospects from precision Higgs measurements and direct searches at current~\cite{CMS:2025hfp} and future colliders~\cite{deBlas:2025gyz}
and the metastability of the EW vacuum at high energy scales~\cite{Gonderinger:2009jp,Elias-Miro:2012eoi,Hiller:2022rla,Hiller:2024zjp}.
Special emphasis is placed on regions of parameter space capable of realising a strong first-order electroweak phase transition (SFOEWPT), see for instance \cite{Profumo:2007wc,Buttazzo:2013uya,Curtin:2014jma,Profumo:2014opa,Hashino:2016xoj,Kurup:2017dzf,Niemi:2024axp,Zhang:2023jvh,Ramsey-Musolf:2024ykk,Braathen:2025svl}, while simultaneously maintaining vacuum stability~\cite{Gonderinger:2009jp,Hiller:2024zjp}. We reiterate that a strong first-order phase transition can also generate gravitational-wave signatures~\cite{Hindmarsh:2013xza,Hashino:2016xoj,Espinosa:2018eve,Ellis:2018mja,Ramsey-Musolf:2024ykk}, potentially detectable by LISA~\cite{Caprini:2019egz}, Taiji~\cite{Hu:2017mde}, and TianQin~\cite{TianQin:2015yph}.
To facilitate the study of its collider phenomenology, the real singlet model is implemented in \textsc{MadGraph5\_aMC@NLO}~\cite{Alwall:2014hca} via \textsc{FeynRules}~\cite{Degrande:2011ua,Darme:2023jdn} and \textsc{NLOCT}~\cite{Degrande:2014vpa}, enabling the computation of cross sections at NLO QCD accuracy, as well as loop-induced processes. For the BSM renormalisation-group evolution analysis, we employ the precision tool \textsc{ARGES}~\cite{Litim:2020jvl}.

Our study combines indirect constraints from Higgs precision measurements, EW precision observables (EWPOs), requirements arising from a strong first order electroweak phase transition (SFOEWPT), and direct collider searches. In addition, we assess constraints from vacuum stability at high scales and cosmological considerations in order to delineate the viable parameter space. We demonstrate that, while current data already exclude  a fraction of the parameter space, the upcoming high-luminosity phase of the LHC and future collider projects will be able to probe it much more comprehensively. While the enhanced xSM reach of the high-luminosity LHC and prospective future colliders has been analysed previously, we highlight the novelty of our present study and its complementarity with recent work:

\begin{itemize}
\item To identify the cosmologically viable parameter space, we combine a state-of-the-art treatment of the EWPT, utilising the dimensionally reduced high-temperature effective theory and non-perturbative results, with a vacuum-stability analysis based on three-loop gauge and two-loop scalar and Yukawa RG evolution.
\item We show in Figs.~\ref{fig:minbrrun2}--\ref{fig:mhl00_indirect} the interplay between these cosmological considerations and distinct collider probes: resonant di-Higgs production; heavy (singlet-like) Higgs production in the $ZZ$ final state; deviations of the Higgs triple self-coupling from its SM value; and the corresponding deviations of the Higgs-$Z$ boson coupling. In doing so, we overlay the constraints from different phenomenological probes rather than performing a statistical combination. 

\item Our results complement Ref.~\cite{Zhang:2023jvh}, which combined resonant di-Higgs and heavy-Higgs searches at the HL-LHC but did not include their interplay with precision Higgs probes. They also update and extend Ref.~\cite{Kotwal:2016tex}, which studied the complementarity between precision Higgs measurements and resonant di-Higgs searches at the HL-LHC and a prospective $\mathcal{O}(100~\text{TeV})$ proton collider. We further include electroweak precision observables in the assessment of future lepton-collider reach. Resonant di-Higgs production is treated in the narrow width approximation; resonant--non-resonant interference effects, which can be relevant especially at low singlet-like Higgs masses, are left for future dedicated work.

\item We identify a phenomenologically challenging ``funnel region'' in the EWPT-viable xSM parameter space, 
clarify its connection with distinct early-Universe thermal histories, and discuss possible handles through loop-induced effects and 
complementary collider probes.
\end{itemize}

The structure of this paper is as follows. In Section~\ref{sec:generalmodel}, we introduce the real scalar singlet extension of the SM, define our parametrisation, and fix the notation used throughout the paper. In Section~\ref{sec:theory}, we discuss the theoretical requirements imposed by the electroweak phase transition and by vacuum stability up to high scales. This includes a review of possible xSM thermal histories, their connection with strong first-order electroweak symmetry breaking, and the role of Higgs--singlet mixing and portal interactions. Section~\ref{sec:constraints} is devoted to experimental probes. We discuss gravitational-wave signals from the electroweak phase transition, indirect constraints from Higgs precision measurements, direct searches for the heavy scalar, and the current combined constraints from LHC Run~II together with the theoretical requirements. In Section~\ref{sec:impact}, we assess the impact of future colliders, focusing on the HL-LHC and on the FCC-ee+FCC-hh programme. We conclude in Section~\ref{sec:con}. In the Appendix we give two equivalent and common parametrisations of the scalar sector, together with expressions for trilinear scalar couplings. We attach to the paper additional material, containing the UFO model implementation and testing, along with some instructions on its use, and the list of Feynman rules.

\section{The minimal scalar extension of the SM}\label{sec:generalmodel}

The minimal extension of the SM by a real scalar singlet constitutes one of the simplest, yet theoretically rich, modifications of the Higgs sector. The addition of a real singlet field $S$ to the SM particle content introduces new interactions through the scalar potential, which can substantially modify the vacuum structure and the nature of the EWPT.

The most general renormalisable scalar potential contains linear, cubic and quartic terms in the singlet field $S$, in addition to Higgs--singlet interaction terms. 
In the literature, two main parametrisations of the singlet-extended scalar potential are commonly employed:
\begin{itemize}
\item
\emph{Explicit linear term:} the singlet field does not acquire a vacuum expectation value (vev), and 
a term linear in  $S$
is included explicitly in the scalar potential; 
\item
\emph{Explicit vev:} spontaneous symmetry breaking induces a non-vanishing vev for the singlet field, which in turn generates linear terms in $S$ dynamically.
\end{itemize}

Although the formulations described above encode the same physical content, they correspond to different choices of input parameters. These choices can impact the transparency and convenience of specific calculations, such as the derivation of Feynman rules and the evaluation of collider observables. In particular, introducing both an explicit linear term and a non-vanishing singlet vev is redundant, since one can always be absorbed into the other through an appropriate field redefinition.

In this section, we introduce the {explicit linear term} parametrisation, which will be adopted throughout the remainder of this work. The alternative {explicit vev} parametrisation, together with its relation to our chosen formulation, is discussed in Appendix~\ref{app:vev}. We derive the relevant expressions for the scalar potential, the mass matrix, the mixing angle, and the mapping between parameters, providing the basis for subsequent discussions of phenomenology, vacuum stability, and collider searches. 

We emphasise the practical advantages of working in the {explicit linear term} formulation, which typically leads to simpler analytical expressions and facilitates the extraction of physical predictions. In particular, as shown in Eq.~\eqref{eq:b4bar}, within the {explicit vev} parametrisation the relation between the singlet vev $v_S$ and the  BSM quartic coupling $b_4$ involves a cumbersome cubic equation. The origin and implications of this redundancy are discussed at the end of Appendix~\ref{sec:paramet}.

In the absence of an explicit $\mathbb{Z}_2$ symmetry, which would forbid odd powers of $S$,
the most general renormalisable scalar potential reads~\cite{Profumo:2007wc,Ramsey-Musolf:2024ykk} 
\begin{equation}\label{eq:xSM}
V(\Phi,S) = - {\mu}^2_H |\Phi|^2 + \lambda_H |\Phi|^4 +   b_1 S - \frac{\mu_S^2}{2} S^2  +     \frac{ {b_4}}{4} S^4 +\frac{ {a_2}}{2}  |\Phi|^2 S^2 + \frac{ {b_3}}{3}S^3  + \frac {a_1}{2} |\Phi|^2 S\,.
\end{equation}
Denoting the neutral Higgs and singlet scalar vacuum expectation values as $v_H$ and $v_S$, respectively, one has
\begin{equation}\label{eq:fieldexp}
\Phi=\begin{pmatrix}
- iG^+ \\
(v_H + H + iG^0)/\sqrt{2}
\end{pmatrix}, \qquad
S = (v_S + s).
\end{equation}
As shown in Ref.~\cite{Profumo:2007wc} and in the Appendix, retention of non-vanishing $v_S$ and the $b_1$ term entails redundancy, as one may perform a linear shift $S\to S-\beta$ to eliminate $b_1$ in terms of $v_S$ or vice-versa for suitable choice of $\beta$. In what follows we keep $b_1$ explicit and set $v_S=0$. 

The Goldstone bosons $G^0$, $G^\pm$ are unaffected, as in the SM, while the physical Higgs sector experiences mixing between $H$ and $s$. After electroweak symmetry breaking (EWSB), tadpole terms for both $H$ and $s$ appear:
\begin{equation}
 \mathcal{L}\supset   \mathcal{I}_s s + \mathcal{I}_H H\,,
\end{equation}
which are eliminated by imposing $\mathcal{I}_s = \mathcal{I}_H = 0$, leading to the relations:

\begin{equation}\label{eq:tadpolea2}
b_1 = -\frac{1}{4} v_H^2 a_1, \qquad \mu_H^2 =  \lambda_H v_H^2\,. 
\end{equation}
The resulting mass matrix in the $(H,s)$ basis becomes:
\begin{equation} \label{eq:M2}
\mathcal{M}^2 = \begin{pmatrix} 2\lambda_H v_H^2 & \frac{a_1}{2} v_H \\ \frac{a_1}{2} v_H & \frac{a_2}{2} v_H^2 - \mu_S^2 \end{pmatrix}.
\end{equation}
Defining:
\begin{equation}\label{eq:masspotential}
m_H^2 \equiv 2\lambda_H v_H^2, \quad m_S^2 \equiv \frac{a_2}{2} v_H^2 - \mu_S^2, \quad m_{HS}^2 \equiv \frac{a_1}{2} v_H,
\end{equation}
the mass matrix can be written in a compact form
\begin{equation}
\begin{pmatrix}
m_H^2 & m^2_{HS} \\
m^2_{HS} & m_S^2 \label{eq:massmatrix}
\end{pmatrix} \,,
\end{equation}
and diagonalised via a rotation:
\begin{equation}
\begin{pmatrix} h_1 \\ h_2 \end{pmatrix} = \begin{pmatrix} \cos\theta & -\sin\theta \\ \sin\theta & \cos\theta \end{pmatrix} \begin{pmatrix} H \\ s \end{pmatrix}, \quad \tan 2\theta = \frac{2 m_{HS}^2}{m_S^2 - m_H^2}.
\end{equation}
The physical masses of the scalar eigenstates read:
\begin{align}
m^2_{h_{1,2}} &= \frac{1}{2} \left[ m_H^2 + m_S^2 \mp \sqrt{(m_H^2 - m_S^2)^2 + 4 m_{HS}^4} \right]\\
&=\frac{1}{2}\left(m_H^2+m_S^2\pm \frac{m_H^2-m_S^2}{\cos 2 \theta}\right)\label{eq:massdiag} \,. 
\end{align}
The parameters can be re-expressed in terms of the physical masses and the mixing angle $\theta$:
\begin{align} \label{eq:a1vstheta}
a_1 &= \frac{\mtwo^2-\mone^2}{v_H}\sin 2\theta\,,\\ \label{eq:lHtheta}
\lH &= \frac{1}{4v_H^2} \left[\mone^2+\mtwo^2-(\mtwo^2-\mone^2)\cos 2\theta \right]\,,\\
\mu_S^2 &= \frac{1}{2}\left[a_2 v_H^2-(\mone^2+\mtwo^2)-(\mtwo^2-\mone^2)\cos 2\theta \right] \,.\label{eq:musa2}
\end{align}
The sign convention for $\theta$  may differ in the literature. For example, a different convention is used between the expression obtained here and in  \cite{Niemi:2024axp} and those of Refs. \cite{Huang:2016cjm,Profumo:2014opa}.
Throughout, we assume that the observed scalar boson is the lighter one, that is
$m_{h_1} \simeq 125 \, \text{GeV}$. Numerically, $v_H \simeq 246 \, \text{GeV}$ and
$\lambda_{H,\text{SM}} =m_{h_1}^2/(2 v_H^2) \simeq 0.13$.

As mentioned above, a presentation of the {explicit vev} parametrisation, together with a detailed discussion of its relation to the {explicit linear term} formulation, is provided in Appendix~\ref{app:vev}.

\subsection{SM Higgs  coupling modification}\label{sec:Higgs_coupling_mod}
A direct consequence of the mixing between the two scalar states is a universal modification of the Higgs couplings to SM particles: both Yukawa couplings and couplings to EW gauge bosons are rescaled by the same factor:
\begin{equation}\label{eq:SMcouplingmod}
y_f \rightarrow y_f \cos\theta, \quad g_{hVV} \rightarrow  g_{hVV} \cos\theta\,.
\end{equation}
We note that, at leading order, Eq.~\eqref{eq:SMcouplingmod} allows one to trade the mixing angle for the deviation of the Higgs coupling to SM particles. In particular, focusing on the coupling to the $Z$ boson, one finds
\begin{equation}\label{eq:delta_gz}
\abs{\delta g_{hZZ}} = 1- \cos\theta = \frac{\theta^2}{2}+\mathcal{O}(\theta^4)\,. 
\end{equation}
Not only the Higgs coupling to other particles is modified, but also the trilinear Higgs self-coupling:
\begin{align}
\kappa_3 = \frac{g_{h_1 h_1 h_1}}{g_{h_1 h_1 h_1}^{\mathrm{SM}}} &= \cos^3\theta + a_2 \frac{v_H^2}{\mone^2} \cos \theta \sin^2\theta - 2  \sin^3\theta\frac{b_3}{3} \frac{v_H }{m_{h_1}^2}\,.
\label{eq:k3lin}
\end{align}
We  examine the behaviour in the small-mixing regime by expanding $\kappa_3$ in powers of the mixing angle:
\begin{align} \kappa_3 = & 1-\left(\frac{3}{2} - a_2 \frac{v_H^2}{\mone^2}\right) \theta ^2-\left(2\frac{b_3}{3}\frac{v_H}{\mone^2}\right)\theta^3+O\left(\theta ^4\right)\,.\label{eq:k3_linearapprox} 
\end{align}
We learn that in this {explicit linear term} formulation $\kappa_3$ depends on a single parameter up to  order $\theta^2$. This should be contrasted with the {explicit vev} formulation, cf.~Eq.~\eqref{eq:k3_vsapprox}, which involves additional parameters. This provides further motivation for our choice of parametrisation.

Vacuum stability requirements suggest that the portal coupling $a_2$ is generically positive.
While this requirement is not strict, negative values are limited in magnitude and require tuning \cite{Hiller:2024zjp}, which will be discussed further in Section \ref{sec:stability}.
Values of $\kappa_3$ smaller than unity can hence be obtained for sufficiently small portal couplings, see Eq.~(\ref{eq:k3_linearapprox}), as
\begin{equation} \kappa_3<1 \, \Rightarrow \, a_2 \le 0.4 +O\left(\theta ^3\right). \end{equation}
Since the $\theta^3$ term depends exclusively on $b_3$, sufficiently large values of $b_3$ can  enlarge or reduce, depending on the relative sign of $\theta$ and $b_3$, the range of $a_2$ for which $\kappa_3 < 1$.
On the other hand, increasing the portal coupling $a_2$ can enhance the cubic Higgs coupling
relative to the SM, $\kappa_3 >1$.

The alignment limit ($\theta \to 0$) also  enables  a simple relation between the two Higgs couplings. 
We obtain from Eq.~\eqref{eq:delta_gz} 
\begin{equation}
\label{eq:k3dgz}
\delta\kappa_3 =  \kappa_3-1=\left(  \frac{a_2}{\lambda_{H,{\rm SM}}}-3\right)\delta g_{hZZ} +\mathcal{O}(\theta^3)\,,
\end{equation}
which shows that deviations in the Higgs trilinear self-coupling can naturally be $\mathcal{O}(10)$ times larger than 
the corresponding deviations in the Higgs--$Z$  coupling once the $a_2$ value is larger than $\sim 2$ (condition valid for all of the points inducing a SFOEWPT once $\mtwo$ is large enough).

We stress that all relations presented in this section are derived at leading order (LO) in the EW couplings.  In the following, whenever $\kappa_3$ or $\delta g_{hZZ}$ appear in plots, they are understood to be LO defined quantities.
We focus only on a heavy singlet-like scalar, $m_{h_2}>m_{h_1}$, that does not couple to additional BSM lighter particles. Nevertheless, if one of these two assumptions is removed more searches can be performed at colliders \cite{ATLAS:2022vkf,CMS:2022dwd,ATLAS:2021hza,CMS:2025gzf,ATLAS:2022abz,CMS:2026zsp}. Of particular interest for the EWPT is the $\mone> 2 \mtwo$ regime wherein exotic Higgs decays provide a novel xSM EWPT signature. See Refs.~\cite{Carena:2019une,Kozaczuk:2019pet,Carena:2022yvx,Wang:2022dkz} for detailed analyses of this regime.

\subsection{The alignment limit}\label{sec:alignment}

We have previously introduced the concept of the alignment limit in which $\theta\rightarrow 0$. 
In this limit, one finds from Eqs.~(\ref{eq:tadpolea2}), (\ref{eq:a1vstheta}) that $a_1 \to 0$ and $b_1 \to 0$, so that the only remaining $\mathbb{Z}_2$-breaking term in the Lagrangian is $b_3$.

The alignment limit is therefore closely related to the $\mathbb{Z}_2$-symmetric limit, in which the number of independent parameters in the Lagrangian~\eqref{eq:xSM} is reduced: imposing a $\mathbb{Z}_2$ symmetry under which the singlet field $S$ is odd, whereas all other fields are even, forces $b_1 = b_3 = a_1 = 0$ leading to $\theta=0$.
On the other hand, unlike the $\mathbb{Z}_2$ symmetric case, in the alignment limit of the general model ($\theta$=0) the $b_3$ term remains. (In the 
explicit vev-parametrisation, this would correspond to a finite singlet vev, see Eq.~(\ref{eq:b3andvS}).) As can be verified from the Feynman rules given in the supplementary material, the $b_3$ contribution drops out of all LO amplitudes except for the trilinear $h_2 h_2 h_2$ interaction, this statement, however, holds only at leading order, as the set of operators arising in the alignment limit does not close under renormalisation. The tadpole diagram arising from the $b_3 S^3$ interaction will generically induce the $b_1 S$ operator, while the combination of $b_3 S^3$ and $a_2 |\Phi|^2 S^2$ will generate $|\Phi|^2 S$ at one-loop order. In principle, one may choose a renormalisation condition that enforces vanishing finite parts of $b_1$, $a_1$, and $\theta$ to all orders, a choice that will be implicit in the following discussion.

At first sight, the alignment scenario might appear phenomenologically trivial, since $\delta g_{hZZ} \rightarrow 0,\,  \delta \kappa_3 \rightarrow  0$ and no deviations in Higgs SM couplings are generated at leading order. The apparent triviality however disappears due to the presence of the portal interaction governed by the parameter $a_2$. The interactions arising from this term, in particular the $h_1 h_2 h_2$ and $h_1 h_1 h_2 h_2$ vertices, remain non-vanishing even for $\theta=0$.\footnote{This can be explicitly verified, again, by taking $\theta \to 0$ in the Feynman rules given in the additional material.} 
As a consequence, non-trivial effects arise at the loop level, including contributions to the running of couplings, where the quartic singlet coupling $b_4$ also plays a crucial role \cite{Hiller:2024zjp}, as will be discussed later in Section \ref{sec:funnel}.

\section{Theoretical requirements: EWPT and vacuum stability \label{sec:theory}}

Beyond its collider phenomenology, the real scalar singlet extension of the SM is theoretically motivated by its impact on EW symmetry breaking and the stability of the Higgs potential. In particular, the singlet can qualitatively alter the dynamics of EW symmetry breaking, allowing for a strong first-order EW phase transition, and can simultaneously stabilise the SM vacuum through renormalisation-group (RG) effects at high scales. These features address two of the main shortcomings of the SM when extrapolated to early-Universe cosmology and high energies.

In this section, we systematically analyse the theoretical requirements that delineate the viable regions of the model. We first investigate the conditions under which a SFOEWPT can occur, employing dimensionally reduced effective field theories and non-perturbative insights where available
(Section \ref{sec:EWSB}). We then study in Section \ref{sec:stabilityUV} the stability of the scalar potential under RG evolution up to high scales, assessing whether the presence of the singlet can stabilise the EW vacuum without encountering Landau poles or instabilities. Finally, we discuss in Section \ref{sec:funnel} the structure and physical interpretation of the resulting allowed regions, including the emergence of funnel-like configurations in parameter space.

Together, these theoretical considerations play a central role in shaping the phenomenology of the model and provide a theoretically well-motivated target for the experimental probes discussed in the subsequent  sections.

\subsection{EWSB phase transition \label{sec:EWSB}}

The nature of the EWPT plays a crucial role in understanding the early universe: a strong first-order transition is required for electroweak baryogenesis and can be a source of gravitational waves. Here, we review the theoretical description of the EWPT using dimensionally reduced 3D effective field theories and lattice methods, following Refs.~\cite{Zhang:2023jvh,Niemi:2024axp,Ramsey-Musolf:2024ykk, Niemi:2024vzw}, and summarise current results relevant for probing the dynamics of the transition.

\begin{figure}
\begin{center}
\includegraphics[width=0.8\textwidth]{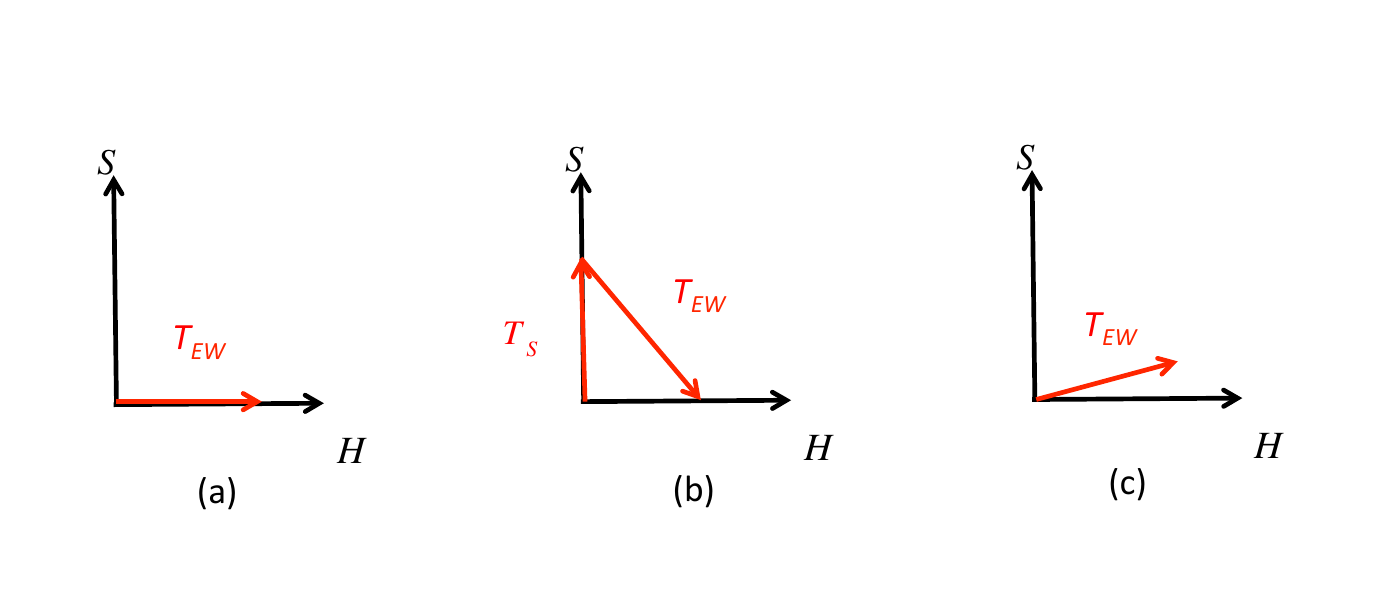}
\end{center}
\vskip-.5cm
\caption{Representative thermal histories of EWSB in the presence of the Higgs field $H$ and the singlet scalar $S$. (a) A one-step transition to the pure Higgs phase at the electroweak temperature $T_{EW}$; (b) a two-step transition, with a first step to the singlet vacuum at $T_S > T_{EW}$ followed by a second step to the pure Higgs phase; (c) a one-step transition at temperature $T_{EW}$ to a mixed phase in which both $H$ and $S$ have non-zero vacuum expectation values. Note that in our parametrisation, the singlet has no vev at $T=0$ but it may have a non-zero vev at $T>0$. \label{fig:scenarios}}
\end{figure}

To provide context for the following analysis, results, and discussion, it is useful to review the thermal histories that could have occurred in an xSM universe. To that end, we refer to Fig.~\ref{fig:scenarios} (reproduced from Ref.~\cite{Ramsey-Musolf:2019lsf}). Scenario (a) denotes a one-step transition at the electroweak temperature $T_\mathrm{EW}$ to a vacuum in which only the Higgs field has non-vanishing vev. In scenarios (b) and (c), the singlet  field may also obtain a non-zero vev at $T>0$, even though in our parametrisation it has no vev at $T=0$. Thus, in the \lq\lq two-step\rq\rq\ scenario of panel (b) the first step involves a transition at temperature $T_S$ to a phase in which only $S$ has a non-zero vev, followed by a second step at $T_\mathrm{EW}$ to a vacuum in which the doublet vev is non-vanishing. At this step the $S$ vev in general may be non-zero, eventually relaxing to a vanishing vev at zero temperature. The one-step transition in panel (c) leads to a vacuum in which both $S$ and $h$ obtain non-zero vevs, with the former relaxing to zero at $T=0$.


In the SM, only the one-step history (a) occurs, and the corresponding transition is a smooth crossover. In the presence of the portal interactions $|\Phi|^2S$ and $|\Phi|^2 S^2$,  thermal loops may render the transition first-order. Scenarios (b) and (c) require the explicit presence of a BSM scalar. In the two-step scenario (b), thermal loop effects may allow the first step at $T_S$ to be either first order or crossover, while the second step is generically first order due to a zero-temperature, tree-level barrier induced by the cross-quartic interaction $|\Phi|^2 S^2$ (with coefficient $a_2$). For scenario (c), the cubic interaction $|\Phi|^2 S$ may also give rise to a zero-temperature, tree-level barrier if the coefficient $a_1$ is negative.\footnote{Cosmological impurities may also catalyse a FOEWPT, as illustrated recently in Refs.~\cite{Blasi:2022woz,Agrawal:2023cgp}.}

One may now draw the connections between collider phenomenology and the EWPT scenarios, as first delineated in \cite{Profumo:2007wc} and again in \cite{Ramsey-Musolf:2019lsf}. For scenario (a), either the magnitudes of the portal couplings $a_{1,2}$ or the number of additional scalars must be sufficiently large to yield a sizeable thermal loop-induced barrier. Since in our parametrisation a non-vanishing doublet-singlet mixing angle requires $a_1\not=0$, a large $|\Phi|^2 S$ thermal loop effect would nominally imply sizeable mixing. On the other hand, in the limit of vanishing $a_1$, the cross-quartic interaction could yield a sizeable barrier effect if the value of $\mu_S^2$ is chosen to cancel the thermal loop corrections to the singlet scalar mass at $T\sim T_\mathrm{EW}$. In this case, the $|\Phi|^2 S^2$ interaction yields a $-T H^3 $ contribution to the potential that acts as a thermal barrier. The cancellation requirement on $\mu_S^2$ then leads to an approximate upper bound on the $T=0$ singlet mass of a few hundred GeV, in principle putting it well within the reach of the LHC and future colliders. This cancellation scenario was originally realised in the context of the minimal supersymmetric standard model and applied to the top squark thermal mass contribution, leading to the so-called \lq\lq light stop scenario\rq\rq (LSS) with $m_{\tilde t}\sim 100$ GeV \cite{Carena:1996wj,Delepine:1996vn,Carena:1997ki}. While the LSS has been ruled out by non-observation of light stops at the LHC, the cancellation mechanism for thermal history (a) remains viable for singlet or electroweak multiplet scalars~\cite{Ramsey-Musolf:2019lsf}.

For the two-step scenario (b), the requirement that $T_S>T_\mathrm{EW}$ again leads to a rough upper bound on the BSM scalar mass of several hundred GeV, a generic feature realised in a wide variety of concrete model scenarios~\cite{Ramsey-Musolf:2019lsf}. The first-order transition to the Higgs phase at $T_\mathrm{EW}$ arises from the $|\Phi|^2 S^2$-induced tree-level barrier. This scenario has been validated at the non-perturbative level for the real triplet extension of the SM~\cite{Niemi:2020hto}. While the analogous non-perturbative study of the xSM did not explicitly delineate this scenario~\cite{Niemi:2024axp}, we have done so as part of the present work. As we discuss below, we find ample choices of the corresponding portal coupling $a_2$ that yield a strongly first-order second-step transition in scenario (b). Since in our parametrisation the singlet receives no $T=0$ vev, this interaction yields no doublet-singlet mixing and, therefore, no tree-level modifications of the $hZZ$ or $hhh$ couplings. Thus, one may obtain a SFOEWPT to the Higgs phase in either the second step of scenario (b) or the cancellation scenario of scenario (a) without appreciable effects in precision studies of associated production ($g_{hZZ}$) or non-resonant di-Higgs production ($\kappa_3$). The resulting \lq\lq funnel regions\rq\rq\, in Figs.~\ref{fig:FOPTpoints},\ref{fig:stabilityZ2},  \ref{fig:stability_full}, and \ref{fig:FOPTgw} below correspond to this situation, which has not been previously appreciated in the literature. On the other hand, the tree-level barrier-induced SFOEWPT of scenario (b) may yield a potentially observable signal in gravitational-wave searches.  

A SFOEWPT in scenario (c) generally requires a non-zero $a_1$ with negative sign and sufficiently large magnitude. The magnitude of the associated non-zero doublet-singlet mixing angle is generically bounded below: $|\sin\theta|\gtrsim 0.01$~\cite{Ramsey-Musolf:2019lsf}. While fine-tuned cancellations may circumvent this bound, one would then expect $|\delta g_{hZZ}|$ and $|\kappa_3 - 1|$ to be larger than a few times $10^{-4}$ and $10^{-3}$, respectively (see Eqs.~(\ref{eq:delta_gz}) and (\ref{eq:k3dgz})). Again, for choices of $|\sin\theta|$ near this generic lower bound, the currently envisioned precision Higgs probes may not exhibit appreciable sensitivity. We also find below that gravitational-wave searches with LISA-level sensitivity would not provide access to this region of xSM parameter space. More extensive studies will be required to assess the sensitivity of prospective beyond-LISA generation gravitational-wave searches to this scenario (c) region. 

Overall, with these general features of the thermal history-collider phenomenology connection in the xSM in mind, we find that the vast majority of SFOEWPT viable parameter space could be probed with a combination of LHC, next-generation collider, and gravitational-wave searches. The challenging funnel region notwithstanding, the experimental testability and discovery potential for the SFOEWPT-viable xSM are significant. We now turn to a detailed discussion of the thermal quantum field theory and collider phenomenology that supports these generic conclusions.

\subsubsection{Review of 3D effective theory}
The effective potential $V_\text{eff}$ governs EWPT thermodynamics, encoding quantities such as the critical temperature $T_c$, transition strength $\Delta v/T_c$, and latent heat $L/T^4$. However, perturbative calculations of $V_\text{eff}$ in non-Abelian gauge theories at finite temperature suffer from infrared (IR) issues~\cite{Linde:1980ts,Gross:1980br}. In the symmetric phase, all fields except the Higgs doublet are massless, leading to IR divergences in bosonic propagators with small momenta in loop integrals. The high-$T$ expansion parameter is formally $g^2 n_B$ \cite{Laine:2016hma}, where $n_B(E) = \frac{1}{e^{E/T}-1}$ is the Bose distribution function and $g$ is a generic weak coupling. For $T\gg E$, $n_B \approx T/E$, and with the dispersion relation $E = \sqrt{\omega_n^2 + m^2+\mathbf{k}^2}$ at zero Matsubara mode and zero mass, it becomes $T/|\mathbf{k}|$. Perturbation theory requires $|\mathbf{k}|\gtrsim g^2T$ to ensure the expansion parameter remains below $1$; modes with $|\mathbf{k}|\lesssim g^2 T$ are non-perturbative, and convergence is slow even slightly above this scale. In contrast, fermionic and non-zero bosonic Matsubara modes acquire thermal masses $\sim \pi T$, enabling reliable perturbative treatment. Thermal resummation generates dynamical thermal masses that dampen these IR divergences \cite{Pisarski:1988vd,Braaten:1989mz}. However, it has drawbacks: higher-loop calculations are cumbersome, and if the loop-corrected mass approaches zero (e.g., Higgs mass near $T_c$), IR divergences persist. The 3D effective theory approach is often more practical, naturally resumming IR divergences in its construction procedure.

The EWPT exhibits a scale hierarchy: the hard scale ($|\mathbf{k}|\sim\pi T$), the soft scale ($|\mathbf{k}|\sim gT$) at which temporal gauge fields acquire Debye masses, and the ultrasoft scale ($|\mathbf{k}| \sim g^2 T$). Dimensional reduction exploits this hierarchy. For static thermodynamics, hard modes ($|\mathbf{k}|\geq\pi T$) are treated perturbatively, with their contributions matched to bosonic zero modes, resulting in a 3D bosonic effective theory. Specifically, a general renormalisable effective theory is formulated, and its parameters are determined by matching perturbatively calculated two-, three-, and four-point functions between the effective theory and the original four-dimensional theory \cite{Kajantie:1995dw,Schicho:2021gca,Niemi:2021qvp}. The effective theory then has the same infrared behaviour as the original theory. Similarly, the effective theory can be further simplified by integrating over soft scales $gT$. 

We point out that lattice simulations of the 3D effective theory are much easier because: i) the purely bosonic theory avoids the fermion sign problem on the lattice; ii) the super-renormalisability of the 3D theory allows exact analytical relations between continuum and lattice actions in the continuum limit \cite{Niemi:2024axp,Laine:1995np}; iii) handling multiple EWPT energy scales in 4D lattice numerically is much more demanding than in 3D simulations. The downside is that the accuracy of the perturbative 4D$\to$3D matching sets an upper bound on the reliability of the full analysis, though this is well controlled provided the zero-temperature couplings remain perturbative. The super-renormalisability of the 3D theory further renders two-loop calculations feasible, making perturbative analyses within the 3D EFT a useful improvement over calculations based on the four-dimensional effective potential.
Lattice simulations give direct access to the electroweak phase transition. They sample field configurations at each temperature without relying on an effective potential or loop integrals, so they are free from infrared divergences. The gauge-invariant Higgs condensate $\langle \Phi^\dagger \Phi \rangle$ serves as the order parameter. Near the critical temperature $T_c$, its probability distribution develops two peaks. The peak at the smaller value of $\langle \Phi^\dagger \Phi \rangle$ corresponds to the symmetric phase, while the peak at the larger value corresponds to the broken phase. At $T_c$ the two peaks carry equal weight, meaning both phases are equally likely. The strength of the transition is measured by $\sqrt{2\,\Delta \langle \Phi^\dagger \Phi \rangle}/T$.

Previous work~\cite{Niemi:2024axp} has shown that two-loop perturbative calculations within the 3D effective theory agree well with lattice results for strong first-order phase transitions, while they fail to capture crossover behaviour in weak transitions. Successful electroweak baryogenesis further requires sufficient suppression of the sphaleron rate in the broken phase~\cite{Li:2025kyo, Annala:2025aci}; since the sphaleron energy scales roughly with the Higgs condensate, the condition $v/T_c > 1$ provides a conservative and widely used proxy for sufficient suppression. We therefore focus exclusively on strong transitions ($v/T_c > 1$) and employ the two-loop 3D EFT to scan the singlet-extended SM parameter space, expecting results that are reliable at the level of lattice simulations.

\subsubsection{DRxSM and EWPT}

The starting point of dimensional reduction in the  xSM (DRxSM) is the scalar potential of the singlet-extended SM given in Eq.~\eqref{eq:xSM}. In the fermionic sector, we retain only the top quark, as Yukawa couplings of the lighter fermions are negligible and the top quark provides the dominant SM contribution to the thermal Higgs mass.

Our EW inputs follow the conventions of Refs.~\cite{Niemi:2021qvp,Niemi:2024axp}, adopting the PDG values~\cite{ParticleDataGroup:2024cfk} for the gauge boson masses, the top-quark pole mass, the SM Higgs pole mass, Fermi's constant, and the strong coupling constant. For the BSM singlet sector, the singlet scalar pole mass $m_{h_2}$ and the mixing angle $\sin\theta$ are inputs, while $b_3$, $b_4$ and $a_2$ are chosen freely in the $\overline{\text{MS}}$ scheme. One-loop corrections to the four-dimensional zero-temperature parameters are required, as they are parametrically of the same order as the two-loop thermal effects in the three-dimensional effective theory. Accordingly, the SM and singlet-sector input parameters are converted to $\overline{\text{MS}}$-renormalised quantities at one-loop order. We then verify the absolute stability of the electroweak vacuum at the input scale by numerically minimising the zero-temperature Coleman--Weinberg potential, ensuring that the global minimum is located at $v \simeq 246\,\mathrm{GeV}$.

The dimensional reduction proceeds in two steps (see Ref.~\cite{Niemi:2021qvp} for a detailed derivation). We first integrate out the hard modes with $k \gtrsim \pi T$. In practice, this is achieved by perturbatively matching static two-, three-, and four-point correlation functions between the original four-dimensional theory and a super-renormalisable three-dimensional effective theory, thereby fixing the parameters of the latter. During this step, the four-dimensional parameters are evolved using one-loop RG equations to the scale $\tilde{\mu} = 4\pi T e^{-\gamma} \simeq 7T$, which removes logarithmic contributions arising from thermal sum-integrals. Any residual scale dependence in the matching procedure enters only at higher orders. After the hard modes have been removed, the temporal components of the gauge fields acquire Debye masses at the soft scale $gT$, leading to the screening of electric fields at finite temperature. Secondly, these soft modes with momenta $k \gtrsim gT$ are integrated out in an analogous manner. For the 3D $\overline{\text{MS}}$ renormalisation scale we set $\tilde{\mu}_3 = T$ \cite{Niemi:2021qvp} and use RG-evolved parameters in loop corrections \cite{Farakos:1994kx}.

The resulting high-$T$ effective theory is the super-renormalisable SU(2) $\times$ U(1) + Higgs + singlet theory
\begin{equation}
    S_{3\mathrm{D}}=\frac{1}{T}\int d^3x\left\{\frac{1}{4}F_{ij}^aF_{ij}^a+\frac{1}{4}B_{ij}B_{ij}+|D_i\Phi|^2+\frac{1}{2}(\partial_iS)^2+V_{\mathrm{3D}}(\Phi,S)\right\}\, ,
\end{equation}
with
\begin{equation}
    V_{\mathrm{3D}}(\Phi,S)=-\tilde{\mu}_H^2\Phi^\dagger\Phi+\tilde{\lambda}_H(\Phi^\dagger\Phi)^2+\tilde{b}_1S-\frac{1}{2}\tilde{\mu}_S^2S^2+\frac{1}{3}\tilde{b}_3S^3+\frac{1}{4}\tilde{b}_4S^4+\frac{1}{2}\tilde{a}_1S\Phi^\dagger\Phi+\frac{1}{2}\tilde{a}_2S^2\Phi^\dagger\Phi.
\end{equation}
Here, $F_{ij}$ and $B_{ij}$ are the SU(2) and U(1) field strengths and $D_i \Phi = \left( \partial_i + i \tilde{g} A_i + \tfrac{1}{2} i \tilde{g}^{\prime} B_i \right) \Phi$. The prefactor $T^{-1}$ arises from the integration over imaginary time and is often absorbed by rescaling fields and couplings. The tilded coefficients are functions of the four-dimensional renormalised parameters and $T$, fixed by one- and two-loop matching relations.

It is now straightforward to calculate the resummed effective potential for $S$ and the Higgs. For this, we shift 
\begin{equation}\Phi\to\Phi+\frac{1}{\sqrt{2}}
\begin{pmatrix}
0 \\
v
\end{pmatrix}\quad\mathrm{~and~}\quad S\to S +w,
\end{equation}
where $v$ and $w$ are constant background fields in the 3D EFT, not to be confused with the four-dimensional zero-temperature vevs $v_H$ and $v_S$ defined in Section~\ref{sec:generalmodel}. The effective potential is formally 
\begin{equation}
V_{\mathrm{eff}}(v,w)=-\frac{T}{V}\ln\intop D\Phi \exp\left(-S_\mathrm{3D}^\prime\right),
\end{equation}
where $S_\mathrm{3D}^\prime$ is the effective 3D action after field shifts, $T$ is the temperature and $V$ the spatial volume. The two-loop expressions for $V_{\text{eff}}$ can be found in \cite{Niemi:2021qvp}. We numerically minimize $V_\text{eff}$ and track its global minimum $(v_{\mathrm{min}},w_{\mathrm{min}})$ as functions of temperature. A discontinuous jump in these order parameters signals a first-order phase transition, from which $T_c$ and $\Delta v/T_c$ are extracted.

\begin{figure}[H]
    \centering
    \includegraphics[width=0.48\linewidth]{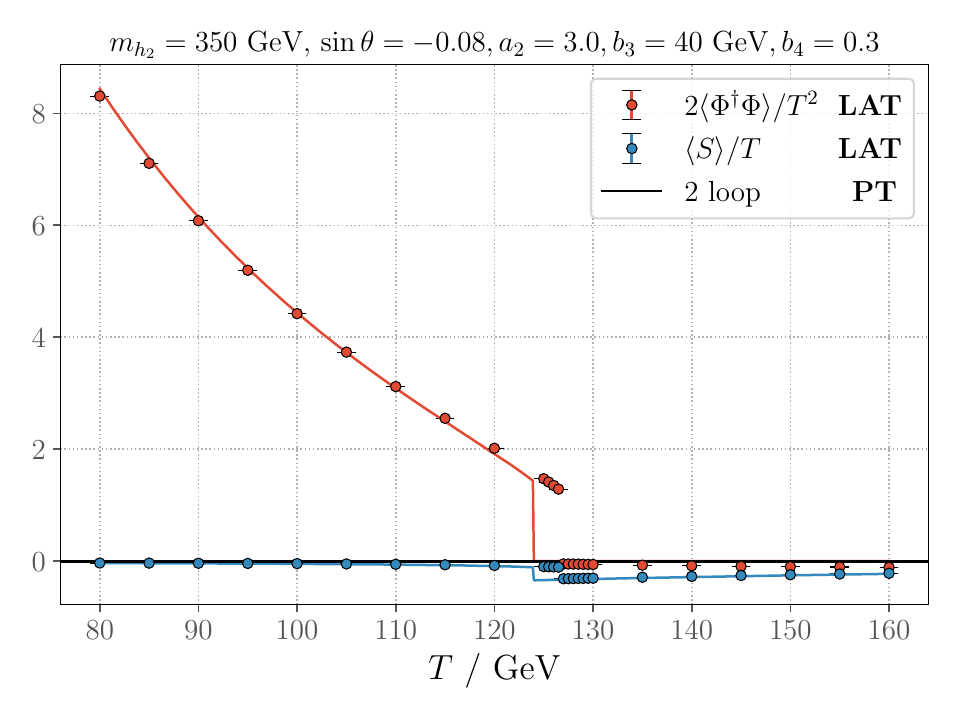}
    \includegraphics[width=0.48\linewidth]{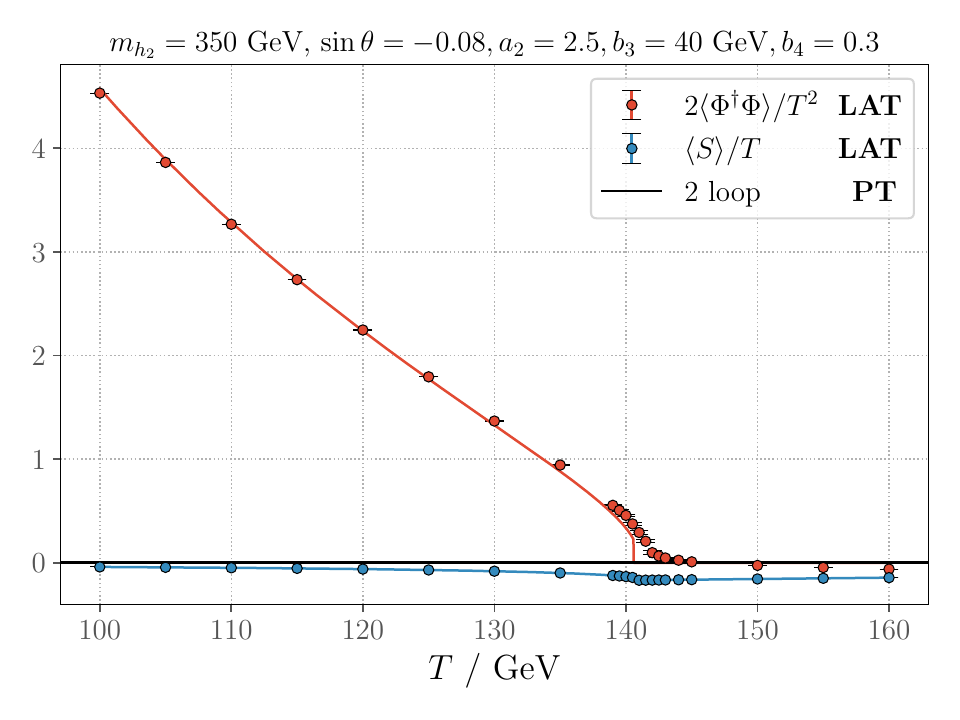}
    \caption{Thermal history for a SFOEWPT benchmark (left) and a crossover benchmark (right). Points are scalar condensates from lattice simulations \cite{Niemi:2024axp}. Solid lines are 2-loop perturbative Higgs $(v/T)^2$ and singlet $w/T$. 
    They are obtained by calculating the effective potential and finding its global minimum at each temperature.}
    \label{fig:thermal_history}
\end{figure}

Figure~\ref{fig:thermal_history} illustrates this procedure for a SFOEWPT benchmark (left) and a crossover benchmark (right). For the strong first-order case, two-loop perturbative results closely track the lattice data~\cite{Niemi:2024axp}. For the crossover benchmark, however, two-loop perturbation theory still predicts a first-order transition where the lattice finds a smooth crossover, a known limitation also observed in the real triplet model~\cite{Niemi:2020hto,Niemi:2024axp}.

A systematic limitation of the perturbative approach described above is its residual gauge dependence. The 4D$\to$3D matching relations are gauge-invariant at the order we work~\cite{Niemi:2021qvp}; gauge dependence enters only through the effective potential of the 3D EFT. In principle, the strict $\hbar$-expansion yields gauge-invariant results order by order via the Nielsen identities~\cite{Fukuda:1975di,Patel:2011th}. However, this expansion is not applicable in our regime: near the $Z_2$-symmetric limit the tree-level condition for the EWPT is $\bar{m}^2_{\phi,3}(T_c) \approx 0$, and the two-loop potential develops an IR divergence at the tree-level minimum~\cite{Niemi:2021qvp,Laine:1994zq}. Numerical minimisation avoids this divergence by resumming a subset of one-particle-reducible corrections~\cite{Kajantie:1995kf}, at the price of a residual gauge dependence that enters at $\mathcal{O}(g^5)$ and beyond~\cite{Niemi:2024vzw,Niemi:2024axp}. We work in Landau gauge ($\xi = 0$), following Refs.~\cite{Niemi:2021qvp,Niemi:2024axp}. In practice, this residual dependence is small: the dominant perturbative uncertainty comes from scalar-loop corrections, not from gauge loops~\cite{Niemi:2021qvp,Chiang:2018gsn}. As a cross check, the gauge-invariant condensate $\Delta v_{\text{phys}} = \sqrt{2 \Delta \langle \Phi^\dagger \Phi \rangle}$ agrees well with the Landau-gauge vev jump $\Delta v_{\text{min}}$ across our parameter space~\cite{Niemi:2024axp}, confirming that gauge artefacts do not significantly affect our numerical results.

\subsubsection{Numerical results}
We focus on the region where $m_{h_2} > m_{h_1}$, i.e.\ the heavy singlet-like scalar regime. The parameter space is explored within the following ranges:

\begin{equation}\begin{array}
{ll}m_{h_2}   \in\left[130,800\right]\,\mathrm{GeV},\,\,
\sin\theta  \in[-0.3,0.3], \,\,
a_{2}       \in[0, 12], \,\,
b_{3}       \in[-200,200]\,\mathrm{GeV}, \,\,
b_{4}       \in[0,2.0].
\end{array}\end{equation}
The mixing angle $\sin\theta$ is experimentally bounded by direct Higgs measurements (see above) and also EW precision observables \cite{Robens:2016xkb, Profumo:2014opa, Li:2019tfd, Zhang:2023jvh, Huang:2017jws}.  

\begin{figure}[ht]
    \centering
    \includegraphics[width=0.48\linewidth]{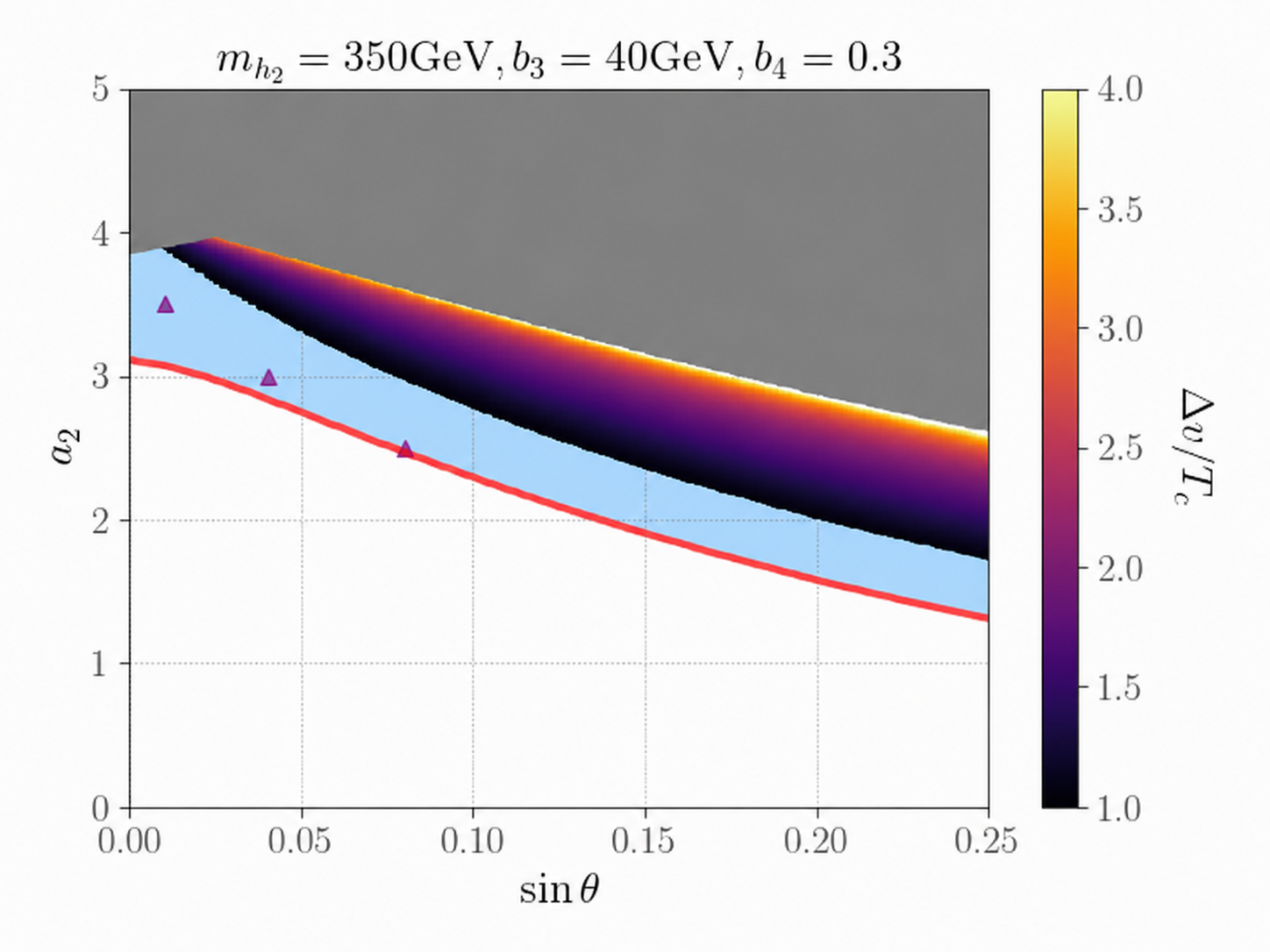}
     \includegraphics[width=0.48\linewidth]{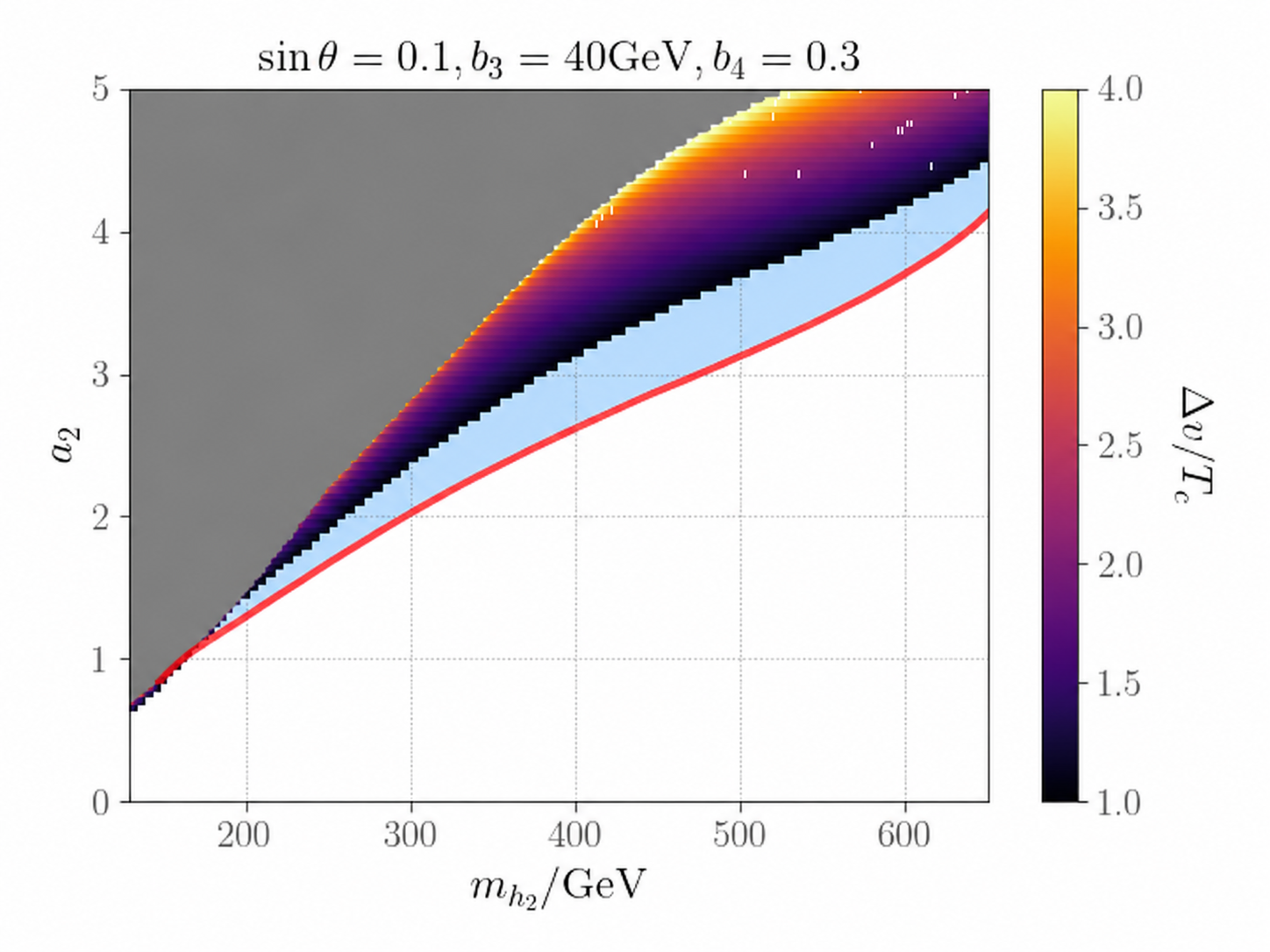}
    \caption{
    Two dimensional projections of the three-dimensional parameter space. The left panel fixes $(m_{h_2}=350\,\mathrm{GeV},b_3=40\,\mathrm{GeV},b_4=0.3)$, and the right panel fixes $(\sin\theta=0.1,b_3=40\,\mathrm{GeV},b_4=0.3)$. The heatmap indicates regions with strong first-order phase transition $\Delta v/T_c >1$. The red solid line separates first-order EWPT (above) from crossover (below). The light blue region between the red curve and the heatmap corresponds to a weak first-order phase transition with $\Delta v/T_c < 1$. The grey region is excluded since the EW minimum is not the global minimum at zero temperature in the one-loop effective potential.} 

    \label{fig:phase_diagram}
\end{figure}

Figure~\ref{fig:phase_diagram} shows the distribution of strong first-order electroweak phase transitions in the parameter space. The boundary separating first-order transitions from crossover behaviour (red solid line) is determined as follows. When the singlet is sufficiently heavy compared to the soft scale $gT$, it can be perturbatively integrated out of the 3D EFT, yielding an SU(2) + Higgs effective theory whose phase diagram has been mapped non-perturbatively on the lattice~\cite{Kajantie:1996mn,Gould:2019qek}. The transition is first-order for $x_c \equiv \tilde{\lambda}_3(T_c)/\tilde{g}_3^2(T_c) \lesssim 0.11$ and a crossover above this value~\cite{Kajantie:1996mn}. We perform this additional reduction at one-loop order using the matching relations of Ref.~\cite{Ramsey-Musolf:2024ykk}; two-loop matching is currently available only in the $Z_2$-symmetric limit~\cite{Niemi:2021qvp}, and we leave its generalisation to future work.

The crossover points (purple dots in the left panel) are identified on the lattice but lie inside the region where two-loop perturbation theory predicts a first-order transition. This is a known limitation: at any finite loop order, the one-loop gauge boson contribution generates a cubic term $\sim g^3 v^3$ that creates a barrier between the symmetric and broken phases~\cite{Niemi:2024axp}. For weak transitions this barrier is an artefact, and the lattice finds a smooth crossover instead. The same behaviour was observed in the real triplet model~\cite{Niemi:2020hto}, where two-loop perturbation theory failed to reproduce the crossover found on the lattice.

\begin{figure}[h!]
    \centering

\includegraphics[width=0.65\linewidth]{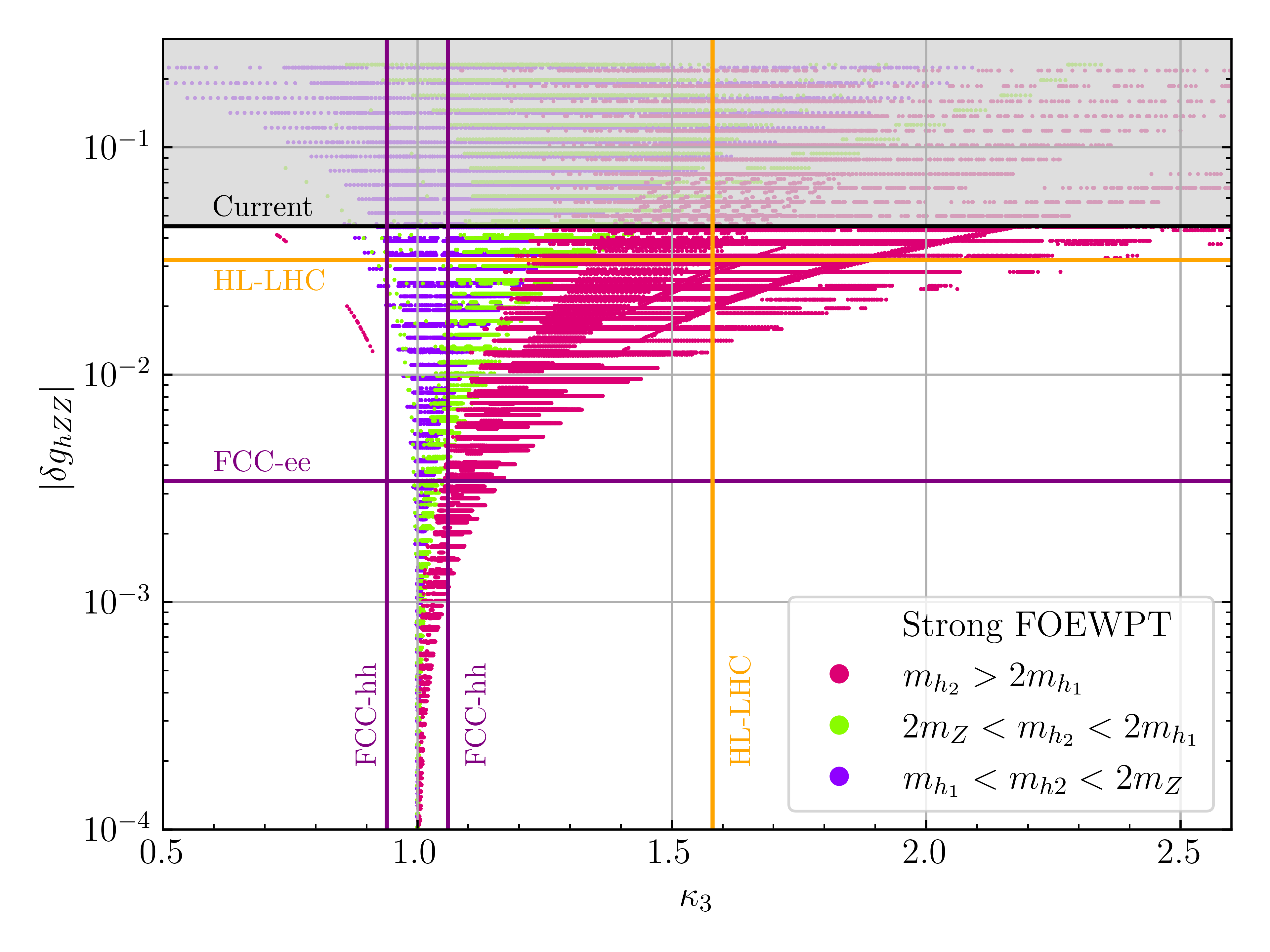}
\caption{Region in the plane $(\kappa_3,\delta g_{hZZ})$ where a strong first-order transition can occur. The colour of the points identifies different mass ranges for the $h_2$. The light/dark shading of the blue and orange bands indicate $68\%$ and $95\%$ exclusions.}\label{fig:FOPTpoints}
\end{figure}

We collect the results of the scan and project them onto the plane of the two key observables: the deviation in the Higgs--$Z$ coupling $\delta g_{hZZ}$ and the Higgs trilinear modifier $\kappa_3$. These two quantities are the primary experimental handles on the xSM: $\delta g_{hZZ} = 1 - \cos\theta$ directly measures the singlet--Higgs mixing at tree level, while $\kappa_3$ probes the shape of the scalar potential. In the alignment limit, Eq.~\eqref{eq:k3dgz} shows that deviations in $\kappa_3$ can be even more than ten times larger than those in $\delta g_{hZZ}$ for $a_2 \gtrsim 2 $, a condition satisfied by all SFOEWPT points at sufficiently large $m_{h_2}$. This makes the Higgs self-coupling a particularly sensitive probe of this scenario. Figure~\ref{fig:FOPTpoints} shows the distribution of parameter points yielding a strong first-order electroweak phase transition in this plane. We discuss below in Sec.~\ref{sec:funnel} the connection between the \lq\lq funnel region\rq\rq\ associated with ($\kappa_3\sim 1$, $|\delta g_{hZZ}|\ll 1$) and the thermal histories of Fig.~\ref{fig:thermal_history}.

\subsection{Stability of the vacuum in the UV \label{sec:stabilityUV}}

Here, we analyse  the effects of the xSM extension on electroweak vacuum stability, reviewing the  basics of the stability of the SM vacuum in Section \ref{sec:stability}
and discussing in Section \ref{sec:Higgsportal} how Higgs portal couplings can improve stability.
The interplay of vacuum stability and the phenomenology of Higgs-singlet mixing is analysed in Section \ref{sec:safe-mixing}.

\subsubsection{Essentials of vacuum stability \label{sec:stability}}

Since the discovery of the Higgs boson and determination of its mass \cite{ATLAS:2012yve, CMS:2012qbp}, it has become possible to assess the stability of the electroweak vacuum within the SM \cite{Buttazzo:2013uya}. Current analyses continue to indicate that the measured values of the SM parameters place the Higgs potential close to the boundary between absolute stability and metastability \cite{Hiller:2024zjp}. This behaviour originates from the renormalisation group evolution of the Higgs quartic coupling $\lambda_H$, which is driven to negative values at scales of order $10^{11}\,\mathrm{GeV}$ for the central values of SM parameters. 
This triggers the occurrence of a secondary minimum in the Higgs effective potential into which the electroweak vacuum may decay. While a theory of nature with a decaying ground state would be unacceptable, vacuum metastability due to a lifetime sufficiently large compared to the age of the universe has become a widely accepted narrative.
The precise location of the stability boundary is primarily controlled by the top Yukawa coupling and therefore by the value of the top-quark mass $m_t$, while the strong coupling constant $\alpha_s$ also plays an important role through its impact on the running of the Yukawa coupling. Consequently, small variations of these parameters can shift the theory from absolute stability to metastability or instability: 
if the top quark were lighter by $1.9\,\sigma$, or the strong coupling larger by $3.5\,\sigma$, the parameters would satisfy the condition for SM-stability, 
$\alpha_s^{(5)}(m_Z) \geq 2.5 \cdot 10^{-3} m_t/\text{GeV} -0.31$  \cite{Hiller:2024zjp} in the relevant window of parameters 
\cite{ParticleDataGroup:2024cfk}; $m_t$ denotes the pole mass. This constraint also highlights that correlations between $\alpha_s$ and $m_t$ are  important.
Note that the Higgs mass is already measured with sufficient precision such that its uncertainty has negligible impact on the stability analysis: while the Higgs mass fixes $\lambda_H$ at the matching scale, the  RG evolution of  $\lambda_H$ is predominantly controlled by the top Yukawa and gauge couplings.

This near-critical behaviour has motivated numerous studies of the Higgs potential and suggests that the metastability of the SM vacuum may provide valuable guidance for physics beyond the SM, e.g.~\cite{Hiller:2022rla,Hiller:2023bdb,Hiller:2024zjp}.
Scalar singlet extensions of the SM have recently been studied in the context of vacuum stability~\cite{Hiller:2024zjp}.\footnote{The correspondence with the notation of Ref.~\cite{Hiller:2024zjp} is as follows: 
$\alpha_{X}=\frac{X}{(4 \pi)^2} , ~~~X=\lambda, v, \delta,  \, 
\lambda=\lambda_H, \, v=\frac{b_4}{4},\,
\delta=\frac{a_2}{2}$.}

\subsubsection{Higgs portal mechanics \label{sec:Higgsportal}}
Stability is lost in the SM at around $\mu \approx 10^{11}$ GeV, where the Higgs quartic $\lambda_H$ turns negative and remains so until far beyond the Planck scale, assuming that quantum-gravity effects do not substantially alter the SM running. The loss of stability can be promoted to a task for model-building.
Positive quartics up to the Planck scale or at least at the Planck scale can be achieved by the gauge portal  \cite{Gopalakrishna:2018uxn,Hiller:2022rla},
based on vector-like quarks or leptons, or the Higgs portal. The latter uses the combined RG-evolution of the Higgs quartic with the portal coupling 
of the SM Higgs with a scalar singlet. At leading order the beta-function $\beta_{\lambda_H}$ of the Higgs quartic reads
\begin{align}\label{eq:betalambda}
\frac{\text{d} \alpha_{\lambda_H}}{\text{d} \ln \mu} =\beta_{\lambda_H}=\beta_{\lambda_H}^{\text{SM}} 
+ \frac{1}{2} \alpha_{a_2}^2 \, , \quad \alpha_{X}=\frac{X}{(4 \pi)^2} , ~~~X=\lambda_H, a_2 \, ,
\end{align}
where $\beta_{\lambda_H}^{\text{SM}}$ denotes the SM-like contribution: it has the same functional form as the one in the SM but 
can assume different values. We learn that any non-zero portal coupling $a_2$ thus increases $\lambda_H$ towards higher energies beyond its SM value.
New instabilities may arise from the presence of the singlet direction in the potential. Necessarily, the potential must be bounded from below in all field directions, which requires
\begin{align} \label{eq:stab}
    \lambda_H >0 \, , ~~b_4 >0 \, , ~~a_2>- 2 \sqrt{\lambda_H b_4 } \, . 
\end{align}
In particular, negative values of the portal coupling are constrained by the mixed Higgs--singlet direction, which will be discussed in more detail. 
For the moment, we focus on positive $a_2$, before returning to the full parameter range in the general scan.

\begin{figure*}
  \centering
  \renewcommand*{\arraystretch}{0}
    \includegraphics[width=.45\columnwidth]{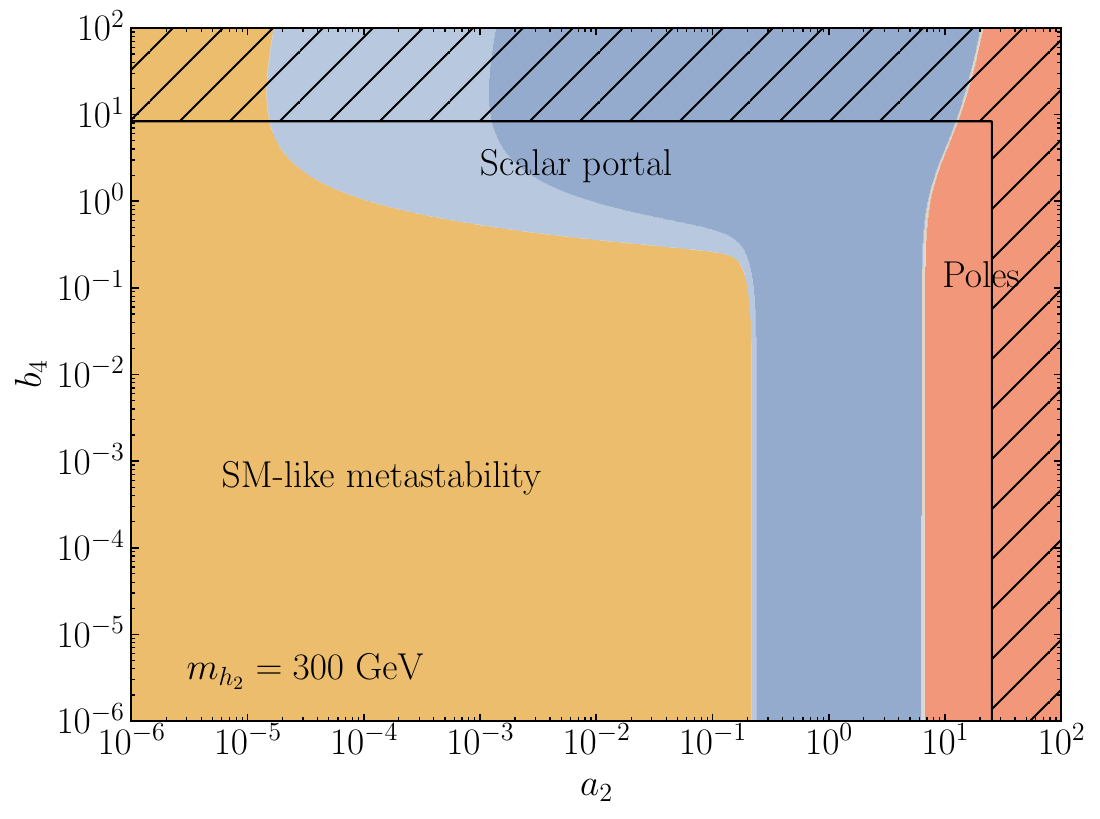} 
       \includegraphics[width=.45\columnwidth]{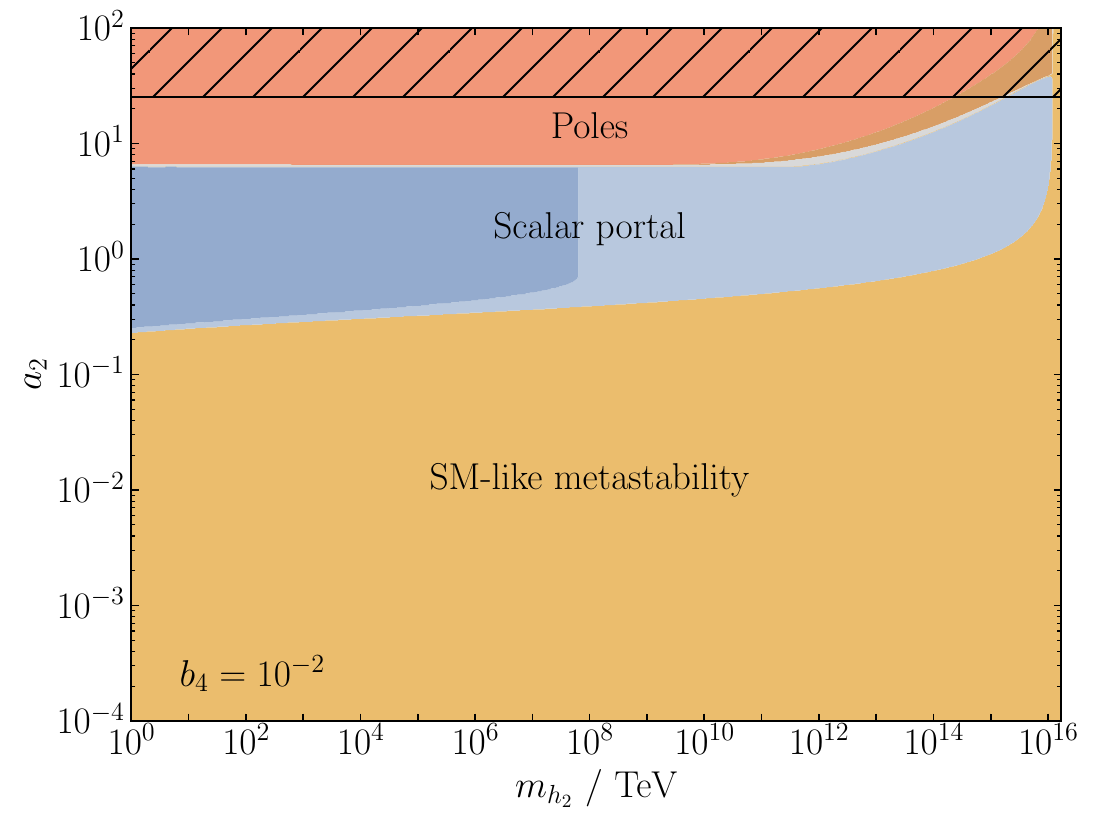}
\caption{BSM critical surface for the real scalar singlet extension \cite{Hiller:2024zjp,Bosse2025real} spanned by the pure BSM quartic $b_4(\mu =m_{h_2})$ and the Higgs portal $a_2(\mu=m_{h_2})$.
Red areas correspond to an RG evolution featuring Landau poles below the Planck scale. Brown indicates instabilities in the Higgs potential ($\min \alpha_{\lambda_H}(\mu) \leq -10^{-4}$). Yellow regions correspond to a SM-like RG evolution with $-10^{-4} \leq \min \alpha_{\lambda_H}(\mu),\,\alpha_{\lambda_H}(M_\text{Pl}) \leq 0$ indicating a metastable Higgs. Dark (light) blue areas  correspond to a strictly (softly) Planck-safe RG evolution with a positive quartic $\lambda_H$ all the way up to (at) $M_\text{\text{Pl}}$. Hatched black areas violate the tree-level perturbative unitarity bounds (\ref{eq:uni}). 
}
\label{fig:BSMsurface}
\end{figure*}

In Fig.~\ref{fig:BSMsurface} (left panel) the portal coupling versus the BSM quartic $b_4$ with their RG-fate colour-coded is shown for a real scalar singlet with mass $m_{h_2}=300$ GeV in the limit of vanishing Higgs–singlet mixing, i.e. in the alignment limit~\cite{Hiller:2024zjp,Bosse2025real}. 
Light blue (dark blue) regions correspond to RG flows where the Higgs quartic remains positive at $M_{\text{Pl}}$ (all the way up to $M_{\text{Pl}}$), while red ones encounter sub-Planckian Landau poles and  yellow denotes SM-like metastability

Black hatched areas violate perturbative unitarity conditions \cite{Gonderinger:2009jp,Dawson:2021jcl}
\begin{align} \label{eq:uni}
   \lambda_H \lesssim \frac{8 \pi}{3} \, , ~~b_4 \lesssim \frac{8 \pi}{3}  \, , ~~|a_2| \lesssim 8 \pi \, ,
\end{align}
somewhat more restrictive than NDA perturbativity.
In Fig.~\ref{fig:BSMsurface} these conditions are merely superimposed, to also indicate the onset of limitations of a purely perturbative analysis.
In the subsequent scan they are used to limit ranges of parameters (\ref{eq:ranges}).

We learn that there is a stability funnel for the quartic portal coupling $a_2$ with values of $O(1)$.
For larger values of the BSM quartic $b_4$ above unity, also smaller values of the portal coupling suffice to make the vacuum stable.
The reason is indirect: $b_4$ affects the Higgs quartic only starting at the three-loop order, but enhances the running of the portal coupling, which in turn supports the Higgs quartic~\cite{Hiller:2024zjp}.
In the right panel of Fig.~\ref{fig:BSMsurface}, the mass of the new scalar versus the portal coupling, evaluated at the scale set by the scalar mass, is shown for a fixed value of the BSM quartic $b_4=10^{-2}$. We find that absolute stability (dark blue area) can be achieved for all values of $m_{h_2}$ up to $\sim 10^7$~TeV. For larger masses, although stability may be lost at intermediate scales, the quartic coupling can become positive again near the Planck scale (light blue area) for masses up to $m_{h_2}\sim 10^{16}$~TeV.
Fig.~\ref{fig:BSMsurface} is based on 3-loop RG of gauge and 2-loop Yukawa and quartic couplings and SM input from \cite{Alam:2022cdv}.
Higher-order corrections, in particular three-loop contributions to the Yukawa and scalar beta-functions, affect the precise location of the stability boundaries. Such effects are expected to be most relevant in regions where the couplings approach the limits of perturbative control.

\begin{figure}
    \centering
\includegraphics[width= 0.7\textwidth]{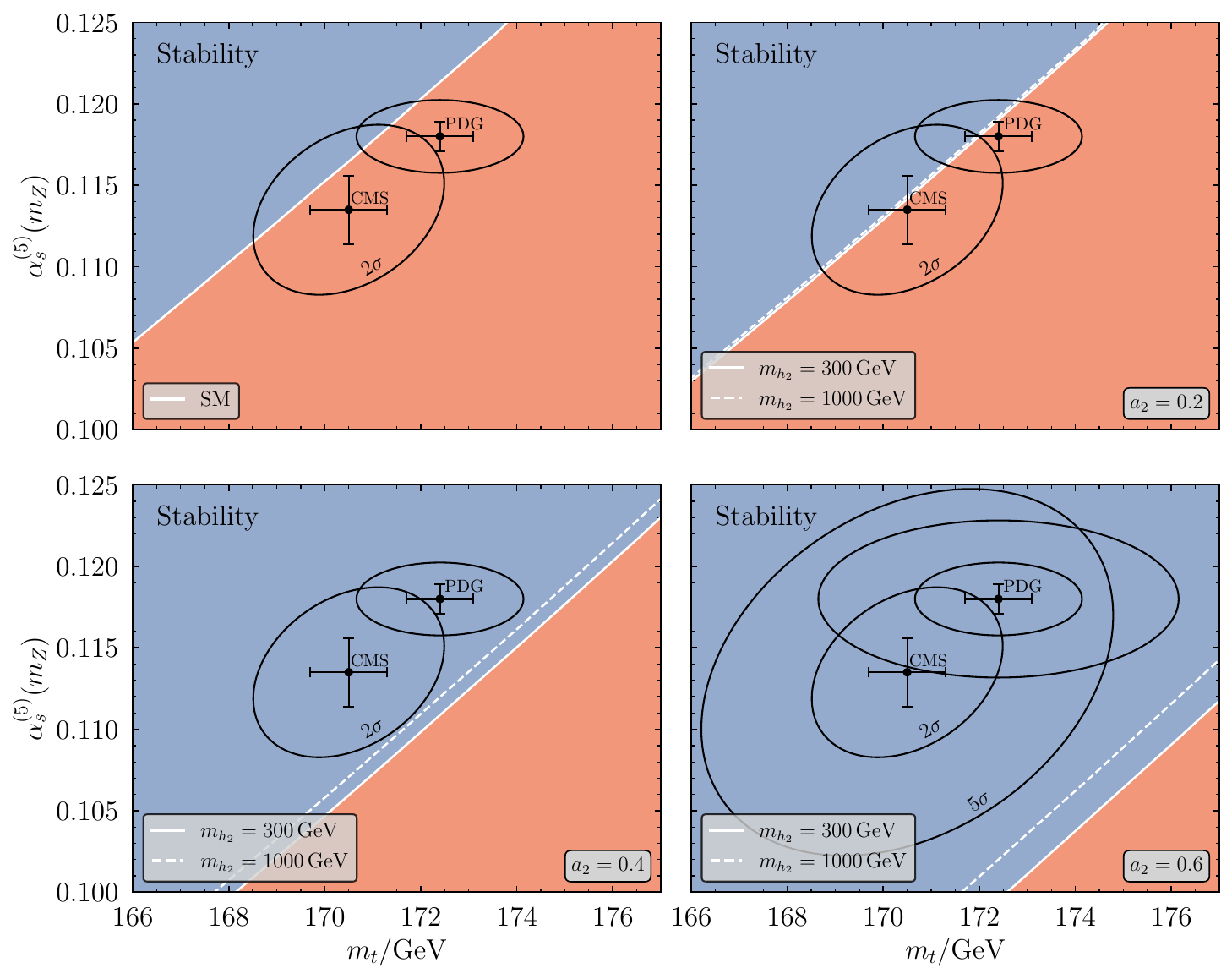}
    \caption{Regions of stability of the SM Higgs potential as a function of the top mass $m_t$ and the strong coupling constant $\alpha_{s}^{(5)}(m_Z)$. The colour coding indicates absolute stability (blue) and metastability/instability (red). The upper-left panel shows the SM stability boundary, calculated following Ref.~\cite{Hiller:2024zjp}. The remaining panels show the corresponding stability boundaries in the singlet extension for fixed portal couplings $a_2=0.2$, $0.4$, and $0.6$. In these panels, the solid and dashed contours correspond to $m_{h_2}=300\,\mathrm{GeV}$ and $1000\,\mathrm{GeV}$, respectively. The current PDG central values with uncorrelated $1\,\sigma$ uncertainties are shown together with the CMS determination \cite{CMS:2019esx}, which includes correlated uncertainties. The ellipses indicate the corresponding $2\,\sigma$ confidence regions for two degrees of freedom (2 d.o.f.); in the $a_2=0.6$ panel, the $5\,\sigma$ contour is also shown.}
    \label{fig:mtvsas}
\end{figure}

Furthermore, we elaborate on the impact of the additional singlet in light of current values and uncertainties on the top mass $m_t$ and the strong coupling constant $\alpha_s$, illustrated in Fig.~\ref{fig:mtvsas}. In each panel, the blue and red regions denote absolute stability and the combined metastable/unstable regime, respectively. The upper-left panel shows the SM stability boundary, while the remaining panels display the corresponding boundaries in the singlet extension for fixed portal couplings $a_2=0.2,\,0.4,\,0.6$. In these BSM panels, the solid and dashed contours correspond to $m_{h_2}=300\,\mathrm{GeV}$ and $1000\,\mathrm{GeV}$, respectively.
Due to the negligible impact of the singlet quartic $b_4$ on vacuum stability for $0 \leq b_4 \lesssim 0.3$ (see the left-hand plot in Fig.~\ref{fig:BSMsurface}), we set $b_4=0$ at the matching scale in Fig.~\ref{fig:mtvsas} for simplicity. Note, however, that a positive $b_4$ is radiatively generated already at one loop, ensuring that the scalar potential remains bounded along the singlet direction. Consequently, the qualitative features of Fig.~\ref{fig:mtvsas} remain unchanged for positive values of $b_4$ in this range.
We show the current PDG central values and uncertainties, using the top mass extracted from cross section measurements, under the assumption of no correlation between $m_t$ and $\alpha_s$. To stress the importance of this correlation, we also show the CMS measurement provided in Ref.~\cite{CMS:2019esx}. 
The SM stability boundary shown in the upper-left panel of Fig.~\ref{fig:mtvsas} is calculated according to Ref.~\cite{Hiller:2024zjp}.\footnote{The $m_t$--$\alpha_s$ correlation  provided by CMS \cite{CMS:2019esx} was erroneously 
displayed in Fig.~2 in  earlier versions of Ref.~\cite{Hiller:2024zjp}, but corrected in v4.
We also show confidence regions for two degrees of freedom (2 d.o.f.), which are slightly larger than the 1 d.o.f. ones of Ref.~\cite{Hiller:2024zjp}.}

Compared to the SM case, the stability boundary in the singlet extension is shifted to the right in Fig.~\ref{fig:mtvsas} for $a_2>0$, 
enlarging the region of absolute stability by including larger values of $m_t$ and smaller ones of $\alpha_s$.
Specifically, we find that a singlet with $m_{h_2} = 300\,\mathrm{GeV}$ (solid line) and portal coupling 
$a_2 = 0.4$ (lower-left panel) shifts the stability boundary towards absolute stability by about $4\,\sigma$, according to the PDG central values and uncertainties. A portal coupling of $a_2 = 0.6$ (lower-right panel) would suffice to render the EW vacuum absolutely stable at more than the $5\,\sigma$ level.
Also shown in Fig.~\ref{fig:mtvsas} are stability boundaries for $m_{h_2} = 1000\,\mathrm{GeV}$ (dashed lines). A heavier BSM scalar mass implies  matching at higher scales, where $\alpha_s$ is lower, and less RG-time is available for the portal mechanism. Therefore, the stability-allowed parameter space for the strong coupling and top mass is reduced with respect to the lighter mass case.

\subsubsection{Higgs-singlet mixing \label{sec:safe-mixing}}
We discuss the case of Higgs--singlet mixing, which gives rise to characteristic collider signatures of Planck-safe models.
Recall that the connection between collider and Planck scale physics is the RG evolution of dimensionless couplings: the scalar quartics, Yukawas and the gauge couplings. From this perspective the main players from the scalar potential (\ref{eq:xSM}) are
$\lambda_H, a_2$ and $b_4$. The other ingredients that determine the UV-fate are the RG-initial conditions at the weak scale.
The set of weak scale BSM parameters that make the model reach the Planck scale safely is selected as viable, and define the BSM critical surface \cite{Hiller:2020fbu}. Parameters from this surface are then used to study the phenomenology.
For the SM couplings the weak scale values are experimentally determined with the notable exception of $\lambda_H$. The latter can be modified from its SM value in scalar singlet extensions for $\theta \neq 0$, see (\ref{eq:lHtheta}), specifically
also by dimensionful parameters of the scalar potential, which then also ultimately impact the UV-fate.

To begin, and to disentangle the two aforementioned effects, we illustrate the interplay of vacuum stability, absence of Landau poles and Higgs phenomenology 
in a simplified scenario in which the Higgs quartic remains close to its SM value
at the low energy matching scale.
The precision SM $\overline{\text{MS}}$ parameters \cite{Alam:2022cdv} are matched onto the real singlet model at the scale $\mu = m_{h_2}$, in accordance with the procedure adopted in Ref.~\cite{Hiller:2024zjp}, see also for studies in further  scalar singlet extensions.
\begin{figure}
\renewcommand*{\arraystretch}{0}
\begin{tabular}{c}
\includegraphics[height=0.25\textheight]{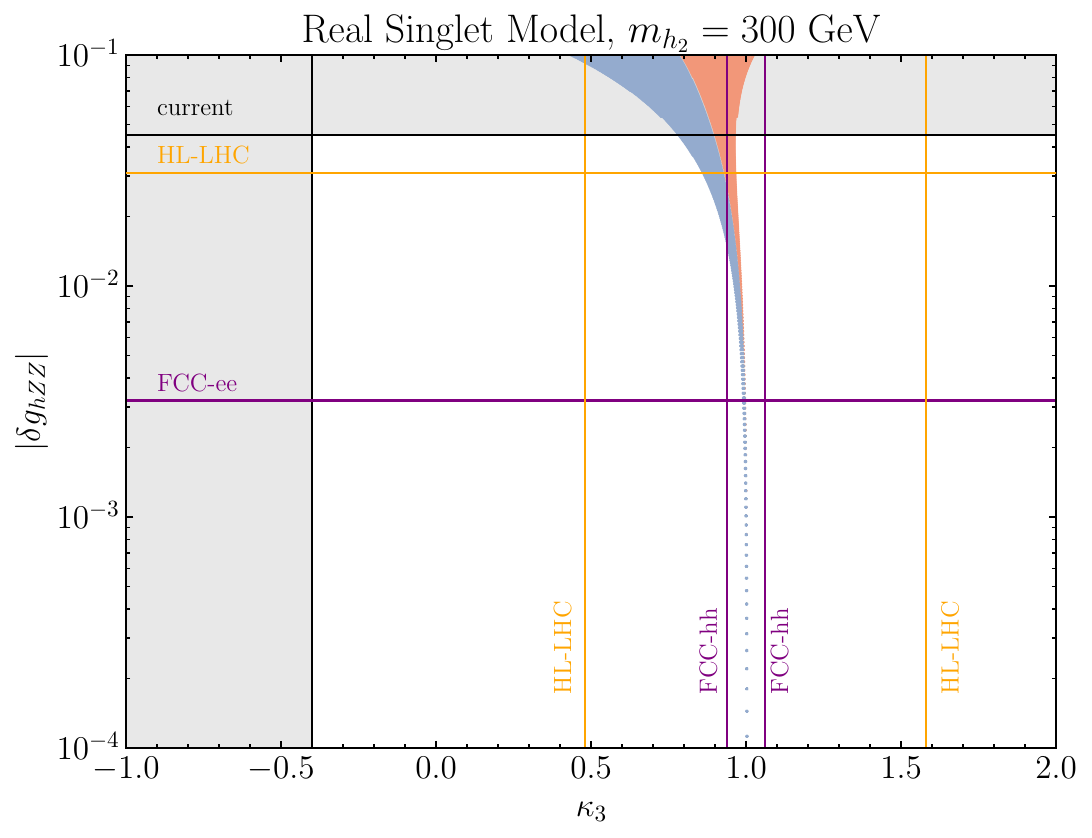}
\includegraphics[height=0.25\textheight]{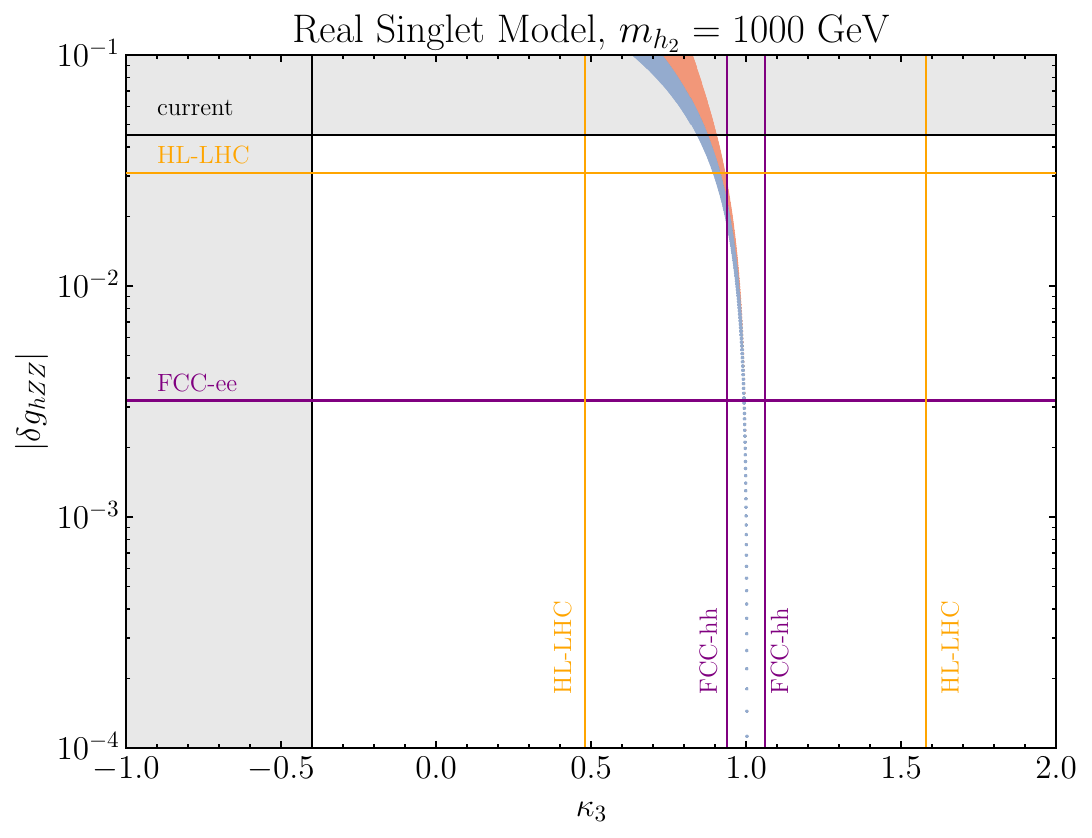}\\
\end{tabular}
\caption{Vacuum stability analysis of the Higgs coupling modifiers $\delta g_{hZZ}$ versus $\kappa_3$ in the simplified real scalar singlet model for $m_{h_2}=300$ GeV (left), $m_{h_2}=1$ TeV (right), see text. The scattered dots correspond to different BSM coupling values, sampled within the unitarity bounds, and their colour indicates the Planck fate as in Fig.~\ref{fig:BSMsurface}. Specifically,
blue means safety and red denote Landau poles below $M_{\text{Pl}}$. Horizontal and vertical lines correspond to experimental sensitivities at current and future colliders. 
Grey regions are  excluded by current data, see \cite{Hiller:2024zjp} for details.
}
\label{fig:stabilityZ2}
\end{figure}
In Fig.~\ref{fig:stabilityZ2}, we show deviations in the Higgs coupling to $ZZ$, $\delta g_{hZZ}$, versus the Higgs trilinear coupling normalised to its SM value, $\kappa_3$, at leading order as given in Eq.~(\ref{eq:k3lin}). 
The plotted points are obtained by sampling the parameter space within the unitarity bounds~\eqref{eq:uni}.
The simplified model predicts suppression with respect to the SM value of the trilinear for BSM scalars heavier than the Higgs \cite{Hiller:2024zjp} and 
larger deviations from the SM arise for lighter BSM scalars. The prospects for near-term HL-LHC probes of $hZZ$-modifications in Planck-safe scenarios appear better than with the trilinear $\kappa_3$.

\begin{figure}
    \centering
    \includegraphics[width= 0.7\textwidth]{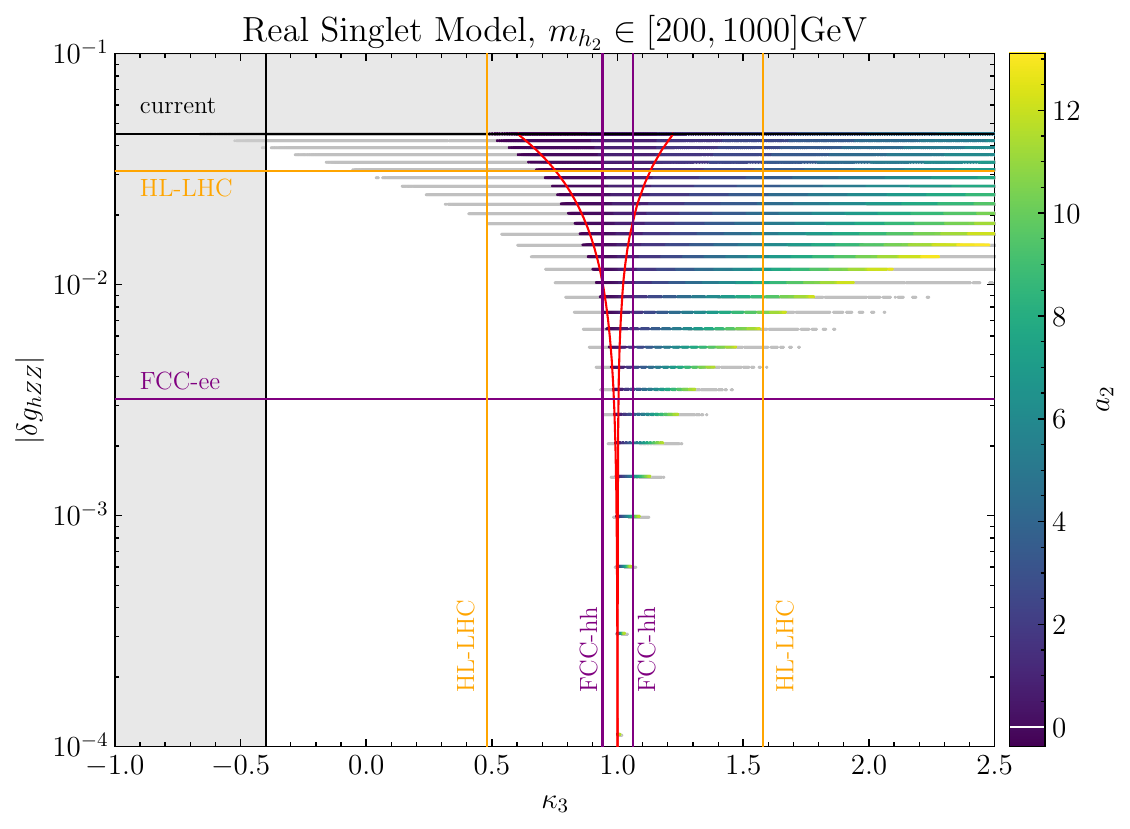}
    \caption{Vacuum stability analysis of the Higgs coupling modifiers $\delta g_{hZZ}$ versus $\kappa_3$ in the general real scalar singlet model \cite{Bosse2025real}.
Horizontal and vertical lines correspond to experimental sensitivities at current and future colliders. 
Gray regions are  excluded by current data. Shown are the Planck-safe points (coloured) as a function of the portal coupling $a_2$. The grey dots are not Planck-safe. Any metastable solution has to be within the region enclosed by the red curves, see text.
}
\label{fig:stability_full}
\end{figure}

In Fig.~\ref{fig:stability_full} we show Higgs couplings $\delta g_{hZZ}$  versus $\kappa_3$ in the general real scalar singlet extension.
Parameters are scanned within the regions 
\begin{align}  \nonumber 
200\,\mathrm{GeV} \;\leq\; m_{h_2} \;\leq\; 1000\,\mathrm{GeV}, \qquad
-1000\,\mathrm{GeV} \;\leq\; b_{3} \;\leq\; 1000\,\mathrm{GeV}, \\ \label{eq:ranges} 
 0 \;\leq\; b_{4} \;\leq\; \frac{8\pi}{3}, \qquad
-8\pi \;\leq\; a_{2} \;\leq\; 8\pi, \qquad
-0.30 \;\leq\; \theta \;\leq\; 0.30, 
\end{align}
and boundaries for the BSM quartics follow from unitarity (\ref{eq:uni}). 
Here, we adopt the renormalisation procedure elaborated in Sec.~\ref{sec:EWSB} for the determination of the phase transition, in order to ensure a consistent parameter definition when combining our stability results with the phase transition study. In this framework, the $\overline{\text{MS}}$ parameters of the real singlet extension are extracted at one-loop order from the loop-corrected relations between the physical inputs and the Lagrangian parameters.
For the present stability analysis, the parameter extraction is performed at $\mu = (m_{h_1}+m_{h_2})/2$, which helps keep logarithmic corrections in the scalar self-energies more moderate when the scalar couplings become sizeable (see also Ref.~\cite{Kainulainen:2019kyp}).

In addition, in Fig.~\ref{fig:stability_full}
all Planck-safe points are colour-coded according to their value of the portal coupling 
\( a_2 \).
Non-Planck-safe points are shown in grey, jointly denoting vacuum instabilities and/or sub-Planckian Landau poles.
We observe that the Planck-safe range of the trilinear $\kappa_3$ is enlarged compared to the one of the simplified model shown in Fig.~\ref{fig:stabilityZ2}. In particular, 
enhancements  $\kappa_3 > 1$ are possible, and also more sizeable suppressions. Larger Planck-safe values of $\kappa_3$ arise for larger values of the portal coupling, with larger $a_2$ required  for smaller $\delta g_{hZZ}$ values, consistent with  expectations (\ref{eq:k3dgz}).
For too large values of the portal coupling sub-Planckian Landau poles appear, for either sign of $a_2$. Negative values are in addition constrained by the boundedness-from-below condition~\eqref{eq:stab}.

All metastable points found in the scan lie within the region marked by the red lines.
Nevertheless, parameter configurations inside this band can remain Planck-safe and therefore appear as coloured points in the figure. Within the explored parameter space, deviations of $\kappa_3$ from its SM value move the model away from the SM-like metastable region and can restore a stable electroweak vacuum. In particular, for $\kappa_3 \gtrsim 1.2$, metastability is typically lifted to stability, unless $\kappa_3$ and hence $a_2$ become too large and develop Landau poles.
By contrast, grey points with $\kappa_3 < 1$ develop instead  an additional runaway direction along the singlet field direction or encounter Landau poles once the magnitude of the portal coupling $|a_2|$ becomes too large.
We observe that Planck-safety is efficiently selecting the viable parameter space of the model.

Let us also comment on negative portals:
Negative portal couplings can also be compatible with Planck-safety; however, this typically requires the magnitude of the portal coupling to remain small. In our scan, viable points are found as low as  $a_2 \simeq -0.38$, close to metastable solutions.
The challenge for negative portals arises because RG effects tend to tighten the boundedness-from-below condition (\ref{eq:stab}) towards  higher scales, making negative $a_2$ increasingly constrained~\cite{Hiller:2024zjp}. In the presence of scalar mixing, however, such points can appear once threshold corrections shift the Higgs quartic $\lambda_H$ relative to its SM value.

One observes from
Fig.~\ref{fig:stabilityZ2}  and Fig.~\ref{fig:stability_full}, that shifts in both couplings, 
$\delta g_{hZZ} \propto |\kappa_3-1|$ can go to zero for model parameters that still maintain Planck safety.
The reason is that safety can already be achieved in the vev-less $Z_2$-symmetric, that is, the no-mixing case since changes to the RG-running suffice \cite{Hiller:2024zjp}.
We also stress that $\kappa_3$ is the leading order Higgs trilinear. In the future an NLO computation of 
the BSM cross section would be desirable.

\subsection{Nature of the funnel points }\label{sec:funnel}

As shown in Figs.~\ref{fig:FOPTpoints},\,\ref{fig:stabilityZ2} and \ref{fig:stability_full}, the funnel region, in which $\theta\rightarrow 0$ and hence $\delta g_{hZZ}^{\rm LO}\rightarrow 0$ and $\delta \kappa_3^{\rm LO}\rightarrow 0$, contains points that both realise a SFOEWPT (Fig.~\ref{fig:FOPTpoints}) and stabilise the SM vacuum (Figs.~\ref{fig:stabilityZ2} and \ref{fig:stability_full}).
 
At first sight, this result may appear puzzling: how can such points arise in a region where the SM Higgs couplings appear unchanged? For the points shown in Fig.~\ref{fig:FOPTpoints}, the transition corresponds to the second step of the thermal history illustrated in Fig.~\ref{fig:scenarios}(b), where the $a_2$ portal operator generates the barrier between the $S$ and $H$ vacua. Since in our parametrisation this operator induces no doublet-singlet mixing, the two-step scenario accommodates a SFOEWPT in the $\theta\to 0$ limit. This situation is similar to the analogous two-step scenario for the real triplet extension~\cite{Niemi:2020hto}. Regarding Figs.~\ref{fig:stabilityZ2} and \ref{fig:stability_full},  the uplift of the Higgs quartic by the portal coupling, also in interplay with $b_4$, suffices to achieve stability \cite{Hiller:2024zjp}
and does not require Higgs-singlet mixing, see e.g.~Fig.~\ref{fig:BSMsurface}.

\begin{figure}
    \centering
    \includegraphics[width=0.48\linewidth]{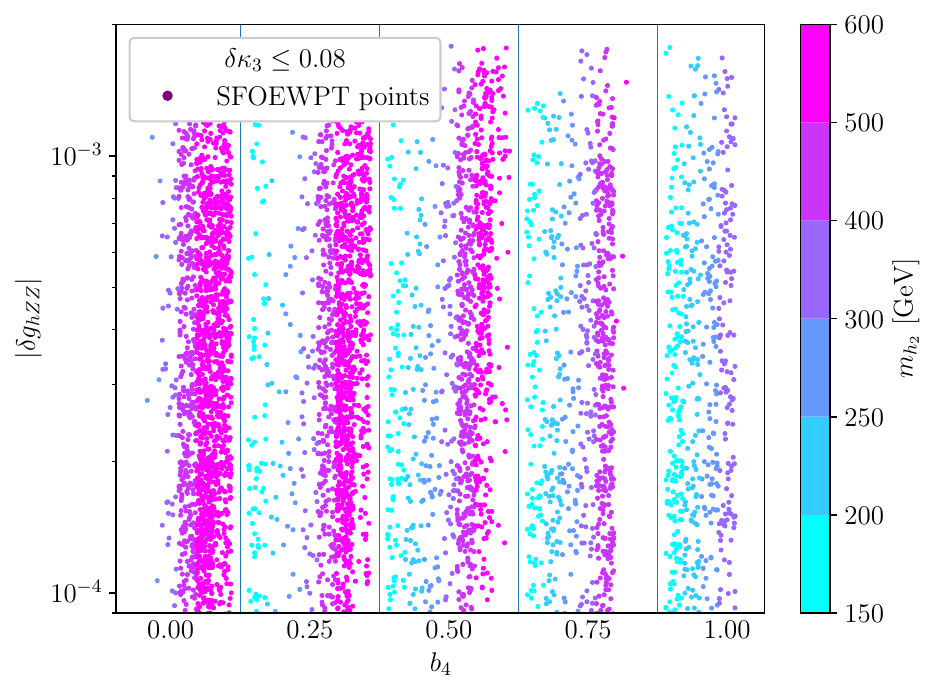}
    \includegraphics[width=0.48\linewidth]{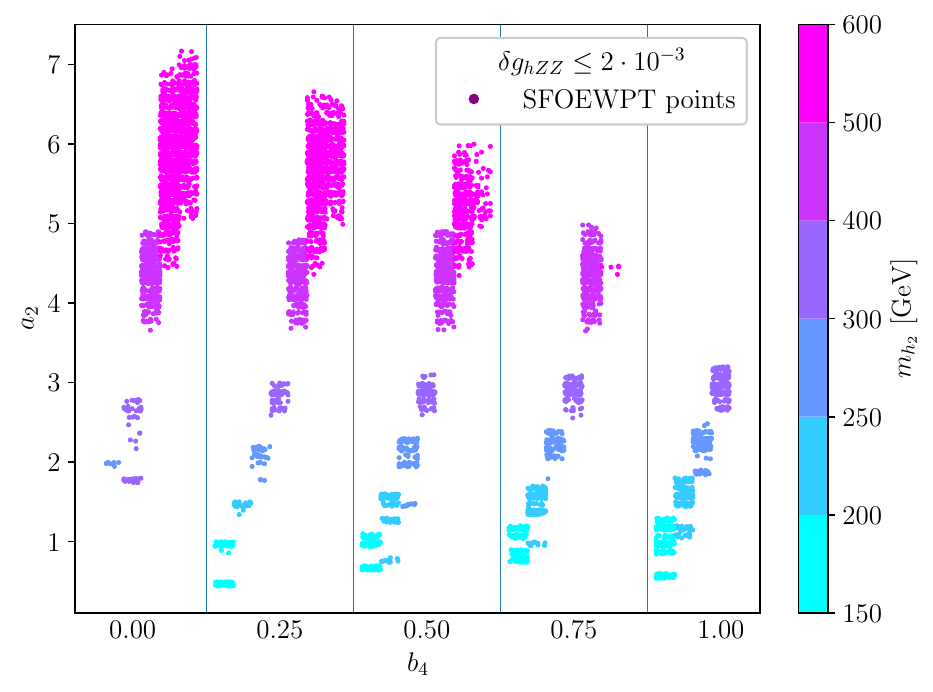}
    \caption{Scan in equally spaced $b_4$ and $a_2$ in the funnel region. Different masses of the scalar $h_2$ are shown in different colours. The value of $b_4$ is constant in every column, but the points are slightly shifted with respect to the value of the mass to increase readability. \textbf{Left:} SFOEWPT points in the $b_4-|\delta g_{hZZ}|$ plane. \textbf{Right:} SFOEWPT points in the $b_4-a_2$ plane.
    }
    \label{fig:funnel_points}
\end{figure}

To better understand the small-$\theta$ funnel, we perform a uniform scan in the $a_2$--$b_4$ plane for different values of the scalar mass $\mtwo$, with the results shown in Fig.~\ref{fig:funnel_points}.
In the left panel we show the distribution of the SFOEWPT points in the $b_4-|\delta g_{hZZ}|$ plane. One might naively expect $b_4$ to play only a minor role, since it enters primarily through the running of $a_2$. However, this is not the case: the points leading to a SFOEWPT exhibit a non-trivial dependence on the combination of $b_4$ and $\mtwo$. Once inside the funnel, larger scalar masses favour smaller singlet self-couplings, and vice versa. This conclusion is largely independent of the value of $|\delta g_{hZZ}|$. In the right panel of Fig.~\ref{fig:funnel_points}, we show the same points in the $b_4$--$a_2$ plane. A clear interplay emerges between the $Z_2$-preserving couplings and the scalar mass: large portal couplings $a_2$ are favoured for larger masses and small singlet self-couplings, while larger self-couplings are favoured for smaller masses and smaller values of $a_2$.
The interplay between the portal and the self-coupling via RGE and their effect on the stability bounds has already been noted in \cite{Hiller:2024zjp}
and is visible in Fig.~\ref{fig:BSMsurface}. 
These effects become particularly clear when plotting the SFOEWPT regions for different benchmark choices, as done in the phenomenological analysis of Sec.~\ref{sec:high-lumi}.

It is natural to ask whether points in the funnel region can nevertheless be tested at colliders.
This is not immediately the case at leading order, as already argued in the context of stability. Indeed, the trilinear Higgs coupling receives one-loop contributions induced solely by $a_2$, which do not vanish in the limit $\theta\rightarrow 0$. The simplest estimate of these contributions can be obtained by integrating the $a_2$ contribution to $\beta_\lambda$ in Eq.~\eqref{eq:betalambda}, leading to the approximate modification of the Higgs trilinear coupling
\begin{equation}\label{eq:deltaNLO}
\Delta_{\rm approx}^{\rm NLO}g_{h_1h_1h_1} =3v_H \frac{a_2^2}{16 \pi^2}\log\left(\frac{\mone+\mtwo}{4\mone}\right).
\end{equation}
More refined quantities could in principle be computed, such as the full BSM contribution to the Higgs quartic coupling $\lambda_H$, using, for instance, the automatic tools of Ref.~\cite{Bahl:2023eau}. However, in the following we will use the simple expression in Eq.~\eqref{eq:deltaNLO} to identify which points would require a full NLO calculation in order to determine whether they can be tested at colliders. 
In general, NLO corrections can be sizeable: the ratio between the NLO-computed Higgs trilinear coupling and its LO value can be as large as a factor of five~\cite{Kanemura:2016lkz,Braathen:2025qxf}.

\section{Experimental probes}\label{sec:constraints}
We discuss gravitational waves from the EWPT in Section \ref{sec:GW}, and collider observables for indirect and direct searches of the model in 
Section \ref{sec:colliders}. In Section \ref{sec:run2} we show the status of current constraints from Run~II, together
with the ones from vacuum stability and a SFOEWPT.

\subsection{Gravitational waves from the EWPT \label{sec:GW}}
A strong first-order EWPT proceeds through bubble nucleation and is accompanied by a stochastic gravitational-wave background, potentially observable by future space-based detectors such as LISA~\cite{LISA:2017pwj}, Taiji~\cite{Gong:2014mca,Hu:2017mde,Ruan:2018tsw}, TianQin~\cite{TianQin:2015yph}, BBO~\cite{Crowder:2005nr,Yagi:2011wg} and DECIGO~\cite{Kawamura:2011zz,Musha:2017usi}. In this section, we compute the bubble nucleation rate and the resulting thermal parameters for each SFOEWPT point identified in Section~\ref{sec:EWSB}, and map them to predicted gravitational-wave spectra.

During a first-order phase transition, the metastable (false) vacuum decays to the stable (true) vacuum through thermal nucleation of critical bubbles. In the 3D EFT, the bubble nucleation rate per unit volume at leading order is~\cite{Ekstedt:2024etx, Gould:2021ccf, Gould:2021oba, Friedrich:2022cak,Croon:2020cgk}:
\begin{equation}
    \Gamma(T)\approx A(T) e^{-S_{\mathrm{eff}}^{\mathrm{LO}}(v, w)},
\end{equation}
where $A(T)$ is the prefactor. At leading order, the dynamical prefactor encoding out-of-equilibrium effects is neglected, and the statistical (functional-determinant) prefactor is estimated on dimensional grounds as $A(T) \sim T^4$~\cite{Gould:2021ccf}.
$S_{\mathrm{eff}}^{\mathrm{LO}}$ is the gauge-invariant 3D Euclidean action at leading order~\cite{Gould:2021oba, Gould:2021ccf, Li:2025kyo, Hirvonen:2022jba}. Its gauge invariance follows from a power counting in couplings~\cite{Gould:2021ccf, Gould:2021oba}: near the phase transition, without significant supercooling, the scalar field values are of order $gT$ and the scalar masses of order $g^2 T$. At this order, the effective potential receives two contributions of the same size $\sim g^6 T^3$: the tree-level scalar terms and the one-loop gauge boson diagrams. Together, they form a gauge-invariant combination at leading order. All other one-loop contributions (from scalars) enter at higher order in the coupling expansion and are omitted~\cite{Gould:2021ccf, Ekstedt:2024etx}. This power counting in $g$ is distinct from the $\hbar$-expansion used in Section~\ref{sec:EWSB} for thermodynamic quantities, which organises the calculation by loop order~\cite{Gould:2021ccf, Gould:2021oba}. The gauge boson cubic terms arise from integrating out the soft gauge bosons in the Higgs phase~\cite{Hirvonen:2021zej, Ekstedt:2022zro}:
\begin{align}
    V_{\mathrm{eff}}^{\mathrm{LO}}(v,w) =&-\frac{1}{2}\tilde{\mu}_{H,3}^2 v^2+\frac{1}{4}\tilde{\lambda}_{H,3} v^4 +\tilde{b}_{1,3}w-\frac{1}{2}\tilde{\mu}_{S,3}^2w^2+\frac{1}{3}\tilde{b}_{3,3}w^3+\frac{1}{4}\tilde{b}_{4,3}w^4 \nonumber \\
    &+\frac{1}{4}\tilde{a}_{1,3}wv^2 +\frac{1}{4}\tilde{a}_{2,3}w^2v^2- \frac{\tilde{g}_3^3}{24\pi}|v|^3-\frac{\left(\tilde{g}_3^2+\tilde{g}_3^{\prime2}\right)^{3/2}}{48\pi}|v|^3\,. 
\end{align} The first line is the tree-level 3D scalar potential, while the second line contains the gauge boson cubic terms proportional to $|v|^3$, arising from the $W$ and $Z$ boson contributions respectively. Together, these two contributions exhaust the leading-order potential.

The gauge-invariant action at leading order then takes the form:
\begin{equation}
    S_{\mathrm{eff}}^{\mathrm{LO}}(\Phi,T)=\int dr \; 4\pi r^2 \left[\frac{1}{2} \left( \frac{d\Phi}{dr}\right)^2 + V_{\mathrm{eff}}^{\mathrm{LO}}(\Phi,T)\right], 
\end{equation}
where the radial profile $\Phi(r)=\{v(r),w(r)\}$ is obtained with \texttt{FindBounce}~\cite{Guada:2020xnz}  by numerically solving the bounce equation:
\begin{align}
    \frac{d^2\Phi}{dr^2}+\frac{2}{r}\frac{d\Phi}{dr}=\frac{dV^\mathrm{LO}_{\mathrm{eff}}(\Phi,T)}{d\Phi}\,,
\end{align}
with boundary conditions, 
\begin{align}
    \lim_{r\rightarrow\infty}\Phi(r)=\{v_\mathrm{false},w_\mathrm{false}\},\quad \frac{d\Phi}{dr}\bigg \vert_{r\rightarrow0}=0.
\end{align}
Here $(v_\mathrm{false}, w_\mathrm{false})$ and $(v_\mathrm{true}, w_\mathrm{true})$ denote the field values at the false and true vacua, i.e.\ the local minima of $V_\mathrm{eff}^\mathrm{LO}$ with higher and lower free energy, respectively.

The gravitational-wave spectrum from a first-order transition is governed by four thermal parameters: the percolation temperature $T_p$, the phase transition strength $\alpha(T_p)$, the inverse duration $\beta(T_p)$, and the bubble wall velocity $v_w(T_p)$. In general, $\alpha$ and $v_w$ depend on the sound speed $c_s$ in the plasma, which is determined by the equation of state. For our purposes, it is sufficient to adopt the leading-order result for a relativistic plasma, $c_s^2=1/3$~\cite{Ai:2023see,Giese:2020rtr, Tenkanen:2022tly}.

Using \texttt{FindBounce}~\cite{Guada:2020xnz}, we evaluate the bounce action $S_\mathrm{eff}^\mathrm{LO}(T)$. The inverse duration of the phase transition is
\begin{equation}
    \frac{\beta}{H}(T)=T\frac{dS_\mathrm{eff}^\mathrm{LO}(T)}{dT}\,.
\end{equation}
The phase transition strength $\alpha$ is defined as the difference of the trace of the energy-momentum tensor between the two vacua, normalised to the radiation enthalpy~\cite{Giese:2020rtr},
\begin{equation}
    \alpha(T)=\frac{T}{4\rho_\mathrm{rad}(T)}\Delta\left[T\frac{dV_{\mathrm{eff}}^\mathrm{LO}}{dT}-3V_{\mathrm{eff}}^\mathrm{LO}\right], 
\end{equation} 
where $\Delta[\cdots] \equiv [\cdots]_\mathrm{true} - [\cdots]_\mathrm{false}$ denotes the difference between the true and false vacua, and $\rho_\mathrm{rad}(T) = \pi^2 g_* T^4/30$ with $g_*=106.75 + 1$ including the additional singlet degree of freedom.

We estimate the bubble-wall velocity under local thermal equilibrium (LTE) using entropy conservation, since out-of-equilibrium effects are subdominant for the transitions considered here~\cite{Laurent:2022jrs}. Hydrodynamic simulations~\cite{Krajewski:2024gma} show good agreement with the analytic LTE method~\cite{Ai:2023see}, and we therefore adopt the fit function of Ref.~\cite{Ai:2023see}:
\begin{equation}
    v_w(T)=\left\{
    \left| \frac{3\alpha + \Psi-1}{2\left(2-3\Psi+\Psi^3\right)} \right|^{p/2}
    +\left|\frac{1}{\sqrt{3}}\frac{1+\sqrt{3\alpha^2+2\alpha}}{1+\alpha}\left(1-a\frac{\left(1-\Psi\right)^b}{\alpha}\right)\right|^p\right\}^{1/p},
\end{equation}
where numerical fit constants $a=0.2233$, $b=1.704$ and $p=-3.433$. 
$\Psi$ is the ratio of the enthalpies in the true and false vacua, 
\begin{equation}
    \Psi(T) =\frac{\omega_{\mathrm{true}}(T)}{\omega_{\mathrm{false}}(T)},\qquad
    \omega_j(T) = \frac{4}{3}\rho_{\mathrm{rad}}(T) - T\,V_\mathrm{eff}^\mathrm{LO}(v_j,w_j) - T^2\frac{dV_\mathrm{eff}^\mathrm{LO}(v_j,w_j)}{dT},
\end{equation}
with $j \in \{\mathrm{true},\,\mathrm{false}\}$.

\begin{figure}[ht]
    \centering

\includegraphics[width=0.7\linewidth]{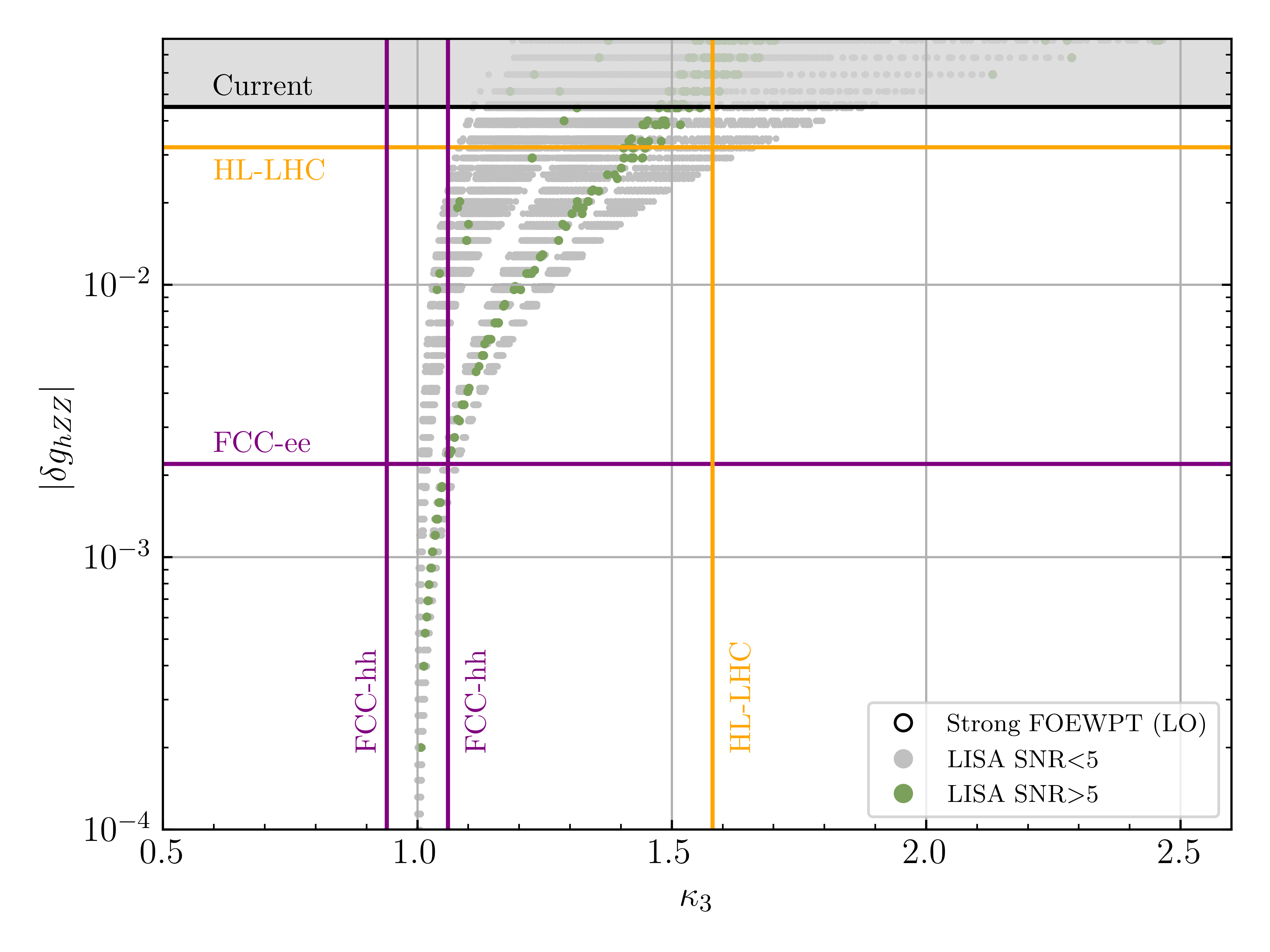}
\caption{Distribution of SFOEWPT points in the $(\kappa_3,\,\delta g_{hZZ})$ plane that yield a detectable gravitational-wave signal. Here $\kappa_3$ is the tree-level value defined in Eq.~\eqref{eq:k3lin}.  Colour coding indicates the singlet mass $m_{h_2}$. The FOEWPT scan is performed at leading order for gauge invariance. All points with $\mathrm{SNR} > 5$ (for LISA with $\mathcal{T} = 3\,\mathrm{yr}$) undergo a two-step phase transition.}\label{fig:FOPTgw}
\end{figure}

For the gravitational-wave signal, the relevant temperature is the percolation temperature $T_p$, at which bubbles fill a sufficient fraction of space for the transition to complete~\cite{Caprini:2019egz}. This differs from the nucleation temperature $T_n$, defined as the temperature at which roughly one bubble nucleates per Hubble volume. In practice the two are often close, but $T_p$ better characterises the bulk dynamics that source gravitational waves, and we adopt it throughout.
The percolation temperature is calculated iteratively using the definition \cite{Enqvist:1991xw, Caprini:2019egz}
\begin{equation}
S_{\mathrm{eff}}^\mathrm{LO}(T_p) \simeq 131 
- 4 \log\!\left(\frac{T_p}{100\,\mathrm{GeV}}\right) 
- 4 \log\!\left(\frac{\beta/H^*}{100}\right) 
+ 3 \log(v_w) \, .
\end{equation}
Here the prefactor contribution $\log(A/T_p^4)$ has been dropped as it enters only at next-to-leading order. Starting from an initial guess, $T_p$ is obtained by iterating this relation until convergence. All thermal parameters --- $\alpha$, $\beta/H$, and $v_w$ --- are then evaluated at $T = T_p$.

Finally, the {\sc PTPlot} package \cite{Caprini:2019egz, Hindmarsh:2013xza} is used to compute the gravitational-wave spectrum. To evaluate the detectability, the Signal-to-Noise Ratio (SNR) is defined as:
\begin{equation}
\mathrm{SNR} = 
\left[ 
  \mathcal{T} \int_{f_{\min}}^{f_{\max}} 
  df 
  \left( \frac{h^2 \Omega_{\mathrm{GW}}(f)}{h^2 \Omega_{\exp}(f)} \right)^{\!2} 
\right]^{1/2} .
\end{equation}
Here $\mathcal{T}$ denotes the observation time, set to $\mathcal{T}=3$ years in this work, $h^2\Omega_{\mathrm{GW}}(f)$ is the gravitational-wave spectrum from the first-order EWPT, and $h^2\Omega_{\exp}(f)$ represents the experimental sensitivity.

Figure~\ref{fig:FOPTgw} shows only a subset of the SFOEWPT parameter space: the points for which the gravitational-wave signal is large enough to be potentially detectable. We have explicitly verified the thermal history for all points with $\mathrm{SNR} > 5$ using the gauge-invariant leading-order effective potential with NLO dimensional reduction. Every one of these points undergoes a two-step phase transition: the singlet field first acquires a non-zero vev at high temperature (symmetric $\to$ singlet phase, with $v = 0$, $w \neq 0$), followed by a first-order electroweak transition at lower temperature (singlet $\to$ EW phase, with $v \neq 0$). No one-step EWPT is found among them.

The dominance of two-step transitions among the detectable points reflects the tree-level barrier mechanism first identified in Refs.~\cite{Profumo:2007wc,Curtin:2014jma}: the portal coupling $a_2$ generates a barrier between the singlet and electroweak minima that is parametrically stronger than the loop-induced cubic term $\sim g^3 v^3$ driving one-step transitions, leading to larger latent heat and enhanced gravitational-wave production. A similar conclusion was reached in Ref.~\cite{Ramsey-Musolf:2024ykk}, where LISA-detectable signals in the xSM were found to require two-step transitions with large portal couplings. We note, however, that the two-step region involves large scalar couplings and significant supercooling, where the validity of the high-temperature expansion underlying dimensional reduction may be reduced~\cite{Niemi:2020hto}. A dedicated non-perturbative lattice study of the two-step regime, extending the work of Ref.~\cite{Niemi:2020hto}, would be valuable to assess the robustness of these predictions.

\subsection{LHC signatures \label{sec:colliders}}
Gravitational-wave observables can probe only a limited, albeit non-negligible, region of the parameter space. Consequently, complementary input from high-energy experiments—most notably collider physics—is essential. Both indirect probes, such as precision measurements of Higgs couplings, and direct probes, including resonant searches, provide powerful and complementary means of constraining the model and identifying deviations from the SM. 
The current situation based on input from Run~II and theory is presented in Section~\ref{sec:run2}.

We reiterate that this study focuses on the case $\mtwo>\mone$ and assumes no portal coupling of $h_2$ to an invisible sector. Relaxing either assumption opens up additional experimental probes. These include invisible Higgs decays~\cite{ATLAS:2022vkf,CMS:2022dwd}, $t\bar{t}+\slashed{E}_{\rm T}$ signatures~\cite{ATLAS:2021hza} if $h_2$ is long-lived, or if it is short-lived but decays into feebly interacting or non-interacting particles, as well as exotic Higgs decays~\cite{CMS:2025gzf} and diphoton resonance searches~\cite{ATLAS:2022abz,CMS:2026zsp}.

\subsubsection{Higgs precision physics}
As already emphasised in Section~\ref{sec:Higgs_coupling_mod}, precision measurements of Higgs boson properties provide a powerful indirect probe of the singlet-extended SM, since both $\delta g_{hZZ}$ in Eq.~\eqref{eq:delta_gz} and $\kappa_3$ in Eq.~\eqref{eq:k3lin} are insensitive to the mass $m_{h_2}$ of the new scalar.

The modification to the Higgs coupling can be obtained by a fit performed in the kappa-framework \cite{deBlas:2019rxi}. Because all coupling modifiers are identical in this model, an even more direct constraint comes from the total Higgs signal strength: 
\begin{equation}
    \mu_{h_1} = \frac{\sigma(pp\rightarrow h_1)_{\rm measured}}{\sigma(pp\rightarrow h_1)_{\rm SM}} = \cos^2 \theta 
\end{equation}
that includes information from all the Higgs production and decay channels. 

The current uncertainty on this quantity can be obtained by combining the full Run~II ATLAS~\cite{ATLAS:2022vkf} and CMS~\cite{CMS:2018uag} results
\begin{equation}
\mu_H ({\rm ATLAS}) = 1.05\pm0.06 \, ,  \qquad \mu_H ({\rm CMS})= 1.02^{+0.07}_{-0.06} \,, 
\end{equation}
obtaining, after symmetrising and combining the uncertainties
\begin{equation}
\mu_H ({\rm ATLAS+CMS, Run\, II}) =  1.036\pm 0.044\,.
\end{equation}

Since we are going to compare this result with the HL-LHC and FCC-ee+FCC-hh projections, we use only the uncertainty from Run~II assuming the central value to be equal one. This will give a 95\% C.L. limit on the angle
\begin{equation} \label{eq:theta-run2}
\theta< 0.30 \Rightarrow \delta g_{hZZ} \le 4.5 \cdot 10^{-2} \,(\,{\rm Run\, II}\, \,95\% {\rm C.L. }\,) \, . 
\end{equation}
The most recent $95\%$ C.L.  on the Higgs trilinear coupling $-1.7<\kappa_3<6.6$, obtained by the ATLAS collaboration~\cite{ATLAS:2025hhd} with 308 fb$^{-1}$ of luminosity, does not allow one to place any relevant exclusion on the model.

\subsubsection{Direct searches}
Direct searches, namely resonant searches, depend on a combination of several model parameters and therefore cannot be straightforwardly translated into universal, mass-independent bounds. Nevertheless, as we will show, they are highly effective in testing the model.
We discuss branching ratios and  the validity of the narrow width approximation (NWA) in the parameter space.

The LO expressions for the decay rates of  $h_2$ into SM vector bosons and fermions read
\begin{align}
\Gamma(h_2\rightarrow f\bar f) &=\sin^2\theta N_c \frac{y_f^2}{16\pi}\mtwo\left(1-\frac{4m_f^2}{\mtwo^2}\right)^\frac{3}{2}\,,\\
\Gamma(h_2\rightarrow ZZ) &=\frac{1}{32\pi} \sin^2\theta \frac{\mtwo^2}{v_H^2}\mtwo \sqrt{1-\frac{4m_Z^2}{\mtwo^2}}\left(1-\frac{4m_Z^2}{\mtwo^2}+3\frac{4m_Z^4}{\mtwo^4}\right)\,,\\
\Gamma(h_2\rightarrow W^+W^-) &=\frac{1}{16\pi} \sin^2 \theta \frac{\mtwo^2}{v_H^2}\mtwo \sqrt{1-\frac{4m_W^2}{\mtwo^2}}\left(1-\frac{4m_W^2}{\mtwo^2}+3\frac{4m_W^4}{\mtwo^4}\right)\,.
\end{align}

Since we are interested in the region of the parameter space where $\mtwo>\mone$ an additional decay channel is available for $\mtwo \geq 2\mone$
\begin{align}
\Gamma(h_2 \rightarrow h_1 h_1) = \frac{1}{32\pi}\frac{1}{\mtwo}g_{211}^2\sqrt{1-\frac{4\mone^2}{\mtwo^2}}\,,\label{eq:211decay}
\end{align}
where 
\begin{align}
g_{211} &= \sin \theta  \left[-\frac{\lM}{2} v_H(1+3\cos 2\theta)+ \mthree \sin 2
   \theta + \frac{2}{v_H} \left(m_{h_1}^2+\frac{m_{h_2}^2}{2}\right) \cos^2\theta\right]\,.\label{eq:g211}
\end{align}

As in the previous section, it is instructive to  expand  in small $\theta$ 
\begin{align}\label{eq:211approx_a2}
g_{211} \simeq& \frac{2}{v_H}\left( m_{h_1}^2+\frac{m_{h_2}^2}{2}- \lM v_H^2\right) \theta
   +2 \mthree \theta ^2+O\left(\theta ^3\right)\,.
\end{align}
We learn that for each value of the mass there exists a non-zero value of $a_2$ for which this decay channel is strongly suppressed, namely

\begin{equation}
   a_2 \simeq \frac{1}{v_H^2}\left(\mone^2+\frac{\mtwo^2}{2}\right) \Rightarrow \Gamma (h_2\rightarrow h_1 h_1) \simeq \mathcal O(\theta^4)\label{eq:Brsuppression}\,,
\end{equation}
leading to an untestable region in the $h_2 \to h_1 h_1$ decay.
Whenever possible, we adopt the BSM scalar decay rates provided in Ref.~\cite{LHCHiggsCrossSectionWorkingGroup:2016ypw}, which include QCD corrections to the decay widths.

\begin{figure}[ht]
    \centering
    \includegraphics[width=0.48\linewidth]{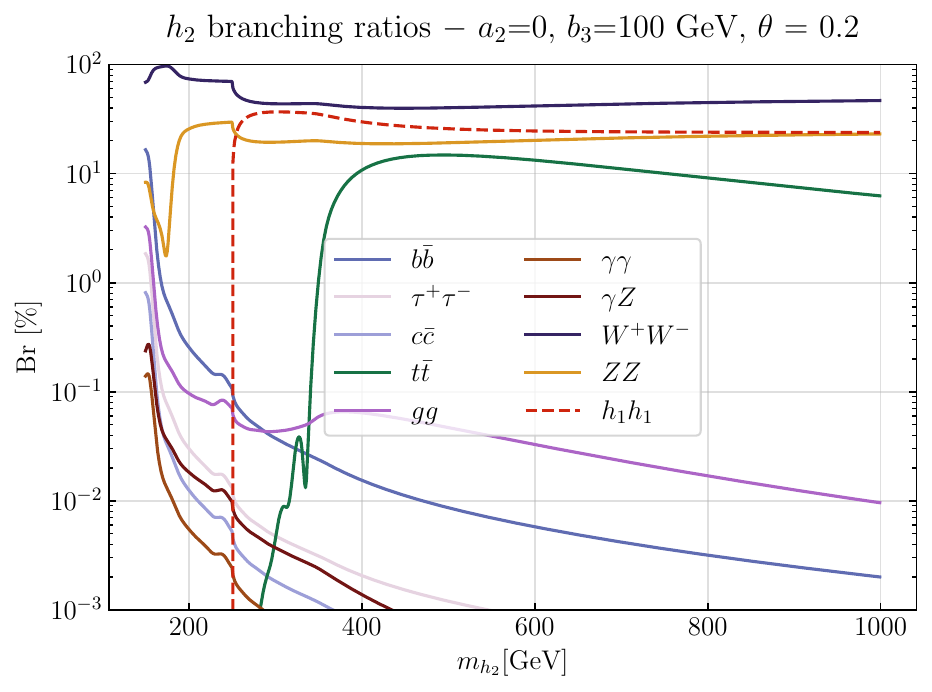}
    \includegraphics[width=0.48\linewidth]{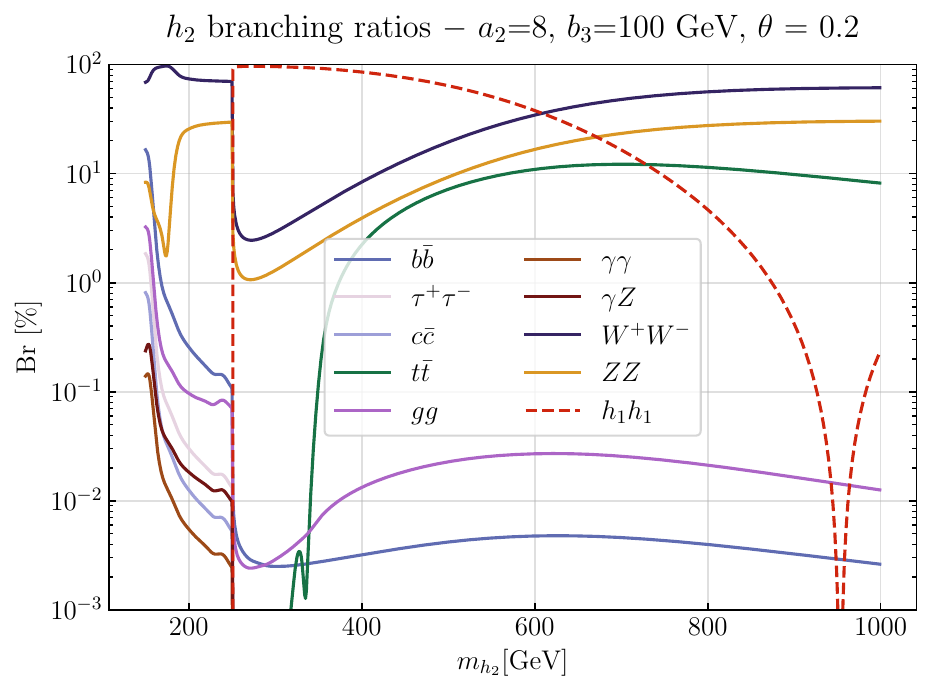}
    \caption{Branching ratios of the BSM scalar $h_2$ as a function of its mass. The left and right panels correspond to different values of $a_2$, chosen at the extrema of the range supporting a strong first-order phase transition. The $h_2\to h_1h_1$ branching-ratio curve is dashed to highlight the strong dependence on the branching ratio minimum on the value of $a_2$.}
    \label{fig:Branchingratios}
\end{figure}

Figure~\ref{fig:Branchingratios} shows the branching ratios of the new scalar $h_2$ into SM particles for representative values of $a_2$ corresponding to the edges of the parameter space compatible with a strong first-order phase transition. As is well known from studies of the SM Higgs boson~\cite{ParticleDataGroup:2024cfk}, the dominant decay modes are into electroweak vector bosons, $h_2 \to ZZ$ and $h_2 \to W^+W^-$, as well as into top-quark pairs. The purely BSM channel $h_2 \to h_1 h_1$ which opens up at $m_{h_2} \geq 250 \, \text{GeV}$ is also phenomenologically relevant; however, it can be significantly suppressed for specific combinations of $a_2$ and $m_{h_2}$, see Eq.~\eqref{eq:Brsuppression}, which is  clearly visible in the right panel of Fig.~\ref{fig:Branchingratios}. We do not display the value of $b_4$, since at leading order it is phenomenologically irrelevant for collider observables.

\begin{figure}[ht]
    \centering
    \includegraphics[width=0.48\linewidth]{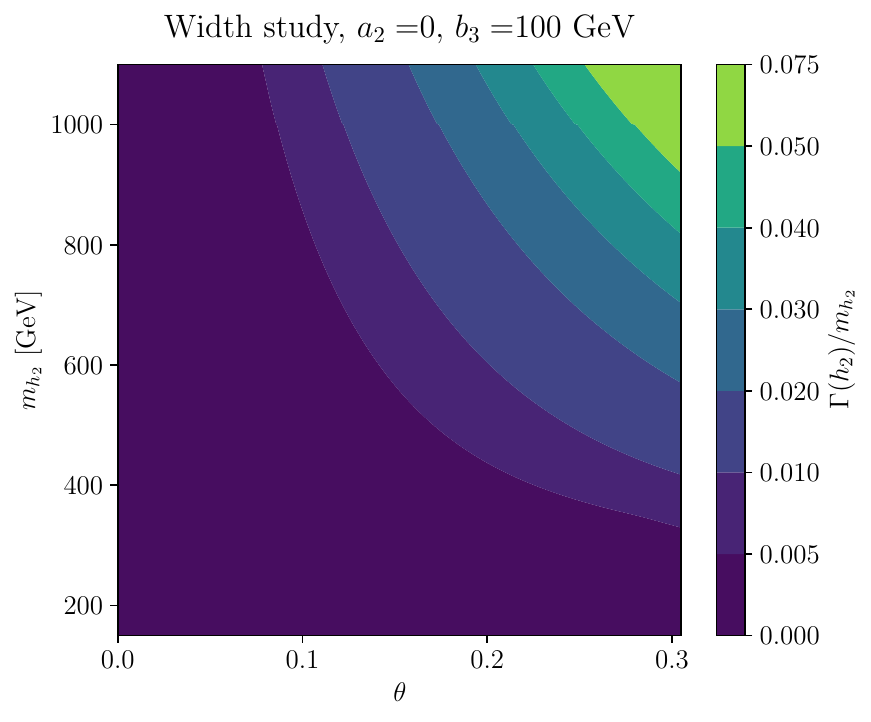}
    \includegraphics[width=0.48\linewidth]{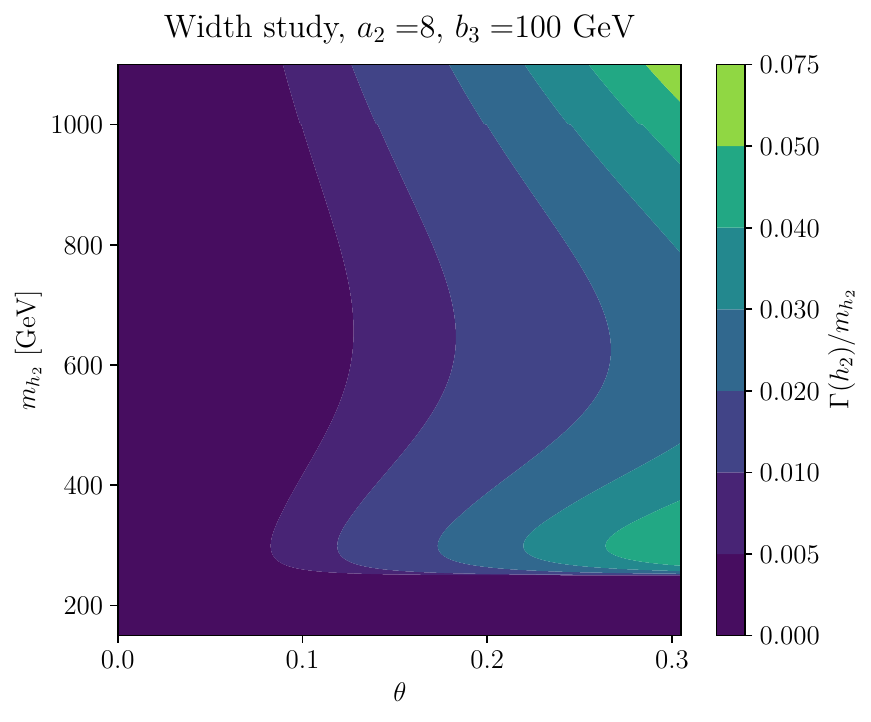}
    \caption{Ratio of the BSM scalar width $\Gamma(h_2)$ to its mass $\mtwo$. The left and right panels correspond to different values of $a_2$, chosen at the extrema of the range compatible with a strong first-order phase transition. }
    \label{fig:NWA}
\end{figure}

In the following, we exploit resonant searches to constrain the model couplings. In order to safely neglect SM-BSM interference and focus exclusively on resonant production, the NWA must be valid throughout the parameter space under consideration. In Fig.~\ref{fig:NWA}, we present the contours of $\Gamma(h_2)/m_{h_2}$ in the $\theta$--$m_{h_2}$ plane. 
Our analysis is restricted to the region where $m_{h_2}$ lies close to the values relevant for a strong first-order phase transition, in particular in our resonant analysis we never go beyond $m_{h_2}=800$ GeV, for which we can see that, already using the limit on the mixing angle $\theta$ coming from Run~II  (\ref{eq:theta-run2})  the ratio between the width and the mass of the new scalar is below $5\%$.

The total width of the new scalar depends on several model parameters
\begin{equation}
    \Gamma_{h_2} = \Gamma_{h_2}(\theta, m_{h_2}, a_2, b_3).
\end{equation}
Therefore, a proper interpretation of the limits derived from resonant searches requires fixing a specific combination of input parameters. Nevertheless, one can make the general observation that the full process of new-scalar production and decay can be factorised as
\begin{equation}\label{eq:decomposition}
\sigma_{pp\rightarrow h_2\rightarrow XX}(\theta,\mtwo,a_2,b_3) = \sin^2\theta \sigma_{pp\rightarrow h_2}(\mtwo)\cdot {\rm Br}_{h_2 \rightarrow XX}(\theta,\mtwo,a_2,b_3)\,, 
\end{equation}
where $\sigma_{pp\rightarrow h_2}(\mtwo)$ is the production for a Higgs-like boson of mass $\mtwo$ in the SM. Writing the cross section as in Eq.~\eqref{eq:decomposition} allows us to study in the $\theta-\mtwo$ plane
which is the minimum branching ratio $ {\rm Br}_{h_2 \rightarrow XX}$ needed for a 95\% C.L. exclusion given the corresponding experimental search.
At the LHC, searches for a heavy scalar boson have already been conducted in both the $h_2 \to h_1 h_1$~\cite{CMS:2024phk} and $h_2 \to ZZ$~\cite{CMS:2024vps,Collaboration:2927677} channels.
\begin{figure}[ht]
    \centering
    \includegraphics[width=0.48\linewidth]{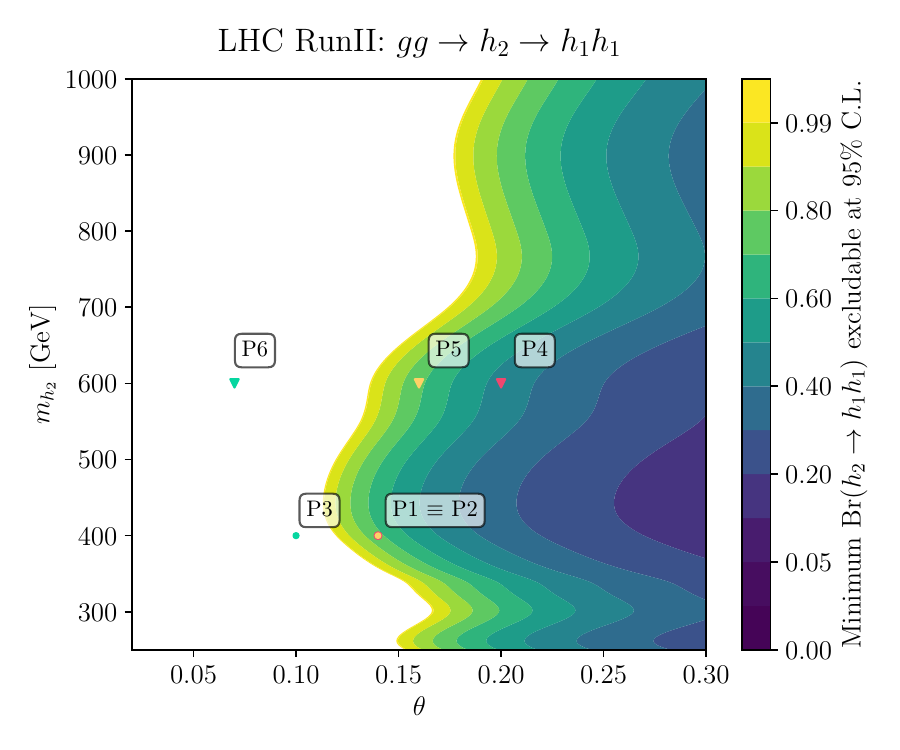}
    \includegraphics[width=0.48\linewidth]{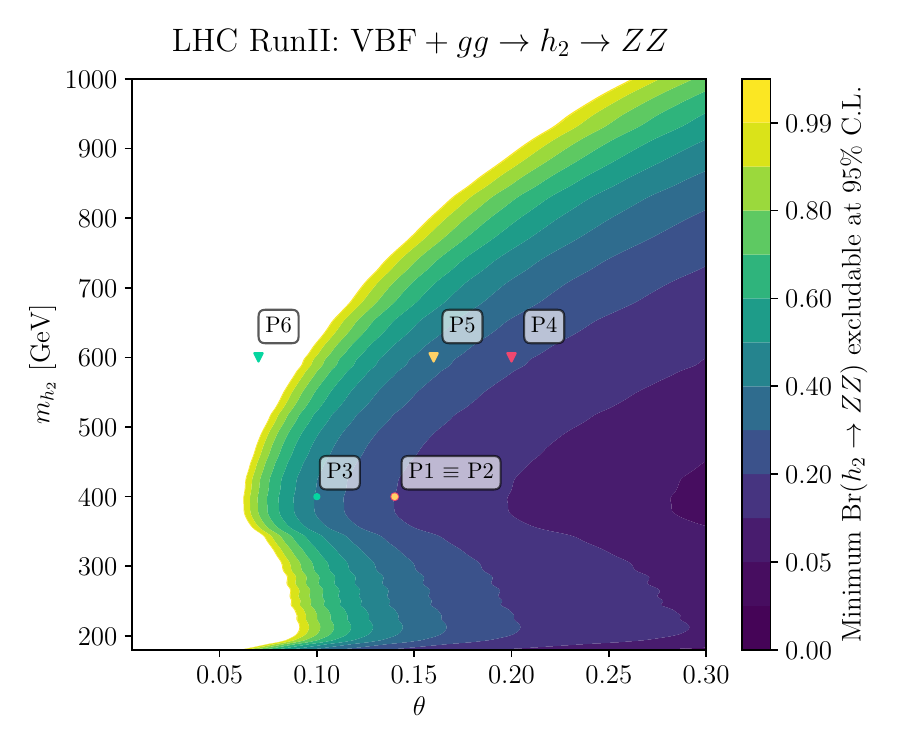}
    \caption{ Minimum branching ratio excludable at $95\%$ C.L. with Run~II LHC data for the $h_2 \to h_1 h_1$ decay channel (left) and the $h_2 \to ZZ$ decay channel (right). The benchmark points P1-P6 appearing here and in Fig.~\ref{fig:run2comparison} are discussed in Sec.~\ref{sec:run2}}.
    \label{fig:minbrrun2}
\end{figure}

 In Fig.~\ref{fig:minbrrun2},  we show the reach of Run~II searches in terms of this simplistic interpretation. The production cross sections have been computed by rescaling the NNLO QCD+NLO EW predictions  by $\sin^2\theta$~\cite{LHCHiggsCrossSectionWorkingGroup:2016ypw}.\footnote{Detailed tables can be found in the related  \href{https://twiki.cern.ch/twiki/pub/LHCPhysics/LHCHWG/Higgs_XSBR_YR4_update.xlsx}{twiki} \cite{TwikiBSM}.} We can already notice that in both channels a large part of the parameter space cannot be tested, even for very large branching ratios.

\begin{figure}
    \centering
\includegraphics[width=0.48\linewidth]{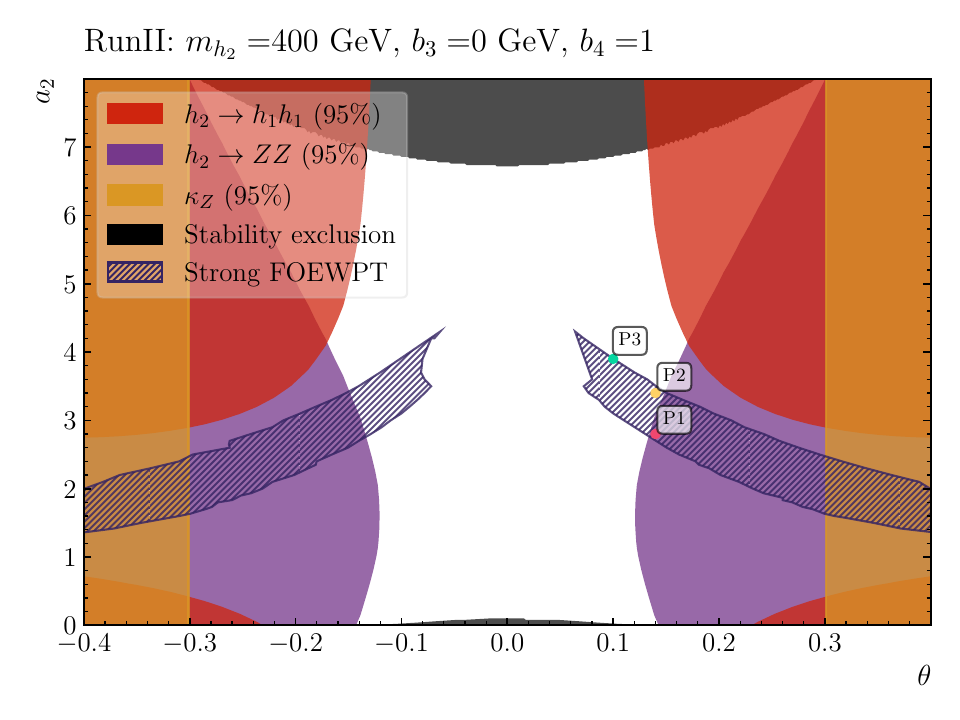}
\includegraphics[width=0.48\linewidth]{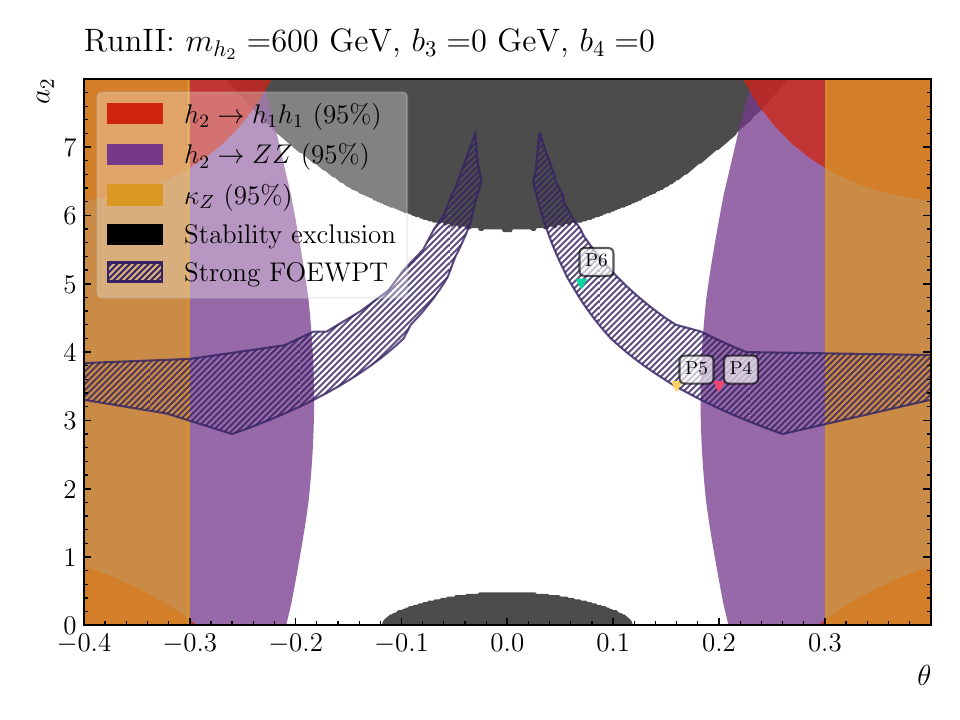}
    \caption{ Exclusion regions for LHC Run~II for different points in the parameter space: left $\mtwo=400$ GeV$,b_3=0,\, b_4=1$, right: $\mtwo=600$ GeV$,b_3=0,\, b_4=0$. Current data on $\kappa_3$ are too weak to relevantly constrain the
    parameter space shown. The benchmark points P1-P6 appearing here and in Fig.~\ref{fig:minbrrun2} are discussed in the main text.}
    \label{fig:run2comparison}
\end{figure}

\subsection{Current combined constraints \label{sec:run2}}
We now present the Run~II status for a set of representative benchmark points, combining direct and indirect collider constraints with the bounds imposed by vacuum stability\footnote{Here, we include metastability as part of the stability exclusion.}.
We then identify which
regions of the $\theta$--$a_2$ plane compatible with a strong first-order phase
transition are excluded by current data. Despite the relative simplicity and
phenomenological accessibility of the model, featuring only one additional scalar
state with a mass within the LHC centre-of-mass energy reach, Fig.~\ref{fig:run2comparison}
shows that a sizeable fraction of the parameter space remains unconstrained by
present LHC searches.
In both Figs.~\ref{fig:minbrrun2} and \ref{fig:run2comparison}, we highlight six benchmark points connecting the exclusion at the branching-ratio level to the exclusion in the $\theta-a_2$ plane.  Points P1 and P2 have the same mass but different values of $a_2$. Although both lie, in principle, in the region that can be probed by resonant $ZZ$ searches, P2 is not excluded because its branching ratio into $ZZ$ is reduced by the increased decay width into Higgs pairs. Conversely, the $h_1h_1$ branching ratio is not large enough to exclude either P1 or P2 in the di-Higgs channel. 
Point P3 behaves similarly to P2 in the $ZZ$ channel. In the $h_1h_1$ channel, however, even assuming a branching ratio equal to one, the production rate is too small to yield an exclusion; correspondingly, P3 lies in the white region of the left panel of Fig.~\ref{fig:minbrrun2}. At $m_{h_2}=600~\mathrm{GeV}$, P4 is the only benchmark within the Run~II reach, and only in the $ZZ$ channel. Point P5 does not have a sufficiently large branching ratio in either of the two channels, while P6 lies in the white ``unreachable'' region of Fig.~\ref{fig:minbrrun2} for both searches.
 
\section{Impact of future colliders \label{sec:impact}}

The limited reach of current data strongly motivates further searches in the Higgs sector, both in the near future at the HL-LHC and on longer timescales at future facilities such as FCC-ee and FCC-hh. In particular, as we will show, the HL-LHC has the potential to exclude a large fraction of the parameter space. However, in the event of an observed excess, a hadron collider operating at the $O(100~\mathrm{TeV})$  energy frontier, such as FCC-hh, would likely be required to achieve a definitive discovery. In the following, we explore this synergy in detail.

\subsection{HL-LHC reach \label{sec:high-lumi}}
Projected constraints at the HL-LHC~\cite{CMS:2025hfp} are expected to reach 
\begin{equation}
\kappa_Z = 1 \pm 0.016 \quad \text{(68\% C.L.)},
\end{equation}
translating into a $95\%$ confidence level limit on the mixing angle,
\begin{equation}\label{eq:thetabound}
|\theta| < 0.254 \quad \text{(95\% C.L.)}.
\end{equation} 

Assuming universality of the Higgs coupling modifications, i.e., $\kappa_i = \kappa_{\mathrm{univ}}$, yields an even stronger limit,
\begin{equation}
|\theta| < 0.18\quad \text{(95\% C.L.)}.
\end{equation}
This would roughly correspond to the bound obtained from the total Higgs signal strength. However, we refrain from including this constraint in our final results, as it is not officially provided by the experimental collaborations.

The HL-LHC is expected to deliver the first determination of the Higgs cubic self-coupling with an uncertainty better than 30\%~\cite{CMS:2025hfp}, namely
\begin{equation}\label{eq:k3bound}
\kappa_3 = 1.00^{+0.29}_{-0.26} \quad (68\%~\text{C.L.}) \,. 
\end{equation}
This measurement provides a new and important probe of the singlet-extended model at the HL-LHC and at future colliders. Although the projected uncertainty is roughly a factor of twenty larger than that expected for the determination of the Higgs coupling to the $Z$ boson, this is not necessarily a limiting factor. Indeed, as discussed in Section~\ref{sec:Higgs_coupling_mod}, in the phenomenologically relevant region of parameter space the deviation in the Higgs self-coupling can be up to about seventy times larger than the corresponding deviations in Higgs couplings to other SM particles,~\eqref{eq:k3dgz}.

We remind the reader that $\kappa_3$ in Eq.~\eqref{eq:k3lin} is defined at LO. Therefore, in the plots presented in this section, any NLO BSM effect that modifies the Higgs trilinear coupling is not included. However, in the alignment limit, $\theta\to 0$, these NLO corrections provide the dominant BSM contribution to non-resonant di-Higgs production. Indeed, at tree level the modification of the Higgs trilinear coupling is suppressed by the small mixing angle $\theta$, namely
\begin{equation}
\Delta^{\rm tree}_{\theta\to 0} g_{h_1h_1h_1}
=
\left(3a_2v_H-\frac{9}{2}\frac{\mone^2}{v_H}\right)\theta^2 \, .
\end{equation}
In the plots of the SFOEWPT points, we highlight in pink those points satisfying
\begin{equation}\label{eq:NLOcond}
\left|\Delta^{\rm NLO}_{\rm approx} g_{h_1h_1h_1}\right|
>
\left|\Delta^{\rm tree}_{\theta\to 0} g_{h_1h_1h_1}\right| \, ,
\end{equation}
where $\Delta^{\rm NLO}_{\rm approx} g_{h_1h_1h_1}$ is defined in Eq.~\eqref{eq:deltaNLO}. This highlights the special role of these points, for which the approximate NLO contribution to the trilinear coupling is larger than the corresponding tree-level contribution in the alignment limit.

Let us first consider the direct searches for resonances at the HL-LHC. 

As shown in the previous section, the region of parameter space that is not accessible through precise measurements of the Higgs couplings can be probed via resonant searches, thanks to the small width-to-mass ratio of $h_2$.  The projected $95\%$ C.L. upper limits on these two cross sections under the NWA have been obtained by combining the expected results from ATLAS and CMS~\cite{CMS:2025hfp} for gluon fusion production (relevant for both $h_2 \to h_1 h_1$ and $h_2 \to ZZ$ channels) and vector boson fusion (applicable to $h_2 \to ZZ$ only). 
\begin{figure}[ht]
    \centering
    \includegraphics[width=0.48\linewidth]{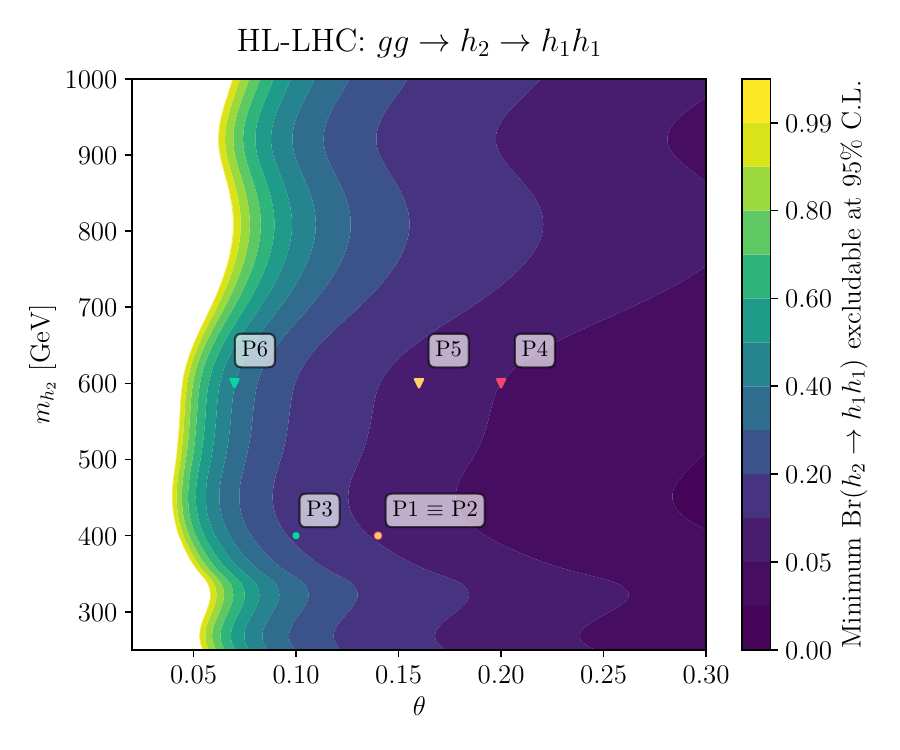}
    \includegraphics[width=0.48\linewidth]{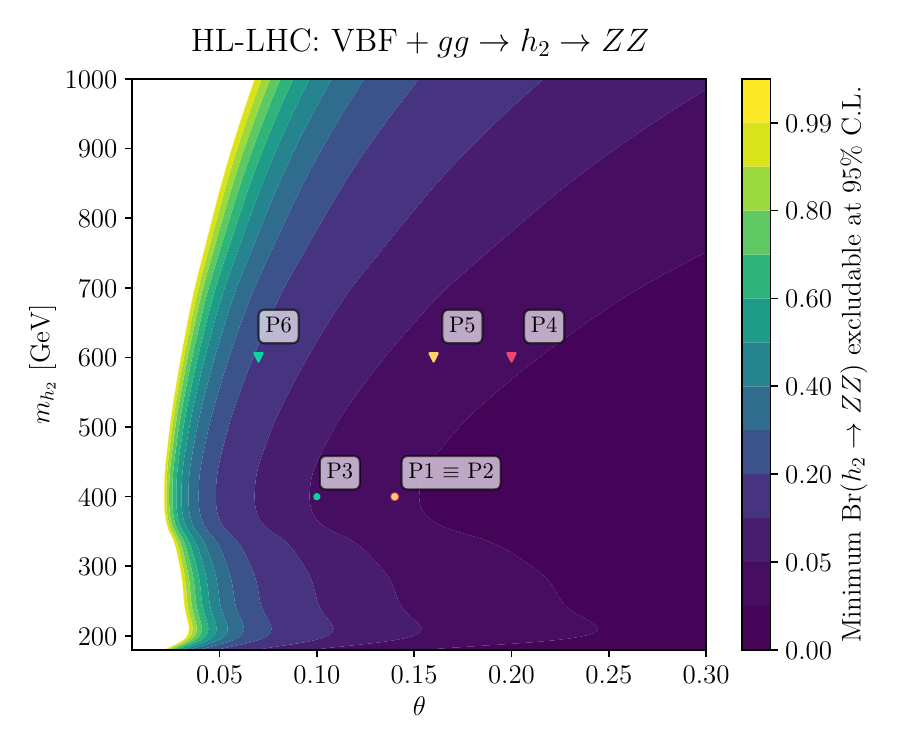}
    \caption{Minimum branching ratio excludable at $95\%$ C.L. at the HL-LHC for the $h_2 \to h_1 h_1$ decay channel (left) and the $h_2 \to ZZ$ decay channel (right). The benchmark points P1-P6 are projected also in the views in  Figs.~\ref{fig:mhl00}-\ref{fig:mhl01_direct} and are discussed in the main text.}
    \label{fig:minbrHL}
\end{figure}
We reproduce the same intuitive analysis of Run~II, showing the minimum branching ratio for each channel in the $\theta-\mtwo$ plane allowing an exclusion. The results of this preliminary analysis are shown in Fig.~\ref{fig:minbrHL}, the HL-LHC is capable of probing almost the entirety of the parameter space, reaching significantly smaller mixing angles and excluding lower branching ratios at each point. 

Focusing on Fig.~\ref{fig:minbrHL}, we observe that the sensitivity of the $h_2 \to h_1 h_1$ and $h_2 \to ZZ$ channels is comparable, with the vector boson channel performing slightly better. Nonetheless, both searches remain crucial, as deviations observed in one channel are expected to manifest in the other, making it possible to combine the two channels in order to reach an even better sensitivity. None of the P1-P6 points lies in the white region and as appears in Fig.~\ref{fig:minbrHL} all points can be tested via the $ZZ$ decay channel while P4-P5-P6 are still branching-ratio limited to be excluded/observed in the $h_2\rightarrow h_1h_1$ channel,
as can be observed in Fig.~\ref{fig:mhl01_direct} where the three points are contained in the $ZZ$ exclusion contour but not in the di-Higgs one.
The two channels can in principle be statistically combined to improve the exclusion and discovery reach, as done in Ref.~\cite{Zhang:2023jvh}. However, we will show that in most cases the $ZZ$ channel alone is sufficient to determine whether a given point in parameter space is excluded at the HL-LHC.

In addition, we decided to keep the two analyses separate. Indeed, a deviation observed in the $ZZ$ channel without a corresponding deviation in the $h_1h_1$ channel, where one would otherwise be expected, could signal the presence of new physics not necessarily associated with this model or, more specifically, not directly linked to modifications of the Higgs potential.

For this reason, later in this section, when presenting the FCC-hh discovery reach, we focus on the $h_1h_1$ channel in conjunction with $\kappa_3$, thus restricting our attention to probes that directly target the Higgs potential. This approach is chosen to specifically target the model under consideration and, more broadly, the possibility of realising a SFOEWPT.

We proceed by analysing  the interplay between collider probes, stability, and SFOEWPT requirements for the new scalar for different points in the parameter space in the HL scenario.

As shown in Fig.~\ref{fig:FOPTpoints}, a strong first-order phase transition can be realised over a wide mass range, extending from just above the Higgs mass up to nearly $m_{h_2} = 1$~TeV. In particular, Fig.~\ref{fig:FOPTpoints} already illustrates how the parameter space can be naturally divided into different mass regions, each suited to distinct resonant search strategies. In this section, for the HL-LHC projections we adopt the upper limits from resonant searches obtained in the NWA as reported in Ref.~\cite{CMS:2025hfp}.

\begin{figure}[ht]
    \centering
    \includegraphics[width=0.65\linewidth]{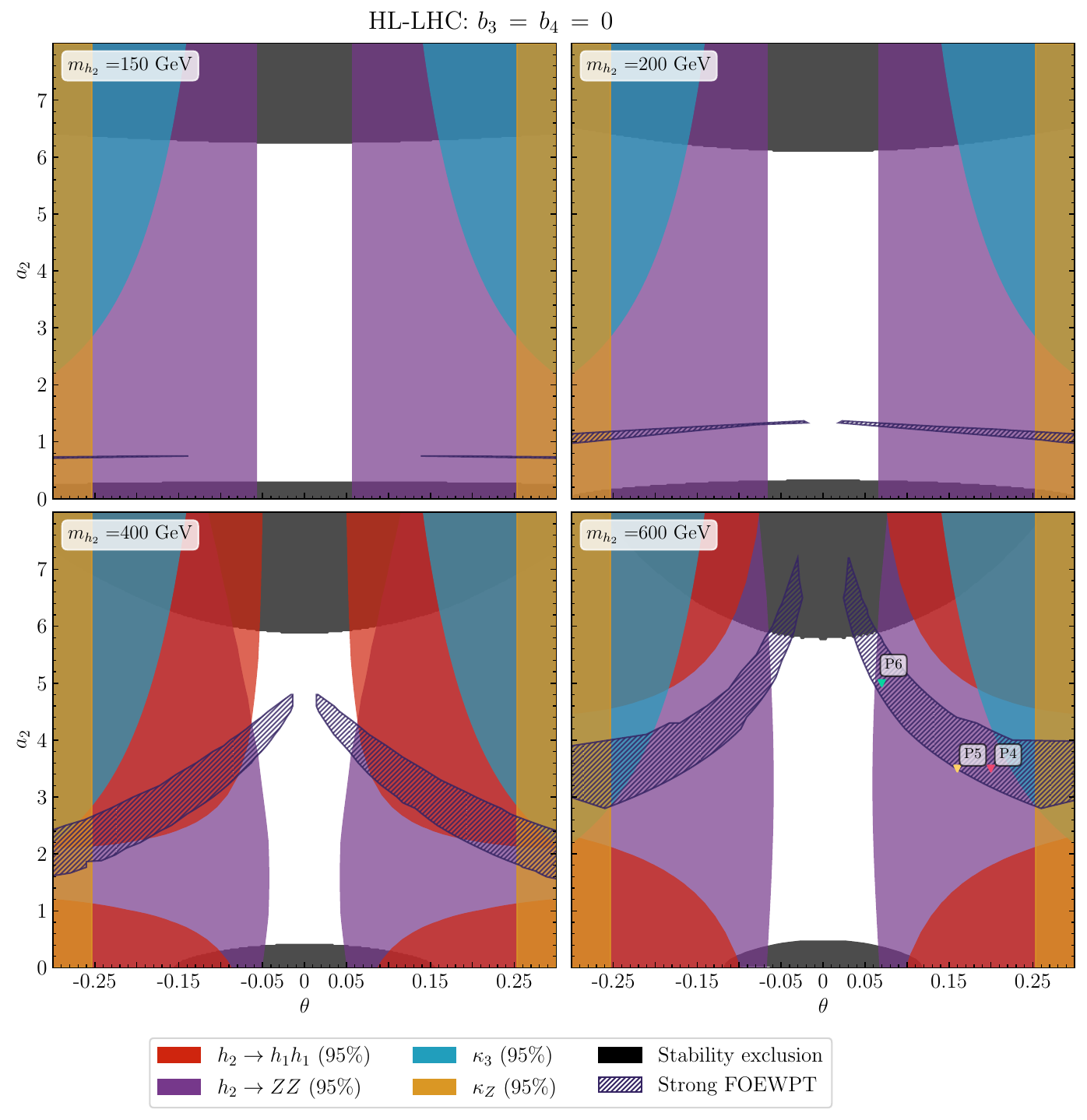}
    \caption{HL-LHC combined constraints for $b_3=b_4=0$ with masses (from left to right, top to bottom) $\mtwo = 150,200,400,600$ GeV in the $\theta-a_2$ plane. The exclusion regions at the 95\% C.L., for resonant searches we used the narrow width approximation. The dashed areas in each plot show the region in the parameter space inducing a SFOEWPT. The black region is excluded by requiring vacuum stability up to the Planck scale and the absence of sub-Planckian Landau poles. The benchmark points P4-P6 and their connection with Fig.~\ref{fig:minbrHL} are discussed in the main text.}
    \label{fig:mhl00}
\end{figure}
At first sight, one might expect most of the region with $\kappa_3 < 1$ to be inaccessible to resonant searches, since the mass of the new scalar lies below twice the $Z$-boson mass, forbidding decays into two on-shell $Z$ bosons.
However, as shown in Fig.~\ref{fig:minbrHL}, exclusions remain possible even below the $2m_Z$ threshold through decays of the scalar into off-shell vector bosons ($ZZ^*$), down to relatively small branching ratios.

\begin{figure}[ht]
    \centering
    \includegraphics[width=0.75\linewidth]{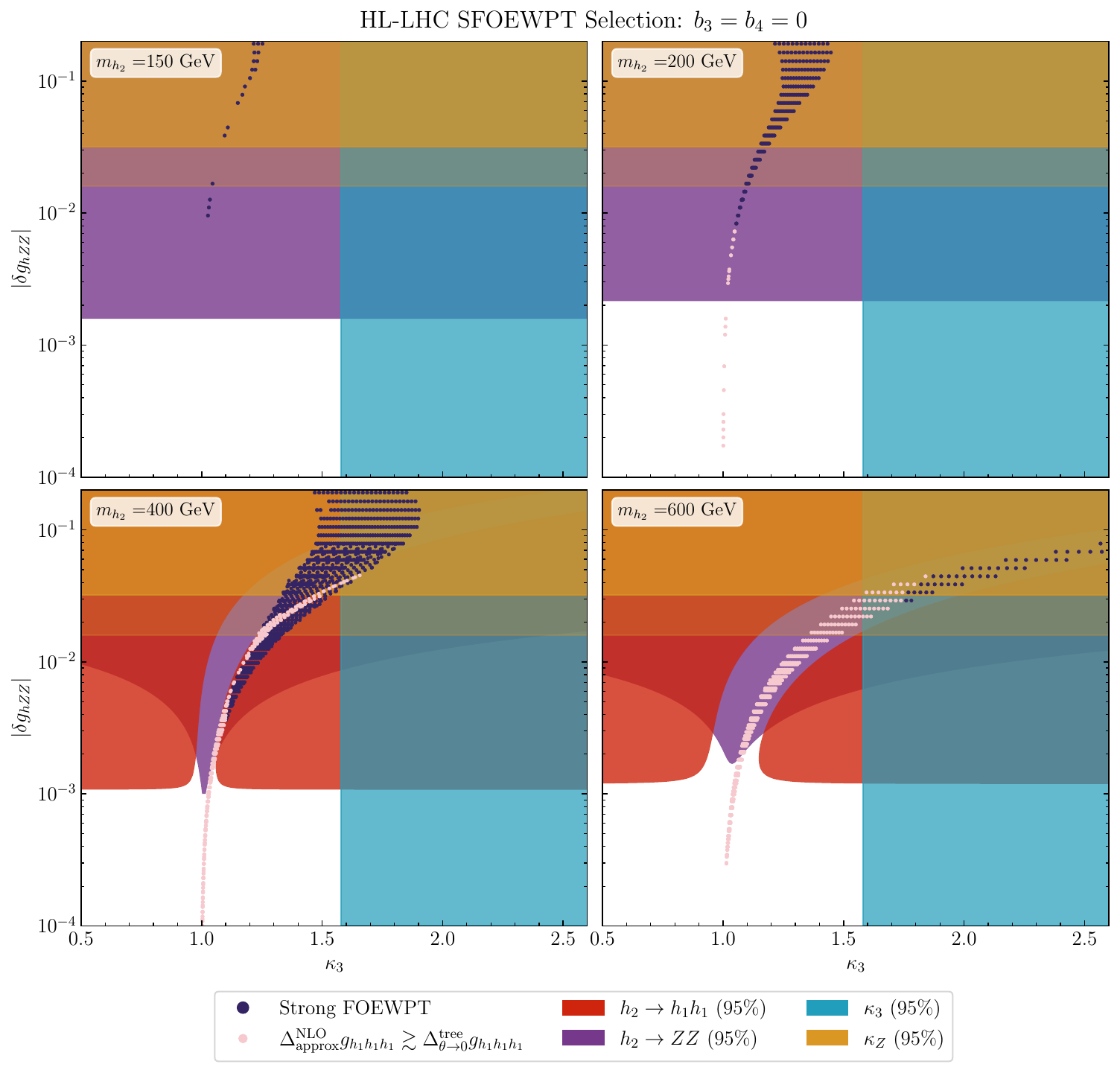}
    \caption{
HL-LHC combined constraints for $b_3=b_4=0$ and masses, from left to right and top to bottom, $\mtwo = 150, 200, 400, 600~\mathrm{GeV}$, shown in the $\kappa_3$--$\delta g_{hZZ}$ plane. The exclusion regions are drawn at 95\% C.L.; for resonant searches, we use the narrow width approximation. The blue markers in each plot denote the points in parameter space that induce a SFOEWPT. The pink points satisfy the condition in Eq.~\eqref{eq:NLOcond}. Although some of them may initially appear to be beyond the HL-LHC reach, this conclusion can be misleading: for these points, the BSM NLO contributions to the Higgs trilinear coupling are larger than the LO ones. Consequently, a complete NLO calculation of di-Higgs production could be decisive for assessing their testability.}
    \label{fig:mhl00_indirect}
\end{figure}

\begin{figure}[ht]
    \centering
    \includegraphics[width=0.65\linewidth]{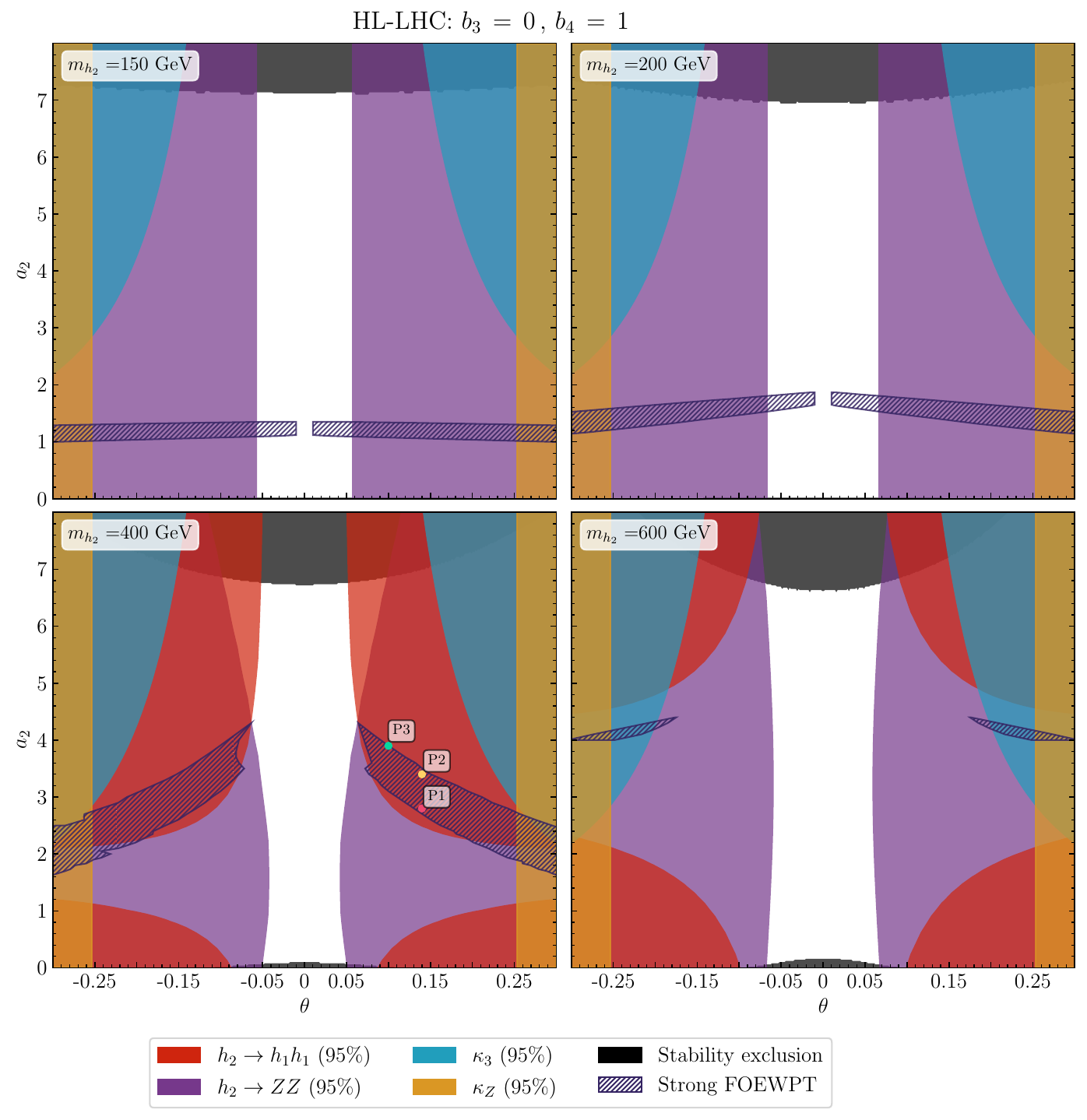}
    \caption{Same as Fig.~\ref{fig:mhl00} for the benchmark point $b_3=0,b_4=1$. The benchmark points P1-P3 and their connection with Fig.~\ref{fig:minbrHL} is discussed in the main text.}
    \label{fig:mhl01_direct}
\end{figure}

This can also be seen in Fig.~\ref{fig:mhl00} where we  present the exclusion curves together with the regions excluded by vacuum stability and the parameter space compatible with a strong first-order phase transition for the parameter choice $b_3=b_4=0$ for four different mass benchmarks $\mtwo\in\{150,200,400,600\} \text{ GeV}$. The top-left panel shows that, for this particular choice of parameters and for $\mtwo<2m_Z$, resonant searches in the $h_2\to ZZ$ channel can still probe the entire SFOEWPT-compatible parameter space. In the top-right panel, for $\mtwo=200~\mathrm{GeV}$, decays into two on-shell $Z$ bosons become accessible. Interestingly, despite crossing the $ZZ$ threshold, the exclusion power is slightly reduced because of  the shape of the expected limits~\cite{CMS:2025hfp,CMS:2024vps,Collaboration:2927677}, with the upper bound being more stringent for lower values of $\mtwo$. We see here that a small part of the parameter space,  characterised by  small mixing-angle $\theta$, contains a slice of the SFOEWPT region that appears unexcluded, however as already mentioned before, for such small angles a higher order computation of the non-resonant di-Higgs  production  would be necessary in order to draw robust conclusions.
For scalar masses above the $2m_{h_1}$ threshold, i.e. $\mtwo \gtrsim 250~\mathrm{GeV}$, decays into pairs of Higgs bosons open up a promising additional search channel. We illustrate this behaviour for two representative scenarios, $\mtwo=400~\mathrm{GeV}$ and $\mtwo=600~\mathrm{GeV}$, shown in the bottom-left and bottom-right panels, respectively.

In the former case, a large portion of the parameter space lies in a region where a deviation from the SM expectation could be observed simultaneously in both the $h_2 \to ZZ$ and $h_2 \to h_1h_1$ channels.
As discussed in Section~\ref{sec:funnel}, larger scalar masses require larger values of the portal coupling $a_2$ in order to realise a strong first-order phase transition. As shown in Eq.~\eqref{eq:k3dgz}, large values of $a_2$ induce sizeable deviations in the Higgs self-coupling $\kappa_3$, which can become more constraining than the precision measurements of the Higgs--$Z$ boson coupling. This behaviour is particularly evident in the $\mtwo = 600$~GeV benchmark shown in the bottom-right panel of Fig.~\ref{fig:mhl00}.

In this case, the decay into a pair of Higgs bosons is not sufficiently enhanced to be observed above the SM background, so that the $ZZ$ final state remains the primary resonant search channel. Nevertheless, a small fraction of the SFOEWPT-compatible parameter space can still produce deviations in $\kappa_3$ within the reach of the HL-LHC. At larger scalar masses and small mixing angles, some regions with large portal couplings $a_2$ remain compatible with a strong first-order phase transition. Here, however, vacuum stability plays a crucial role, excluding parts of the parameter space that would otherwise remain inaccessible to HL-LHC searches.
It is worth noting that the vacuum-stability constraints, shown as black regions in Fig.~\ref{fig:mhl00}, are complementary to both collider constraints and the requirement of a sufficiently strong first-order phase transition. In particular, stability excludes overly large portal couplings $a_2$, which would lead to sub-Planckian Landau poles, as well as too small or negative values, which would fail to stabilise the electroweak vacuum; see Fig.~\ref{fig:BSMsurface}. We also recall that larger values of the BSM quartic coupling $b_4$ allow stability to be achieved with smaller values of the portal coupling $a_2$~\cite{Hiller:2024zjp}.

A complementary view to Fig.~\ref{fig:mhl00} is provided in Fig.~\ref{fig:mhl00_indirect}, where the SFOEWPT points are shown in the $\delta g_{hZZ}$--$\kappa_3$ plane. This representation makes even clearer the feature already highlighted for the $\mtwo=200~\mathrm{GeV}$ benchmark: the HL-LHC can probe essentially the entire parameter space, except for points close to the alignment limit. For these points, however, definitive conclusions require a full NLO computation, since all points in this region satisfy Eq.~\eqref{eq:NLOcond}, as indicated by the pink colouring.

To study the dependence of the SFOEWPT region on $b_4$, we repeat the same analysis for the benchmark point $b_3=0, b_4=1$, using the same set of scalar masses. The results are shown in Fig.~\ref{fig:mhl01_direct}. 
When comparing the two scenarios, it is useful to recall the observation made in Sec.~\ref{sec:funnel}, and in particular in Fig.~\ref{fig:funnel_points}: in the alignment limit, there is a non-trivial correlation among $\mtwo$, $a_2$ and $b_4$. At lower masses, the SFOEWPT is favoured by larger values of $b_4$, so that the SFOEWPT region grows when moving from the top panels of Fig.~\ref{fig:mhl00} to those of Fig.~\ref{fig:mhl01_direct}. In the bottom panels, however, the opposite behaviour is observed: as the scalar mass increases, large values of $b_4$ become disfavoured. This leads to a mild reduction of the SFOEWPT region for $\mtwo=400$ GeV and to a severe reduction for $\mtwo=600$ GeV.

Lastly, we study the effect of the trilinear $b_3$ term. We learn that, for larger scalar masses, its impact is subleading and mainly amounts to a distortion of the SFOEWPT region. For lighter masses, where smaller values of $a_2$ are required to realise a SFOEWPT, $b_3$ can instead play a decisive role in determining the overall sign of the corrections to the Higgs trilinear coupling in Eq.~\eqref{eq:k3_linearapprox}. In particular, large positive (negative) values of $b_3$ can lead to $\kappa_3<1$ for positive (negative) values of the mixing angle $\theta$.

\begin{figure}[t]
    \centering
    \includegraphics[width=0.65\linewidth]{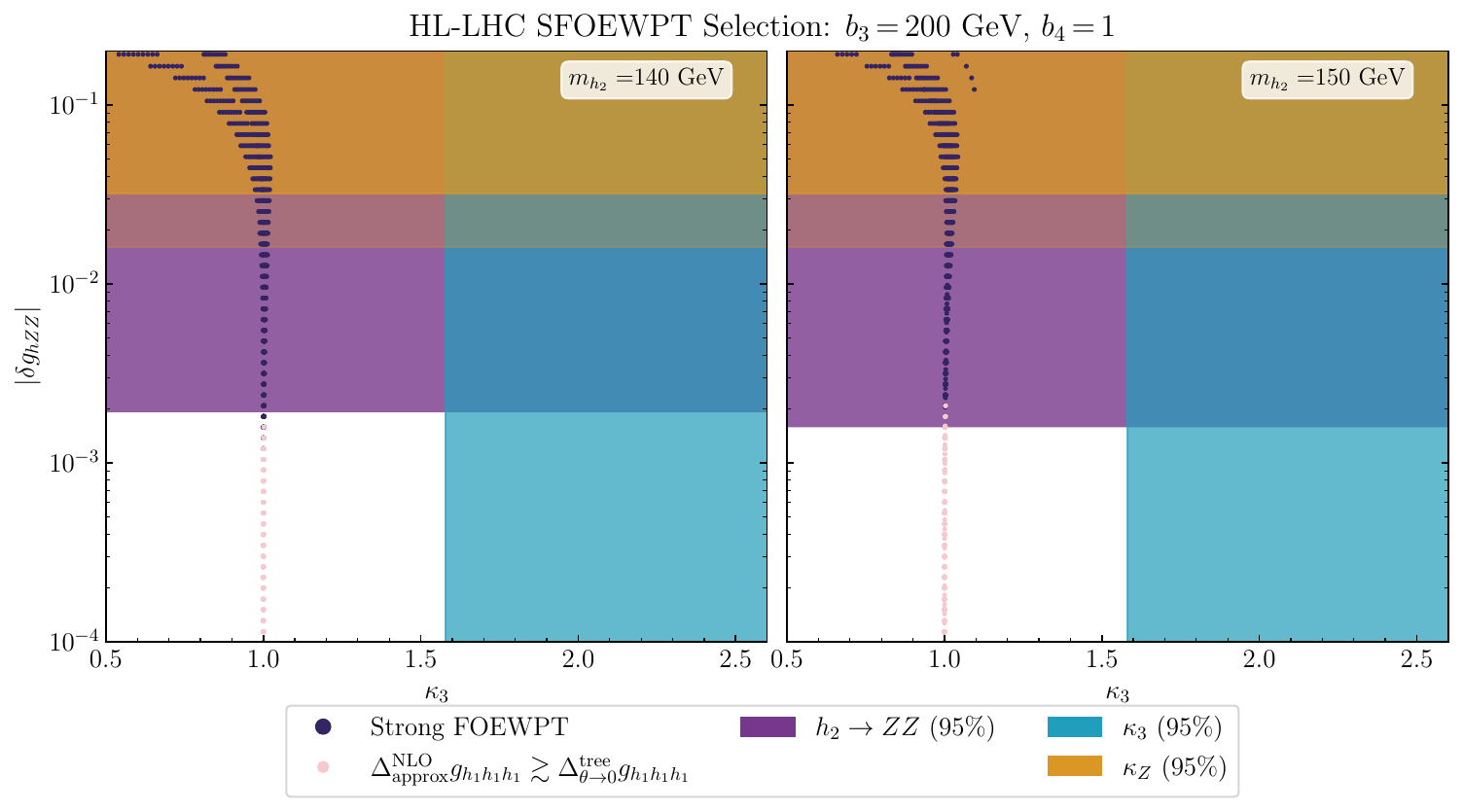}
    \caption{Same as Fig.~\ref{fig:mhl00_indirect} but for $\mtwo=140,150$ GeV and for the benchmark point $b_3=200\, {\rm GeV},b_4=1$.}
    \label{fig:mhl01_b3}
\end{figure}

This possibility is illustrated in Fig.~\ref{fig:mhl01_b3} for $\mtwo=140,150~\mathrm{GeV}$ and $b_4=1$. As shown there, the $\kappa_3<1$ region is largely excluded by the HL-LHC. However, some points survive close to the alignment limit, where the caveats concerning NLO effects discussed above apply. In particular, for such low masses, the finite contributions neglected in Eq.~\eqref{eq:NLOcond} may be numerically relevant, potentially making sizeable NLO corrections to $g_{h_1h_1h_1}$ important for more points than those highlighted in pink.

We close this section with a comment on the use of the NWA. Previous studies \cite{Arco:2025nii,Heinemeyer:2024hxa,Braathen:2025svl} have investigated how the width of a new scalar affects the total $S\to h_1h_1$ cross section, as well as the error induced by neglecting interference terms and retaining only the squared resonant contribution.\footnote{We use $S$ here to indicate a general scalar that could arise in more complicated models.} The exclusions shown in Figs.~\ref{fig:mhl00}-\ref{fig:mhl01_direct} are mainly driven by indirect probes and by resonant $ZZ$ production, making this issue less critical for the exclusion reach discussed here. On the other side an excess in the $ZZ$ channel could manifest itself also in the $h_1h_1$ channel. In order to correctly model this excess the full inclusion of interference terms in $h_1h_1$ production could be important even for $\Gamma_{h_2}/\mtwo\simeq O(1\%)$, while for smaller width to mass ratios (that would characterise a large part of the parameter space) further studies would be needed.

\subsection{FCC reach and constraints \label{sec:FCC}}
In this final section, we analyse the constraints from FCC-ee and FCC-hh on Higgs couplings, together with projected limits from resonant searches and the corresponding discovery reach.

The outstanding discovery potential of an $\mathcal{O}(100~{\rm TeV})$ machine has been explored for a long time~\cite{Kotwal:2016tex} (see Ref.~\cite{Profumo:2014opa} for corresponding early studies for future circular and linear $e^+e^-$ colliders). In this work, we consider the most recent FCC-hh running scenario, with a centre-of-mass energy of $84~{\rm TeV}$ and an integrated luminosity of $30~{\rm ab}^{-1}$, together with the most recent projections for exclusion limits and discovery reach~\cite{Selvaggi:2025kmd,stapf_2025_2kwzz-9dj08}. We further analyse the synergy between direct and indirect probes of the Higgs potential, emphasising how they complement each other.

For the FCC-ee projections, we extract a bound on $\theta$\footnote{As one can observe in~\cite{deBlas:2019rxi}, the $\kappa_Z$ extraction is not significantly improved by FCC-hh measurements.} from the projected sensitivity to $\kappa_Z$~ \cite{Selvaggi:2025kmd,deBlas:2025gyz,Armadillo:2026mvp}, obtaining
\begin{equation} \label{eq:FCCeetheta}
    \theta(\kappa_Z) < 0.066 \,(\rm 95\% \,C.L.)\,
\end{equation}
while for FCC-hh we assume \cite{Gallo:2024lin}
\begin{equation}
    \kappa_3 = 1.00 \pm 0.03 \,(\rm 68\%\, C.L.)\,
\end{equation}
even if more aggressive projections have been presented in the same paper.

\subsubsection{EWPO at FCC-ee}
It is well established that the HL-LHC determination of $\kappa_Z$ imposes a more stringent constraint on the mixing angle $\theta$ than the fit to Electroweak Precision Observables (EWPOs) performed at LEP \cite{Dawson:2021jcl}. The proposed TeraZ run of FCC-ee at a centre-of-mass energy of 91 GeV is expected to achieve unprecedented precision in EWPO measurements, naturally prompting the question of whether these observables could surpass the FCC-ee determination of $\kappa_Z$ itself in constraining $\theta$.

Projections in the $S$--$T$ Peskin--Takeuchi plane \cite{PhysRevD.46.381} are not officially provided for future colliders. However, their uncertainties and correlations can be extracted from the Briefing Book \cite{deBlas:2025gyz} by propagating the projected measurement errors on $M_W$ and $\sin^2 \theta_{w,\rm eff}$,
\begin{align}
\Delta M_W & = \frac {\alpha}{c_w^2 - s_w^2 } \frac{M_W}{2}\left (- \frac {\Delta S}{2}+ c_w^2 \Delta T +\frac {1} {4} \frac{(c_w^2-s_w^2)}{s_w^2} \Delta U\ \right)\\
\Delta\sin^2 \theta_{w,\rm eff} &= \frac{\alpha}{4} \frac{1}{ (c_w^2 - s_w^2)}( \Delta S - 4 s_w^2 c_w^2 \Delta T) 
\end{align}
setting $\Delta U=0$ one obtains,\footnote{We have verified a posteriori for LEP, where the full $STU$ covariance matrix is known, that the inclusion of U does not change sensitively the $\theta_{\rm max}$ bound.}

\begin{equation}
{\rm Cov}^{-1}(S,T) = 
\begin{pmatrix}
\sigma_{M_W} &\sigma_{s_w^2}
\end{pmatrix}
\begin{pmatrix}
\frac{\partial }{\partial S} \Delta M_W &\frac{\partial }{\partial T} \Delta M_W \\
\frac{\partial }{\partial S}\Delta\sin^2 \theta_{w,\rm eff} &\frac{\partial}{\partial T} \Delta\sin^2 \theta_{w,\rm eff} 
\end{pmatrix}^{-1}_{S,T=0}
\begin{pmatrix}
\sigma_{M_W} \\\sigma_{s_w^2}
\end{pmatrix},
\end{equation}
where $\sigma_{M_W}$ and $\sigma_{s^2_w}$ denote the uncertainties on $\Delta M_W$ and $\sin^2 \theta_{w,\rm eff}$, including the statistical, systematic and theoretical components presented in Ref.~\cite{deBlas:2025gyz}. The theoretical uncertainties are considered in two error-reduction scenarios, labelled ``conservative'' and ``aggressive''.
The projected future uncertainties on the $S,T$ parameters and their correlation matrix in the two scenarios result
\begin{align}
\begin{cases}
\sigma(\Delta S)_{\rm cons} &= 4.6\cdot 10^{-3} \\
\sigma(\Delta T)_{\rm cons} &= 4.8\cdot 10^{-3}
\end{cases}&&
{\rm Corr}(S,T)_{\rm cons}=
\begin{pmatrix}
1 & 0.926 \\
0.926  & 1
\end{pmatrix}\,,\\
&&\nonumber\\
\begin{cases}
\sigma(\Delta S)_{\rm aggr} &= 8.6\cdot 10^{-4} \\
\sigma(\Delta T)_{\rm aggr} &= 1.1\cdot 10^{-3}
\end{cases}&&
{\rm Corr}(S,T)_{\rm aggr}=
\begin{pmatrix}
1 & 0.961 \\
0.961  & 1
\end{pmatrix}\,.
\end{align}

The only modifications to the SM $W$ and $Z$ boson self-energies in the xSM arise from the modified $g_V$ Higgs-boson couplings in~\eqref{eq:SMcouplingmod} and from the additional loop involving the new scalar $h_2$, which has the same functional form as the Higgs-boson contribution, but with different couplings and mass. As already observed in \cite{Profumo:2014opa} the BSM contribution to Peskin-Takeuchi parameters can be expressed as
\begin{equation}
\Delta \mathcal{O} = \sin^2\theta\, (f_{\mathcal O}(\mtwo) -f_{\mathcal O}(\mone))\,.
\end{equation}
We compute the $f_{\mathcal O}$ using the expression for the renormalised weak boson self-energies present in Appendix B in \cite{Denner:1991kt}.\footnote{An overall minus sign must be included in passing from the convention of this reference to those in  Ref.~\cite{PhysRevD.46.381}.}
The analytical forms of the $f_{\mathcal O}(m)$ functions read
{\small\begin{align*}
f_{S}(m) &=\frac{1}{12\pi} \bigg[
(2m^2-10 M^2_Z)B_0'(M_Z^2) -B_0(M_Z^2)
 \nonumber\\
&\left.+\frac{{m^2+M^2_Z}}{(m^2-M^2_Z)^3}\left[m^2 A_0(M_Z^2)-M^2_Z A_0(m^2)\right]-\frac{1}{6} -\frac{2}{3} \frac{   m^6-M_Z^6}{(m^2 - M_Z^2)^3}\right]\,,\\
f_{T}(m) &= \frac{1}{48\pi s_w^2}\frac{1}{M_W^2}\sum_{V=W,Z}(-1)^{i_V}\left[(2m^2-10M^2_V)B_0(M_V^2)-(m^2-M_V^2)^2B_0'(M_V^2)\right] \,, \\
f_{U}(m) &= \frac{1}{12\pi}\sum_{V=W,Z} (-1)^{i_V}\bigg[(2m^2-10M_V^2)B_0'(M_V^2)-B_0(M_V^2)\nonumber\\
&\left.+\frac{m^2+M^2_V}{(m^2-M^2_V)^3}\left[m^2 A_0(M_V^2)-M^2_V A_0(m^2)\right]-2\frac{M_V^2m^2}{(m^2-M_V^2)^2}\right]\,,
\end{align*}}%
 with $i_W=0, \,i_Z=1$ and $B_0(M_V^2):=B_0(0,M_V^2,m^2)$ with the Passarino-Veltman defined as in \texttt{LOOPTOOLS}~\cite{Hahn:1998yk} and evaluated with the same tool. 
 
\begin{figure}
     \centering
\includegraphics[width=0.6\linewidth]{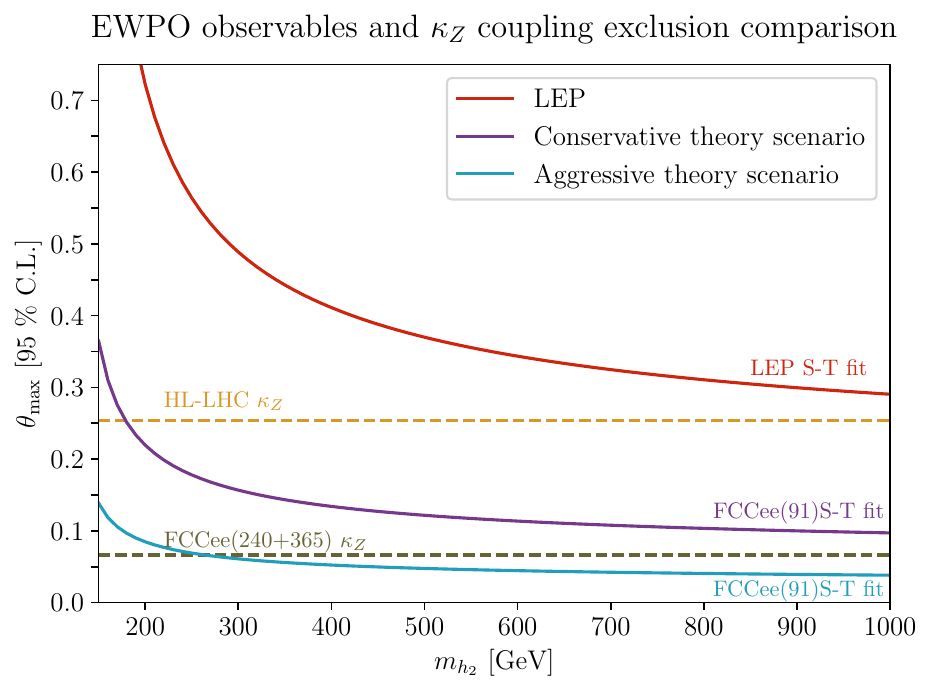}
     \caption{95\% C.L. on the mixing angle $\theta$ using EWPO from LEP fit \cite{10.1093/ptep/ptaa104} and projections for the FCC-ee conservative and aggressive scenario \cite{deBlas:2025gyz}. To facilitate the comparison we included dashed lines showing the upper bound on $\theta$ obtained considering the $Z$-boson Higgs coupling measurements from HL-LHC (\ref{eq:thetabound}) and FCC-ee (\ref{eq:FCCeetheta}). }
     \label{fig:EWPO}
 \end{figure}

 The results of the fit are shown in Fig.~\ref{fig:EWPO}. We successfully reproduce the LEP results of \cite{Dawson:2021jcl} using the same input \cite{10.1093/ptep/ptaa104} once the y-axis of Fig.~\ref{fig:EWPO} is correctly scaled. The HL-LHC precise extraction of $\kappa_Z$ beats LEP sensitivity via the $S-T$ joint fit, however in passing to FCC-ee the situation changes, and depending on the aggressive (conservative) reduction of current theory uncertainties  the FCC-ee measurement of $\kappa_Z$ can be a stronger (weaker) probe than the $ST$ simultaneous fit for $\mtwo>250\, {\rm GeV}$, while for smaller masses the former probe is always better.
 
\subsubsection{FCC results}
\begin{figure}
    \centering
    {\includegraphics[width=0.6\linewidth]{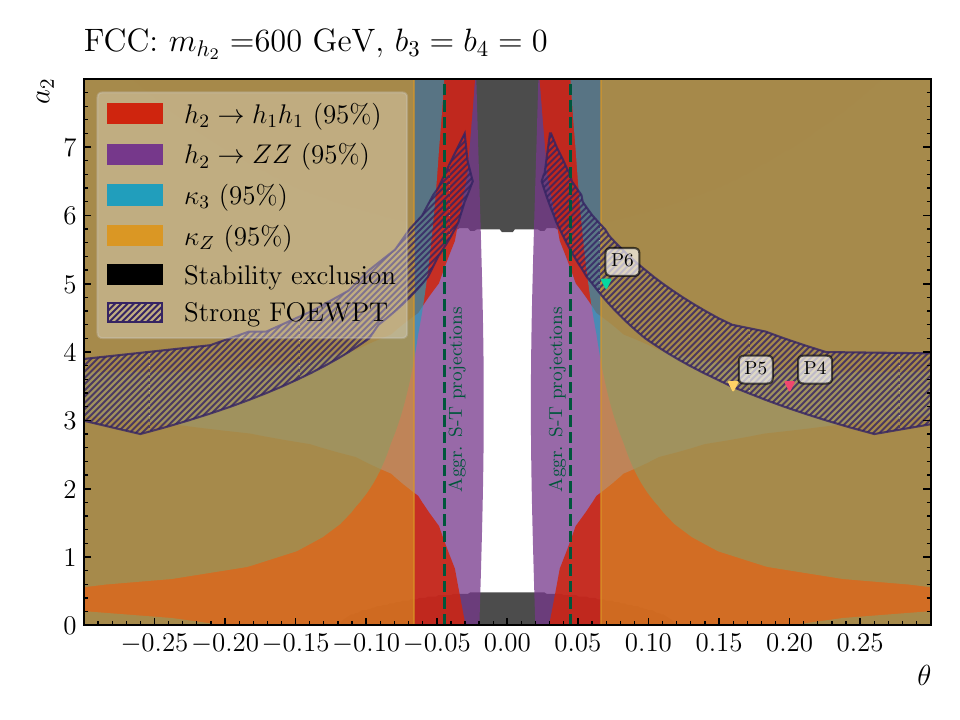}}
    \caption{
    Same as Fig.~\ref{fig:mhl00} but using FCC-ee+FCC-hh projections. The entire SFOEWPT region is completely covered by  resonant searches, resonant di-Higgs searches alone can now also cover the P6 point, unlike at HL-LHC.}
    \label{fig:FCC_exclusion}
\end{figure}

We begin by assessing the exclusion capability of the FCC-ee+FCC-hh in Fig.~\ref{fig:FCC_exclusion}. 
The fate of the benchmark points P4-P6 is clear: all of them can be excluded by precisely measuring the Higgs coupling, both the coupling to the $Z$ boson and the Higgs self-coupling. Furthermore, P6 can now be excluded also by looking at the di-Higgs decay.

\begin{figure}
    \centering
    {\includegraphics[width=0.8\linewidth]{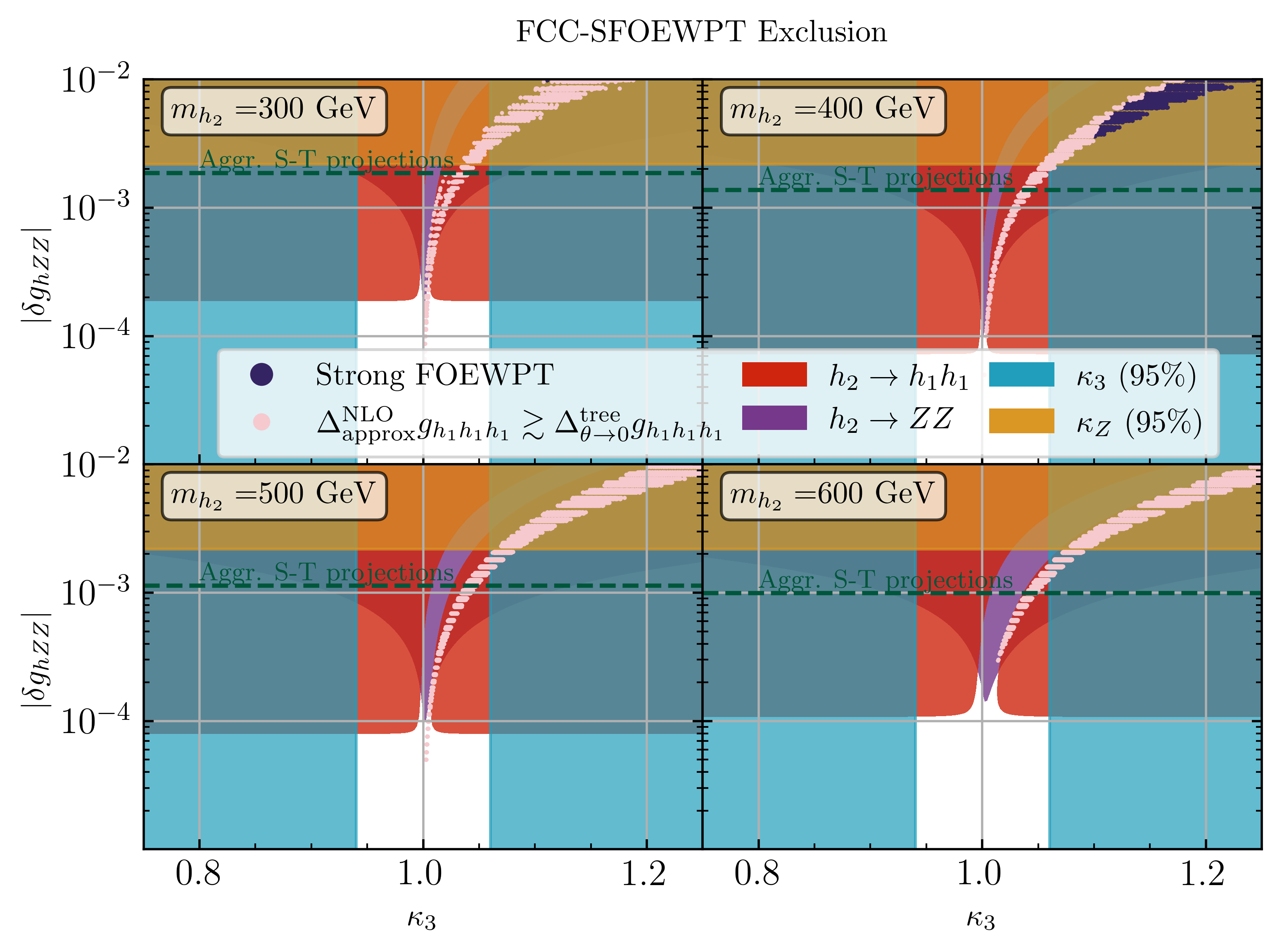}}
    \caption{
    FCC-ee+FCC-hh $95\%$ exclusion in the $\kappa_3-\delta g_{hZZ}$ plane, in each sub-panel the mass of the new scalar is fixed and we plot the SFOEWPT points for all values of $b_3$ and $b_4$ explored. The few regions that remain apparently out of reach are populated only by points satisfying Eq.~\eqref{eq:NLOcond}. For these points, the discovery potential should therefore be assessed through a more precise beyond-LO calculation.} 
    \label{fig:FCC_exclusion-quartet}
\end{figure}

More generally,  comparing the FCC-ee+FCC-hh results with the corresponding HL-LHC ones shown in the bottom-right panel of  Fig.~\ref{fig:mhl00}, we observe that the FCC-ee+FCC-hh programme is able to fully exclude this benchmark point. Interestingly, this additional sensitivity does not originate from FCC-ee. Indeed, a comparison with Fig.~\ref{fig:FCC_exclusion-quartet} shows that HL-LHC resonant searches are already competitive with the extremely precise Higgs coupling measurements expected at FCC-ee.

The same figure highlights the exceptional exclusion power provided by FCC-hh resonant searches. In Fig.~\ref{fig:FCC_exclusion-quartet} we display all SFOEWPT points obtained by varying $b_3$ and $b_4$ in the $(\kappa_3,\delta g_{hZZ})$ plane at fixed mass $\mtwo$.  A large fraction of the parameter space with $\theta \gtrsim 10^{-1}$ remains accessible even when the branching ratio of the new scalar into a Higgs-boson pair is highly suppressed, ${\rm Br}(h_2 \to h_1 h_1) < 2\%$, as can occur for specific combinations of $\mtwo$ and $a_2$.

The FCC-hh reach is enhanced both by the increased signal cross section, driven by the gluon parton distribution function at low momentum fraction, and by the projected luminosity, which is an order of magnitude larger than that of the HL-LHC. Only a limited set of points, clustered near the alignment limit and all satisfying Eq.~\eqref{eq:NLOcond}, remains beyond the FCC-hh reach. As already noted in the HL-LHC discussion, however, a higher-order analysis is required to establish whether these points would genuinely evade detection.

As shown in the previous section, the HL-LHC has strong sensitivity to the parameter space of the scalar singlet model. Nevertheless, it is equally important to consider the scenario in which the HL-LHC observes an excess in one or more decay channels, or a significant deviation in precision Higgs measurements. With this motivation, we study the discovery reach of FCC-hh, focusing in particular on the $h_2 \to h_1 h_1$ channel~\cite{Kotwal:2016tex} and on the Higgs self-coupling $\kappa_3$ as an indirect probe. 

This choice is motivated by considerations already emphasised in the previous sections: while the decay of the heavy scalar into two $Z$ bosons and deviations in $\kappa_Z$ constitute powerful probes, they are not unique signatures of this model. In order to directly assess the possibility of electroweak-symmetry-breaking-induced baryogenesis, it is essential to probe the structure of the Higgs potential itself, and hence to focus on Higgs-related observables.

At present, no detailed theoretical predictions analogous to those of Ref.~\cite{TwikiBSM} are publicly available for a centre-of-mass energy of 84~TeV and for a Higgs boson with arbitrary mass. In order to account for higher-order corrections, we therefore extract a $K$-factor,
\begin{equation}
K \;=\; \frac{\sigma_{\rm 80~TeV}^{\rm NNLO~QCD+NLO~EW}(gg\to h)}
{\sigma_{\rm 80~TeV}^{\rm LO}(gg\to h)} \, ,
\end{equation}
using the gluon-fusion cross section at NNLO~QCD+NLO~EW accuracy at 80~TeV from Ref.~\cite{TwikiHES}, and simulating the LO $gg\to h_1$ production with our \texttt{UFO} implementation. We have verified that, in the mass range of interest, this $K$-factor exhibits only a mild dependence on the Higgs-boson mass, by comparing our LO results at 13 and 14~TeV with the corresponding NNLO~QCD+NLO~EW predictions reported in Ref.~\cite{TwikiBSM}.

\begin{figure}
    \centering
    \includegraphics[width=0.48\linewidth]{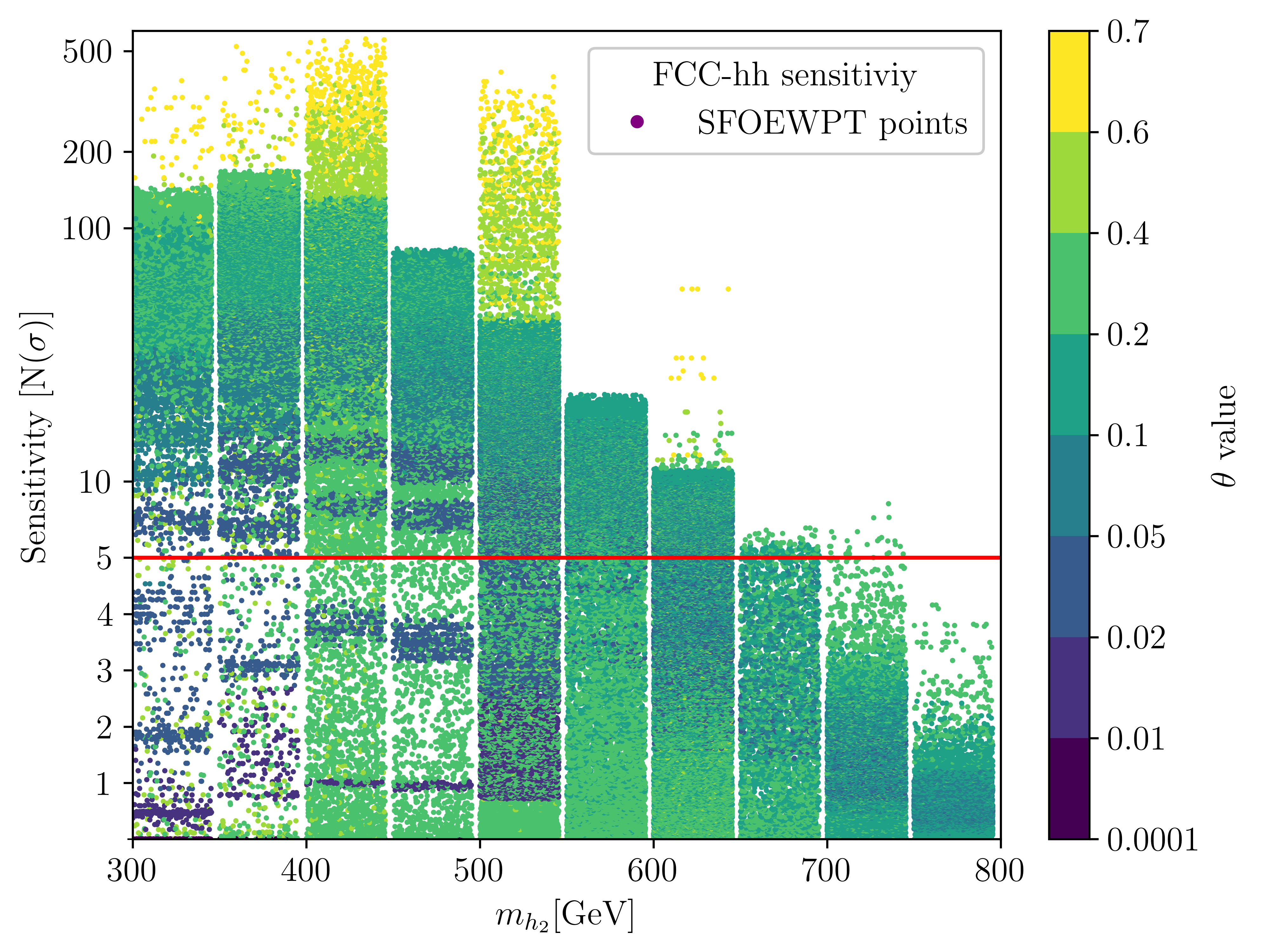}
    \includegraphics[width=0.48\linewidth]{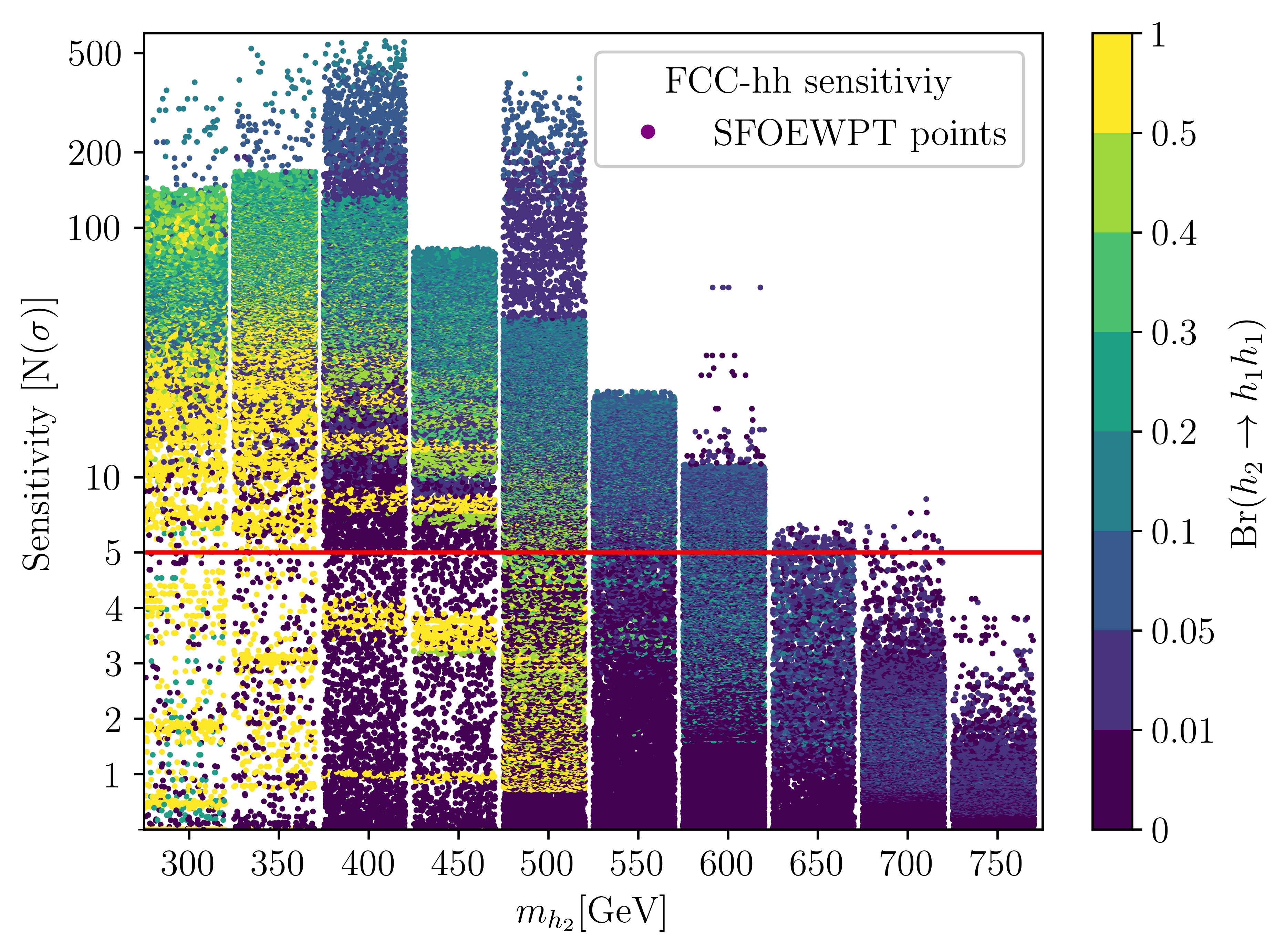}
    \caption{FCC-hh discovery reach for each point giving rise to a SFOEWPT in our scan. The vertical axes show the sensitivity, expressed in units of $\sigma$, for different benchmark points in the scalar mass $m_{h_2}$ on the horizontal axis. \textbf{Left:} points coloured by the mixing angle $\theta$, \textbf{Right:} points coloured by branching ratio Br$(h_2\rightarrow h_1 h_1)$.}
    \label{fig:h2h1h1_reach}
\end{figure}

We present the results of the $gg \to h_2 \to h_1 h_1$ analysis using the latest FCC-hh projections from Ref.~\cite{stapf_2025_2kwzz-9dj08} in Fig.~\ref{fig:h2h1h1_reach}. From the figure, we observe that resonant searches can probe a significant portion of the parameter space; however, their reach can be limited by two distinct effects. First, the branching ratio ${\rm Br}(h_2 \to h_1 h_1)$ can be strongly suppressed, as predicted by Eq.~\eqref{eq:211approx_a2}, and may even approach zero. Second, even for very large branching ratios (yellow points in the right panel of Fig. \ref{fig:h2h1h1_reach}), a strong first-order phase transition can occur for very small values of the mixing angle $\theta$. In this case, the new scalar becomes naturally elusive, since the production cross section $gg \to h_2$ is suppressed as $\theta^2$ for small mixing.

The first limitation can be partially alleviated by noting that large values of $\theta$ typically induce sizeable deviations in $\kappa_Z$, whose absence would therefore exclude this region. However, should a deviation be observed, it could not be uniquely attributed to this model without additional information, as the resonant sensitivity would be lacking. This further underscores the importance of Higgs-related observables in translating potential signals of new physics into a consistent interpretation in terms of UV-complete models.

\begin{figure}
    \centering
    \includegraphics[width=0.48\linewidth]{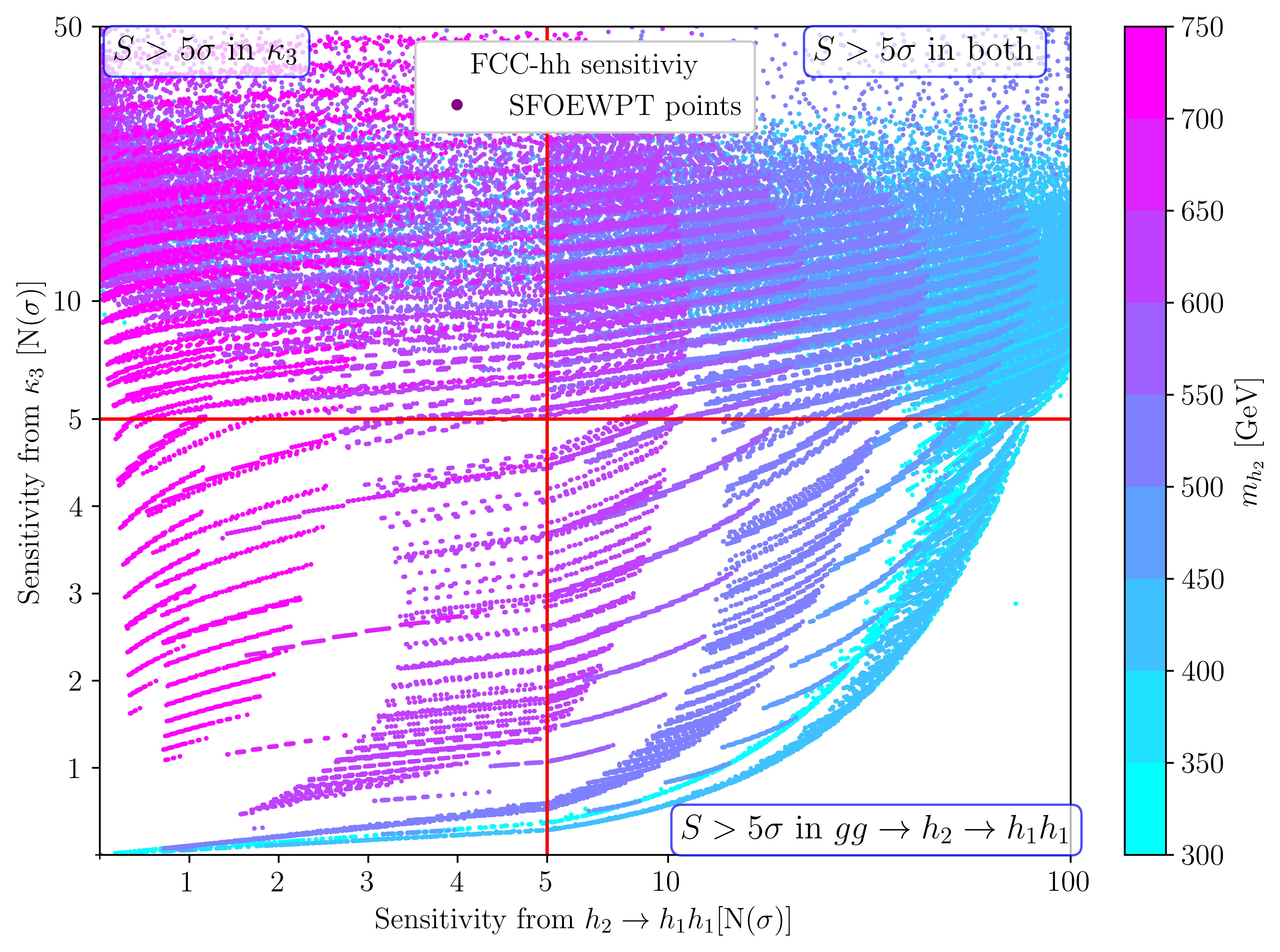}
    \includegraphics[width=0.48\linewidth]{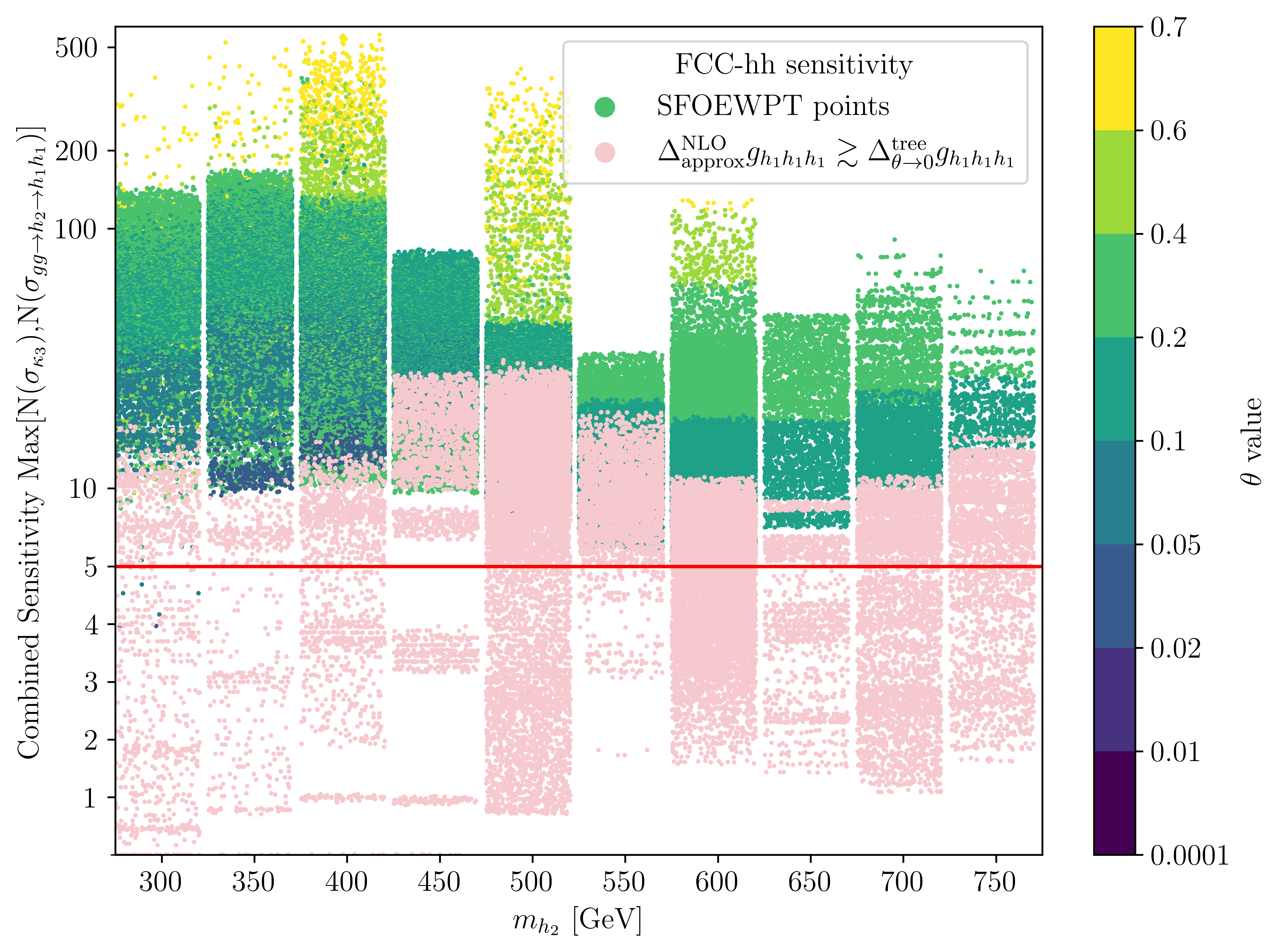}
    \caption{\textbf{Left:} Strong FOEWPT points in the two-dimensional significance  plane for $h_2\rightarrow h_1h_1$ and $\kappa_3$, points are coloured according to the mass of the scalar $m_{h_2}$. \textbf{Right:}  discovery reach of FCC-hh for the scalar singlet combining the different Higgs observables $\kappa_3$ and $h_2\rightarrow h_1    h_1$. We observe that almost all points below the $5\sigma$ discovery threshold satisfy Eq.~\eqref{eq:NLOcond}. For these points, a beyond-LO analysis is therefore required to reliably assess the actual discovery potential.
    }
    \label{fig:combined_reach}
\end{figure}

The sensitivity of resonant searches can be further improved along two main directions. 
First, one may consider the combination of multiple production modes, such as VBF, associated production with a vector boson ($Vh_2$), or with top-quark pairs ($t\bar t h_2$), for which dedicated projections are not yet available. 
Second, as anticipated in the introduction to this section, one can exploit the synergy between direct and indirect probes, in particular through modifications of the Higgs potential encoded in the measurement of $\kappa_3$. 
Assuming again a $3 \%$ sensitivity on this parameter, we show in the left panel of Fig.~\ref{fig:combined_reach} the  points in a two-dimensional sensitivity plane, with the deviation in the resonant $h_2 \to h_1 h_1$ channel on the horizontal axis and the sensitivity to $\kappa_3$ on the vertical axis. The $\kappa_3$ deviations are treated in a simplified way, assuming the standard deviation to remain constant independently of the experimentally measured value. A more realistic approach would be the one followed in \cite{CMS:2026zsp} in which the error on the extraction  of $\kappa_3$ is computed for each $\kappa_3^{\rm true}$ considered.

Not only does a $5\sigma$ deviation appear simultaneously in both channels for a large fraction of the parameter space, but a precise measurement of $\kappa_3$ alone would also probe a wide region where direct searches are limited. 

By combining direct and indirect probes and considering, point by point, the maximal sensitivity between the $\kappa_3$ indirect constraint and the $h_2 \to h_1 h_1$ resonant search, we obtain the results shown in the right panel of Fig.~\ref{fig:combined_reach}. Including information from $\kappa_3$ measurements significantly extends the discovery reach to a visibly larger region of the parameter space, particularly at higher masses. In many cases, these observations would be further corroborated by resonant $ZZ$ searches, enabling both the discovery and identification of this model at FCC-hh. A final comment concerns the points that seem unreachable even at FCC-hh, lying below the $5\sigma$ threshold in both channels. Previous studies \cite{Arco:2025nii,Heinemeyer:2024hxa,Braathen:2025svl} have shown that the inclusion of interference terms in $gg \, ( \,\rightarrow h_2) \rightarrow  h_1 h_1$, neglected in this paper, can enhance the sensitivity in the di-Higgs spectrum and this could therefore be one way to boost the experimental reach. 

Furthermore, as shown in the right panel of Fig.~\ref{fig:combined_reach}, almost all points outside the projected discovery reach lie close to the alignment limit, $\theta \to 0$. In this region, the BSM loop-induced contributions to the Higgs trilinear coupling are larger than the corresponding tree-level ones, according to the criterion in Eq.~\eqref{eq:NLOcond}. As emphasised above, for these points the LO-defined $\kappa_3$ is not a reliable probe by itself. A higher-order calculation of non-resonant di-Higgs production is therefore required to determine the actual discovery potential in this challenging region of parameter space.

\section{Conclusions and outlook \label{sec:con}}

We have revisited the real scalar singlet extension of the Standard Model, focusing on its ability to realise a strong first-order electroweak phase transition, one of the key ingredients for electroweak baryogenesis, while also improving the stability of the Higgs vacuum up to high energy scales. Despite its minimal field content, the model exhibits a rich phenomenology arising from the interplay between scalar mixing, the structure of the Higgs potential, and the dynamics of the electroweak phase transition.

We have systematically analysed the viable parameter space by combining complementary theoretical and experimental constraints. On the theoretical side, we employed a dimensionally reduced three-dimensional effective theory to delineate the regions supporting a SFOEWPT, benchmarked with selected lattice simulations, and imposed vacuum stability up to high scales using renormalisation-group evolution. On the experimental side, we combined indirect probes from precision Higgs measurements with direct searches for a heavy scalar resonance at colliders. For the latter, we verified that, over most of the phenomenologically relevant parameter space considered here, the narrow width approximation remains valid, allowing resonant searches to be consistently interpreted.

Our analysis shows that a SFOEWPT can be realised over a wide range of singlet-like scalar masses, extending from just above the SM Higgs boson up to masses of nearly $1~\text{TeV}$, as expected based on earlier studies and the general arguments given in Ref.~\cite{Ramsey-Musolf:2019lsf}. We have phrased our results mainly in terms of projected exclusions, for simplicity and to ensure a uniform treatment of the different signatures and constraints. A dedicated discovery-reach analysis would be required for a quantitative assessment, but our results already point to clear discovery potential. In particular, an excess in indirect probes could be mapped onto the mass and coupling ranges tested by resonant searches in $pp\to h_2\to h_1h_1,ZZ$. The main exception is the funnel region near alignment, where tree-level Higgs-coupling deviations are suppressed; probing this region will likely require exploiting loop-induced effects in di-Higgs production and deserves a dedicated analysis. Future gravitational-wave searches with LISA-level sensitivity could provide a complementary probe of this region, as illustrated in Fig.~\ref{fig:FOPTgw}.

An important outcome of our study is that deviations in the Higgs self-coupling $\kappa_3$ can be enhanced with respect to those in the Higgs--$Z$ coupling, see Eq.~\eqref{eq:k3dgz}, making $\kappa_3$ a particularly sensitive probe of this model. We find that sizeable deviations in $\kappa_3$ may already be observable at the HL-LHC in regions where deviations in $\kappa_Z$ are too small to be detected. Away from the narrow funnel close to alignment, the combination of direct searches, Higgs precision measurements, and stability requirements provides broad coverage of the SFOEWPT-compatible 
parameter space, see, for instance, Figs.~\ref{fig:mhl00}, \ref{fig:mhl01_direct}, \ref{fig:mhl01_b3}.

While the combination of stability requirements and HL-LHC searches can probe a large part of the parameter space compatible with a SFOEWPT, we have also explored the implications of a potential deviation from SM expectations. In such a case, future facilities such as the FCC and CEPC/SppC would be crucial for discovery and model discrimination. In particular, the synergy between direct resonant searches and indirect probes of the Higgs potential at FCC-hh could allow deviations to be interpreted in terms of a concrete modification of the Higgs sector.

Our analysis also highlights the importance of higher-order effects and effective-field-theory methods for connecting collider observables with early-Universe dynamics as well as the necessity of lattice studies to benchmark state-of-the-art perturbative computations. In particular, dimensional reduction provides a controlled framework for studying the phase transition, while higher-order corrections to di-Higgs production may be essential for assessing the testability of the near-alignment funnel region. Lattice benchmarking is essential for ensuring perturbative computations are applied in regimes where a sufficiently strong first-order EWPT is realised.

The real scalar singlet model studied here therefore provides a useful benchmark for a broader class of scalar extensions of the Standard Model. Natural future directions include a dedicated discovery-reach study, a more detailed investigation of gravitational-wave signatures, higher-order calculations for resonant and non-resonant di-Higgs production, and extensions including dark matter sectors or additional scalar states. 

\section*{Acknowledgements}
The work of Robert Delaunay inspired the choice of the colour palette used in the combination plots. We are grateful to P.~Mastrapasqua, M.~Selvaggi, B.~Stapf, and A.~Taliercio for providing us with the latest FCC-hh projections for resonant searches, and to the ATLAS and CMS collaborations for intense and fruitful interactions, as well as for the motivation for this study. We also thank A.~Mariotti and S.~Blasi for initial discussions, and Tom Steudtner for valuable assistance with the SM stability analysis. We are also grateful to Lauri Niemi, Tuomas Tenkanen, V.Q.~Tran, Yanda Wu, Wenxing Zhang and Jiang Zhu for helpful discussions on the electroweak phase transition and gravitational waves.

This research was supported in part by the National Science Foundation under Grant No.~PHY-2309135 to the Kavli Institute for Theoretical Physics (KITP), and by the Science and Technology Facilities Council (STFC) under the Consolidated Grant T/X000796/1 (DFL). ST is supported by a FRIA grant of the Belgian Fund for Research, F.R.S.-FNRS (Fonds de la Recherche Scientifique--FNRS). MJRM and GX were supported in part under National Natural Science Foundation of China (NSFC) grant No. 12375094. MJRM was also supported in part under NSFC grant No. W2441004.

\begin{appendices}

\section{Parametrisation in terms of the vev}
\label{app:vev}

An alternative form of the potential in (\ref{eq:xSM}) consists of explicitly assigning a vev $v_S$ to the singlet field. The scalar potential becomes:
\begin{equation}\label{eq:UVLagrgeneral}
V(\Phi,S) = -\overline{\mu_H}^2  \abs{\Phi}^2 +\overline\lH \abs{\Phi}^4 - \frac{\overline{\mu_S}^2}{2} S^2  +     \frac{\overline\lS}{4} S^4 +\frac{\overline\lM}{2}  \abs{\Phi}^2 S^2 + \frac{\overline\mthree}{3}S^3  + \overline\mfour \abs{\Phi}^2 S.
\end{equation}
The field decomposition after EWSB reads:
\begin{equation}
\Phi=\begin{pmatrix}
- iG^+ \\
(v_H + H + iG_0)/\sqrt{2}
\end{pmatrix}, \qquad
S = ( \vS + s).
\end{equation}

To avoid confusion, we use the same symbols between Eqs.~\eqref{eq:xSM} and \eqref{eq:UVLagrgeneral}. However, the numerical values of these parameters must be carefully mapped from one Lagrangian to the other (see Section \ref{sec:paramet}). We therefore use overlined symbols to remind the reader of this difference.

In this parametrisation, once the fields acquire a \textit{vev}, the EWSB Lagrangian includes tadpole terms. These can be removed by imposing $\mathcal{I}_s=\mathcal{I}_H=0$. Tadpole conditions now involve $v_S$:
\begin{align}
\omuH^2 &=\olH v_H^2+\frac{\olM}{2}\vS^2+v_S \omfour,\label{eq:uh}  \\
\omuS^2 &=\olS \vS^2+\frac{\olM}{2}v_H^2+\omthree v_S +\omfour\frac{v_H^2}{2 v_S}.\label{eq:us} 
\end{align}
The mass matrix in the $(H,s)$ basis is:
\begin{equation}
\begin{pmatrix}
H & s
\end{pmatrix}  
\begin{pmatrix}
2\olH v_H^2 & \quad v_H(\olM  \vS + \omfour) \\
v_H(\olM   \vS + \omfour) &  \quad2\olS \vS^2 + \omthree v_S - \omfour \frac{v_H^2}{2 v_S}
\end{pmatrix} 
\begin{pmatrix}
H \\ s
\end{pmatrix}  \,,
\end{equation}
as in Eq.~\eqref{eq:M2}. Keeping the form of Eq.~\eqref{eq:massmatrix} we identify
\begin{equation}\label{eq:massmixedTOTAL}
m_H^2 \equiv 2\olH v_H^2,    \qquad m_S^2 \equiv 2\olS \vS^2 + \omthree v_S - \omfour \frac{v_H^2}{2 v_S}, \qquad m_{HS}^2 \equiv\olM v_H v_S  + \omfour v_H\,, 
\end{equation}
The relations for physical masses and mixing angle are obtained analogously, leading to:
\begin{align}
\olM &= \frac{1}{ v_H v_S}\left[( \mtwo^2- \mone^2 ) \sin\theta\cos\theta-\omfour v_H \right] \,,\label{eq:a2bar}\\
\olH &= \frac{1}{4v_H^2} \left[\mone^2+\mtwo^2-(\mtwo^2-\mone^2)\cos 2\theta\right]\,,\label{eq:lamHbar}\\ 
\olS &= \frac{1}{4 v_S^2}\left[\mone^2 + \mtwo^2 - (\mone^2  - \mtwo^2)\cos2\theta\right]- \frac{\omthree}{2 v_S} + \omfour \frac{v_H^2}{4 v_S^3}\,.\label{eq:b4bar}
\end{align}
The input parameters are chosen to yield the simplest possible Feynman rules that are listed in the additional material.

\subsection{Change of parametrisation and redundancies}\label{sec:paramet}

Passing from one convention to the other is not immediate, as expected, the  dimensionless parameters remain the same 
\begin{equation}
\lS = \olS \,,
\lH = \olH\,, 
\lM = \olM\,,
\end{equation}
but the mapping from the scalar \textit{vev} $v_S$ to the linear term $b_1$ is less trivial.
Expanding the Lagrangian in Eq.~\eqref{eq:UVLagrgeneral} after EWSB
we can identify the linear term in $S$ with the corresponding $b_1$ term in Eq.~\eqref{eq:xSM}
\begin{equation}
b_1 =-{\overline {\mu_S} ^2} v_S + \omthree v_S^2 +\olS v_S^3 \,,\label{eq:b1vs}
\end{equation}
and the other terms in the \emph{explicit linear term} parametrisation can be expressed as
\begin{align}
a_1 &= \overline{a_1} + 2 \overline{a_2} v_S\,,\\
\mthree &= \omthree + 3 \olS v_S\,,  \label{eq:b3andvS}\\
\mu_S^2&= \overline{\mu_S^2} - 2 \omthree v_S - 3 \olS v_S^2\,,\\
\mu_H^2&= \overline{\mu_H^2} - \omfour v_S - \frac{\olM}{2} v_S^2\,. 
\end{align}
In reverse, it is possible to express the parameters in the vev parametrisation in terms of those in the explicit linear-term parametrisation, in this case 
\begin{equation}\label{eq:b1}
b_1 = -\mu_S^2 v_S - b_3 v_S^2 + b_4 v_S^3\,,
\end{equation}
while all the other terms translate as 
\begin{align}
\overline{a_1} &= a_1 - 2 a_2 v_S\,,\\
\omthree &= \mthree - 3 \lS v_S \,,\\
\overline{\mu_S^2} &= \mu_S^2 + 2 \mthree v_S - 3 \lS v_S^2\,,\\
\overline{\mu_H^2}&= \mu_H^2 + \mfour v_S - \frac{\lM}{2} v_S^2\,. 
\end{align}
Some comments are in order regarding the above equations and the different choices of free parameters between the two Lagrangian formulations.

To begin with, let us temporarily set aside the distinction between the two parametrisations and focus on Eqs.~\eqref{eq:a2bar}--\eqref{eq:b4bar}. One can, in fact, adopt an alternative set of free parameters by inverting these equations and expressing $v_S$ and $\overline{a_1}$ in terms of $\olM$ and $\olS$. Upon inverting Eq.~\eqref{eq:b4bar}, one encounters a cubic equation for $v_S$, for which, in general, no additional physical condition prevents the existence of three distinct real solutions for the scalar vacuum expectation value. Each solution is associated with a different value of $\overline{a_1}$. Consequently, the Feynman rules given in the additional material can be re-expressed in terms of the new parameters in three apparently distinct forms, depending on which solution for $v_S$ is chosen. However, once numerical values for the physical parameters are assigned, all three formulations yield identical contact interactions and amplitudes, independently of the chosen solution. This demonstrates the presence of a redundancy in the choice of $\overline{a_1}$ and $v_S$.

The origin of this redundancy can be traced back to Eqs.~\eqref{eq:b1vs} or~\eqref{eq:b1}, where the relation between the linear term and the scalar vacuum expectation value is itself dictated by a cubic equation. The presence of the explicit linear term in the Lagrangian highlights the fact that the singlet field $S$ can always be translated by a constant shift. Such a shift leads to a redefinition of the Lagrangian parameters without affecting any physical observables, as Lagrangians themselves are not physical objects. Consequently, different roots of the cubic equation correspond to different parameter choices, yet yield the same physical predictions.

\subsection{Couplings}

The different choices of free parameters, together with the intrinsically different formulations give rise to different expressions for $\kappa_3$ and $g_{211}$ in the \emph{explicit vev}
parametrisation 
\begin{equation}
\kappa_3=\sin ^3\theta  \left(-\frac{ v_H}{v_S}-\frac{1}{4}\frac{  \overline{a_1} v_H^3}{ \mone^2 v_S^2}+\frac{\omthree v_H}{3\mone^2}\right)+\frac{1}{2}\cos \theta  \left[ \cos 2 \theta +1+\frac{  \overline{a_1} v_H^2}{ 2v_S\mone^2}( \cos 2 \theta -1)\right] \label{eq:k3vs}\,. 
\end{equation}
While in the linear-term formalism only three parameters enter the expression for $\kappa_3$, see Eq.~\eqref{eq:k3lin}, we observe that in the \emph{explicit vev} parametrisation an additional parameter appears, namely $v_S$, leading to a dependence on $(\theta, \overline{a_1}, \overline{b_3}, v_S)$. This is not merely a matter of parameter choice. The additional dependence originates from the quartic self-coupling of the singlet scalar, $b_4$, which does not contribute to $\kappa_3$ as long as $S$ does not acquire a \textit{vev}. Conversely, once $S$ develops a vev, a trilinear interaction term proportional to $b_4\, v_S\, h_1^3$ is generated in the Lagrangian. As a result, even if one reparametrises the theory by trading $\overline{a_1} \leftrightarrow b_4$, the expression for $\kappa_3$ inevitably involves one additional parameter compared to the \emph{explicit linear term} parametrisation.
Within this formalism, we can also examine the linearised expression for
 $\kappa_3$
\begin{equation}
\kappa_3 \simeq 1- \left(\frac{3}{2}+\omfour \frac{v_H^2}{v_S \mone^2} \right)\theta ^2+ \left( \frac{v_H}{\mone^2} \left(\frac{\omthree}{3}- \frac{\overline{a_1}}{4} \frac{v_H^2}{v_S^2}\right) - \frac{v_H}{v_S}\right)\theta ^3+O\left(\theta ^4\right)\,, \label{eq:k3_vsapprox} 
\end{equation}
where we note the presence of an additional parameter with respect to Eq.~\eqref{eq:k3lin}.
Going back to the $h_2\rightarrow h_1 h_1$ decay, Eq.~\eqref{eq:211decay}, the $g_{211}$ term reads
\begin{eqnarray}
g_{211} &=& \sin \theta \bigg[\left(m_{h_1}^2+ \frac{m_{h_2}^2}{2}\right)\left(\frac{ \cos 2 \theta +1}{ v_H}+\frac{\sin 2 \theta}{v_S}\right)\nonumber\\
&-&\frac{1}{2} \omthree \sin 2 \theta   + \frac{\overline{a_1}}{8}\frac{ v_H }{v_S}\left(\frac{3 v_H }{ v_S}\sin 2 \theta+6 \cos 2 \theta+2\right)\biggr]\,,
\end{eqnarray}
again a much more cumbersome expression compared with the linear-term formalism, even in its linearised version:
\begin{equation}
  g_{211}   \simeq\frac{2}{v_H}\left( m_{h_1}^2 +\frac{m_{h_2}^2}{2}+\omfour\frac{v_H^2}{v_S}\right) \theta+ \left[\frac{3}{2} \omfour \frac{v_H^2}{v_S^2}-\omthree +\frac{2}{v_S}\left(
   m_{h_1}^2+\frac{m_{h_2}^2}{2}\right)\right]\theta ^2+O\left(\theta ^3\right)\,. 
\end{equation}

    \end{appendices}
\clearpage
\bibliographystyle{JHEP}

\bibliography{bibfr}
\end{document}